\documentclass[aps,reprint,superscriptaddress,notitlepage,longbibliography,nofootinbib]{revtex4-1}

\pdfoutput=1

\usepackage{amssymb}
\usepackage{graphicx}
\usepackage{makecell}
\usepackage{color}
\usepackage{xcolor}
\usepackage{amsmath}
\usepackage{bm}
\usepackage{url}
\usepackage[colorlinks=true, urlcolor=blue, citecolor=blue, anchorcolor=red]{hyperref}

\usepackage[normalem]{ulem}

\usepackage[section]{placeins} 
\usepackage{setspace} 
\usepackage{physics}
\usepackage{tikz}
\usetikzlibrary{quantikz}
\usepackage{adjustbox}

\usepackage{tabularx}
\usepackage{ragged2e} 
\usepackage{booktabs} 
\newcolumntype{L}[1]{>{\hsize=#1\hsize\RaggedRight} X}

\begin{document}

\title{Parameterized quantum circuits as machine learning models}

\author{Marcello Benedetti}
\email{marcello.benedetti@cambridgequantum.com}
\affiliation{Cambridge Quantum Computing Limited, CB2 1UB Cambridge, United Kingdom}
\affiliation{Department of Computer Science, University College London, WC1E 6BT London, United Kingdom}

\author{Erika Lloyd}
\affiliation{Cambridge Quantum Computing Limited, CB2 1UB Cambridge, United Kingdom}

\author{Stefan Sack}
\affiliation{Cambridge Quantum Computing Limited, CB2 1UB Cambridge, United Kingdom}

\author{Mattia Fiorentini}
\affiliation{Cambridge Quantum Computing Limited, CB2 1UB Cambridge, United Kingdom}

\date{October 10, 2019}

\begin{abstract}
Hybrid quantum-classical systems make it possible to utilize existing quantum computers to their fullest extent. Within this framework, parameterized quantum circuits can be regarded as machine learning models with remarkable expressive power. This Review presents the components of these models and discusses their application to a variety of data-driven tasks, such as supervised learning and generative modeling. With an increasing number of experimental demonstrations carried out on actual quantum hardware and with software being actively developed, this rapidly growing field is poised to have a broad spectrum of real-world applications. 
\end{abstract}

\maketitle

\section{Introduction}
\label{s:introduction}

Developments in material science, hardware manufacturing, and disciplines such as error-correction and compilation, have brought us one step closer to large-scale, fault-tolerant, universal quantum computers. However, this process is incremental and may take years. In fact, existing quantum hardware implements few tens of physical qubits and can only perform short sequences of gates before being overwhelmed by noise. In such a setting, much anticipated algorithms such as Shor's remain out of reach. Nevertheless, there is a growing consensus that noisy intermediate-scale quantum (NISQ) devices may find useful applications and commercialization in the near term~\cite{preskill2018quantum,mohseni2017commercialize}. As prototypes of quantum computers are made available to researchers for experimentation, algorithmic research is adapting to match the pace of hardware development.

Parameterized quantum circuits (PQCs) offer a concrete way to implement algorithms and demonstrate quantum supremacy in the NISQ era. PQCs are typically composed of fixed gates, e.g., controlled NOTs, and adjustable gates, e.g., qubit rotations. Even at low circuit depth, some classes of PQCs are capable of generating highly non-trivial outputs. For example, under well-believed complexity-theoretic assumptions, the class of PQCs called instantaneous quantum polynomial-time cannot be efficiently simulated by classical resources (see Lund \textit{et al.}~\cite{lund2017quantum} and Harrow and Montanaro~\cite{harrow2017quantum} for accessible Reviews of quantum supremacy proposals). The demonstration of quantum supremacy is an important milestone in the development of quantum computers. In practice, however, it is highly desirable to demonstrate a quantum advantage on applications.

The main approach taken by the community consists in formalizing problems of interest as variational optimization problems and use hybrid systems of quantum and classical hardware to find approximate solutions. The intuition is that by implementing some subroutines on classical hardware, the requirement of quantum resources is significantly reduced, particularly the number of qubits, circuit depth, and coherence time. Therefore, in the hybrid algorithmic approach NISQ hardware focuses entirely on the classically intractable part of the problem.

The hybrid approach turned out to be successful in attacking scaled-down problems in chemistry, combinatorial optimization and machine learning. For example, the variational quantum eigensolver (VQE)~\cite{peruzzo2014variational} has been used for searching the ground state of the electronic Hamiltonian of molecules~\cite{o2016scalable,kandala2017hardware}. Similarly, the quantum approximate optimization algorithm (QAOA)~\cite{farhi2014quantum} has been used to find approximate solutions of classical Ising models~\cite{Moll2017} and clustering problems formulated as MaxCut~\cite{otterbach2017unsupervised}. The focus of this Review is on hybrid approaches for machine learning. In this field, quantum circuits are seen as components of a \textit{model} for some data-driven task. \textit{Learning} describes the process of iteratively updating the model's parameters towards the goal.

\begin{figure*}
\onecolumngrid
\centering
\includegraphics[width=.85\textwidth]{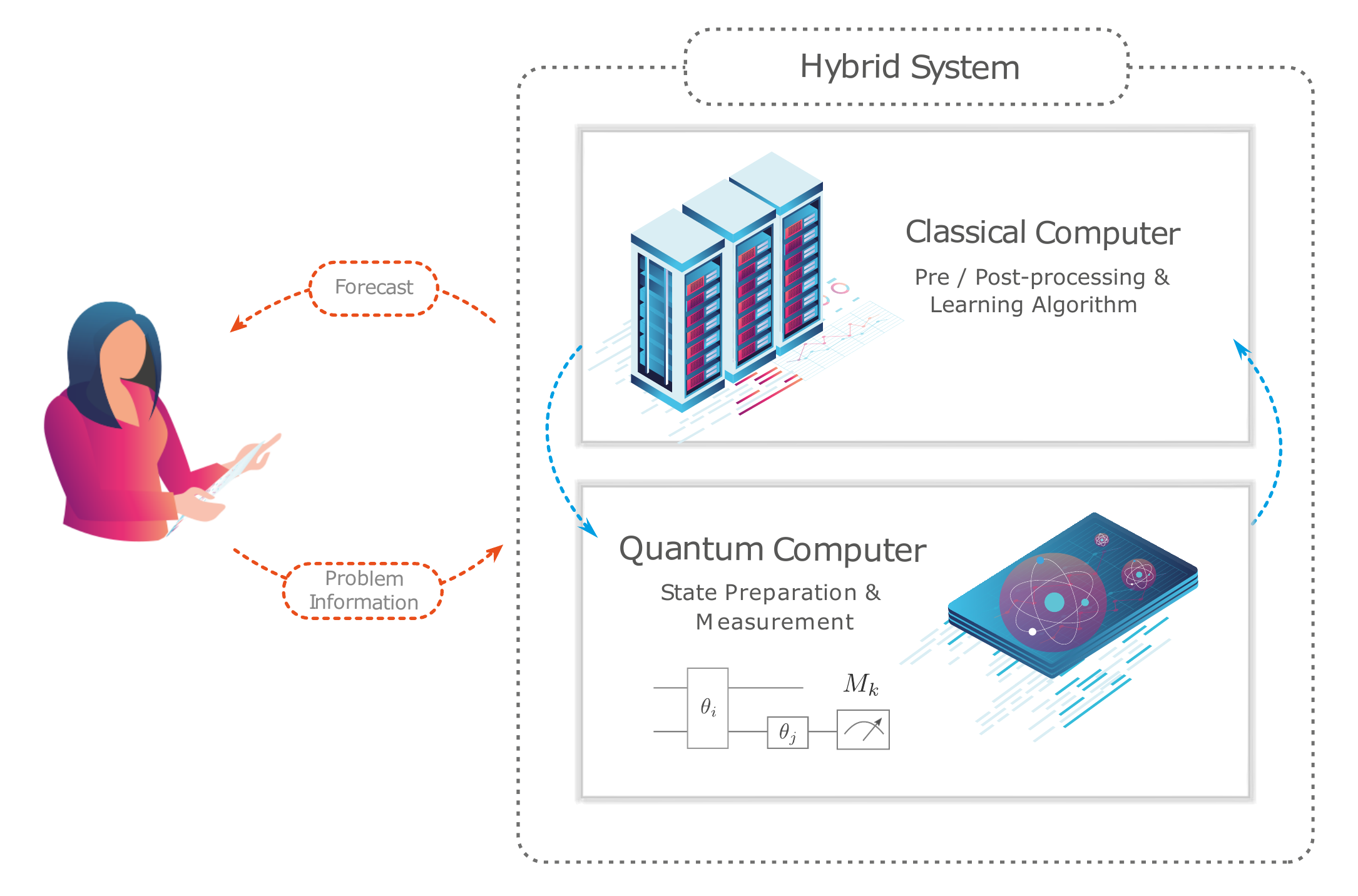}
\vspace{-5pt} 
\caption{High-level depiction of hybrid algorithms used for machine learning. The role of the human is to set up the model using prior information, assess the learning process, and exploit the forecasts. Within the hybrid system, the quantum computer prepares quantum states according to a set of parameters. Using the measurement outcomes, the classical learning algorithm adjusts the parameters in order to minimize an objective function. The updated parameters, now defining a new quantum circuit, are fed back to the quantum hardware in a closed loop.}
\label{fig:1}
\end{figure*}
\twocolumngrid
 
The general hybrid approach is illustrated in Fig.~\ref{fig:1} and is made of three main components: the human, the classical computer, and the quantum computer. The human interprets the problem information and selects an initial model to represent it. The data is pre-processed on a classical computer to determine a set of parameters for the PQC. The quantum hardware prepares a quantum state as prescribed by a PQC and performs measurements. Measurement outcomes are post-processed by the classical computer to generate a forecast. To improve the forecast, the classical computer implements a learning algorithm that updates the model's parameters. The overall algorithm is run in a closed loop between the classical and quantum hardware. The human supervises the process and uses forecasts towards the goal.

To the best of our knowledge, the earliest hybrid systems were proposed in the context of quantum algorithm learning. Bang \textit{et al.}~\cite{bang2008quantum} described a method where a classical computer controls the unitary operation implemented by a quantum device. Each execution of the quantum device is deemed as either a `success' or `failure', and the classical algorithm adjusts the unitary operation towards its target. Starting from a dataset of input-output pairs their simulated system learns an equivalent of Deutsch's algorithm for finding whether a function is constant or balanced. Gammelmark and M{\o}lmer~\cite{gammelmark2009quantum} took a more general approach in which the parameters of the quantum system are quantized as well. In their simulations they successfully learn Grover's search and Shor's integer factorization algorithms.

These early proposals attacked problems that are well known within the quantum computing community but much less known among machine learning researchers. More recently though, the hybrid approach based on PQCs has been shown to perform well on machine learning tasks such as classification, regression, and generative modeling. The success is in part due to similarities between PQCs and celebrated classical models such as kernel methods and neural networks. 

In the following Sections we introduce many of these multidisciplinary ideas, and we direct the Readers towards the relevant literature. Our style of exposition is pedagogical and not overly-technical, although we assume familiarity with basic machine learning definitions and methods (see Mehta \textit{et al.}~\cite{mehta2019high} for a physics-oriented introduction to machine learning), and basic working knowledge on quantum computing (see Nielsen and Chuang~\cite{nielsen2002quantum}, Chapter 2, for an introduction). We make use of several acronyms when referring to models and algorithms; to help the Reader we summarize all the acronyms in Table~\ref{tab:acronyms}. 

The structure of the Review is as follows: in Section~\ref{s:framework} we describe the components of machine learning models based on PQCs and their learning algorithms; in Section~\ref{s:applications} we describe their applications to classical and quantum tasks; and in Section~\ref{s:outlook} we summarize the advantages of this approach and give an outlook of the field. 

\begin{table}[b]
\small
\begin{tabular}{>{\bfseries}l l }
MERA & multi-scale entanglement renormalization ansatz \\
NISQ & noisy intermediate-scale quantum \\
PAC & probably approximately correct \\
PQC & parameterized quantum circuit \\
QAE & quantum autoencoder \\
QAOA & quantum approximate optimization algorithm \\
QCBM & quantum circuit Born machine \\
QKE & quantum kernel estimator \\
QGAN & quantum generative adversarial network \\
SPSA & simultaneous perturbation stochastic approximation \\
TTN & tree tensor network \\
VQM & variational quantum model \\
VQE & variational quantum eigensolver \\
\end{tabular}
\caption{Acronyms used in this Review.}
\label{tab:acronyms}
\end{table}

\section{Framework}
\label{s:framework}

We assume the computer to be a closed quantum system. With $n$ qubits, its state can be described as a unit vector living in a complex inner product vector space $\mathbb{C}^{2^n}$. The computation always starts with a state of simple preparation in the computational basis, for example the product state $\ket{0}^{\otimes n}$ (when clear from the context we often drop the tensor notation and refer to this state simply as $\ket{0}$). A unitary operator $U$ is applied to the initial state producing a new state $U\ket{0}$. Here, the value of an observable quantity can be measured. Physical observables are associated with Hermitian operators. Let $M = \sum_i \lambda_i P_i$ be the Hermitian operator of interest, where $\lambda_i$ is the $i$-th eigenvalue and $P_i$ is the projector on the corresponding eigenspace. The \textit{Born rule} states that the outcome of the measurement corresponds to one of the eigenvalues and follows probability distribution $p(\lambda_i) = \tr( P_i U \dyad{0}{0} U^\dag )$. Plugging this in the definition of expectation values we obtain
\begin{equation}
\label{f:exp_val}
\expval{M} = \sum_i \lambda_i p(\lambda_i) = \tr( M U \dyad{0}{0} U^\dag ) .
\end{equation}

As we will see, one can exploit the probabilistic nature of quantum measurements to define a variety of machine learning models, and PQCs offer a concrete way to implement adjustable unitary operators $U$. 

\begin{figure*}
\includegraphics[width=.95\textwidth]{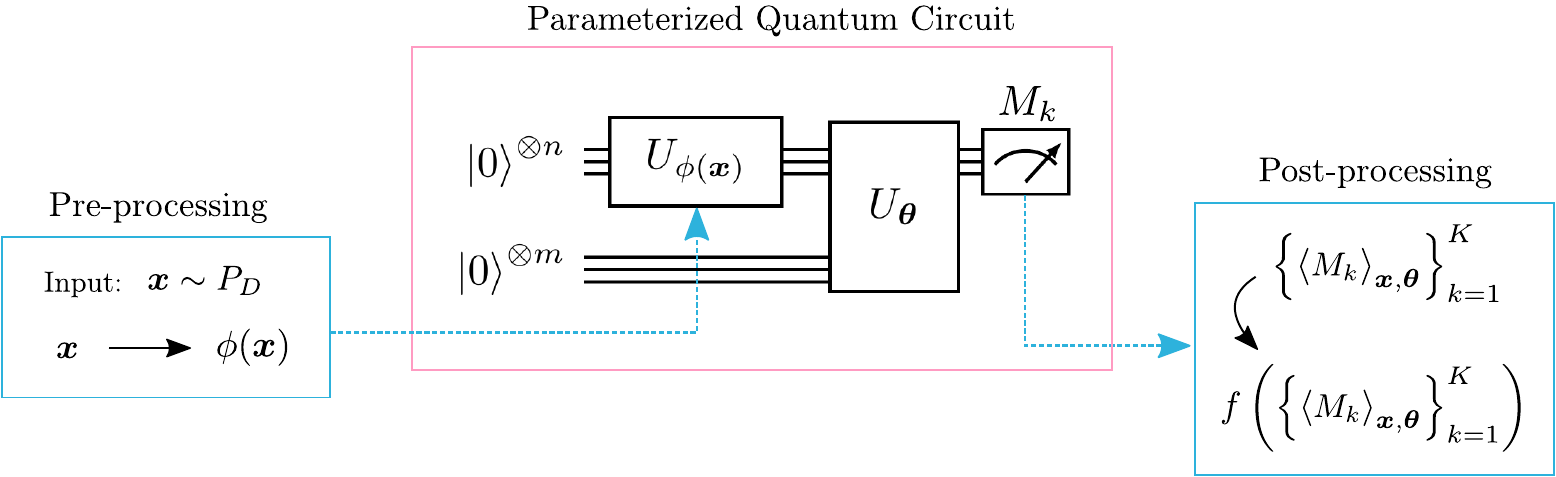}
\caption{A machine learning model comprised of classical pre/post-processing and parameterized quantum circuit. A data vector is sampled from the dataset distribution, $\bm{x}\sim P_D$. The pre-processing scheme maps it to the vector $\phi(\bm{x})$ that parameterizes the encoder circuit $U_{\phi(\bm{x})}$. A variational circuit $U_{\bm{\theta}}$, parameterized by a vector $\bm{\theta}$, acts on the state prepared by the encoder circuit and possibly on an additional register of ancilla qubits, producing the state ${ U_{\bm{\theta}}U_{\phi(\bm{x})}\ket{0}}$. A set of observable quantities ${\{ \expval{M_k}_{\bm{x},\bm{\theta}}\}_{k=1}^{K}}$ is estimated from the measurements. These estimates are then mapped to the output space through classical post-processing function $f$. For a supervised model, this output is the forecast associated to input $\bm{x}$. Generative models can be expressed in this framework with small adaptations.}
\label{f:PQC}
\end{figure*}

Figure~\ref{f:PQC} shows the components of a supervised learning model based on a PQC. First, a data vector is sampled from the training set and transformed by classical pre-processing, for example with de-correlation and standardization functions. Second, the transformed data point is mapped to the parameters of an \textit{encoder circuit} $U_{\phi(\bm{x})}$. Third, a \textit{variational circuit} $U_{\bm{\theta}}$, which possibly acts on an extended qubit register, implements the core operation of the model. This is followed by the estimation of a set of expectation values $\{ \expval{M_k}_{\bm{x},\bm{\theta}} \}_{k=1}^{K}$ from measurements~\footnote{The number of repetitions required for the estimation of each term is determined by the desired precision as well as by the variance $\text{Var}(M_k) = \expval{M_k^2} - \expval{M_k}^2$. In this Review we won't discuss estimation methods.}. 

A post-processing function $f$ is then applied to this set in order to provide a suitable output for the task. As an example, if we were to perform regression, $f$ could be a linear combination of the kind $\sum_k w_k\expval{M_k}_{\bm{x},\bm{\theta}}$, with additional parameters $w_k$. Note that it is possible to parameterize and train all the components of the model, including pre- and post-processing functions. 

Many of the proposals found in the literature fit within this framework with very small adaptation. We now describe the encoder and variational circuits in detail and explain their links to other well-known machine learning models. 

\subsection{The encoder circuit \texorpdfstring{$U_{\phi(\pmb{x})}$}{}}
\label{s:encoder_circuit}

There are several ways to encode data into qubits and each one provides different expressive power. This choice of encoding is related to kernel methods, a well-established field whose goal is to embed data into a higher dimensional feature space where a specific problem may be easier to solve. For example, non-linear feature maps change the relative position between data points such that a dataset may become easier to classify in the feature space. In a similar way, the process of encoding classical data into a quantum state can be interpreted as a feature map ${ \bm{x} \rightarrow U_{\phi(\bm{x})}\ket{0}^{\otimes{n}} }$ to the high-dimensional vector space of the states of $n$ qubits. Here, $\phi$ is a user-defined pre-processing function which transforms the data vector into circuit parameters. 

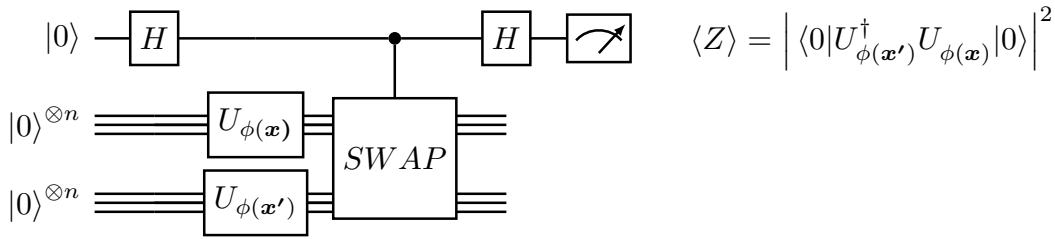
\begin{figure*}
\centering
\begin{adjustbox}{width=.8\textwidth}
\begin{tikzcd}
\lstick[1]{$\ket{0}$} & \gate{H} & \qw & \ctrl{1} & \gate{H} & \meter{} &\rstick{$\expval{Z} = \abs{ \mel{0}{ U^\dagger_{\phi(\bm{x^\prime})} U^{\phantom{\dagger}}_{\phi(\bm{x})} } {0}}^2 $} \\
\lstick[1]{$\ket{0}^{\otimes n}$} & \qwbundle[alternate]{} & \gate{U_{\phi(\bm{x)}}} \qwbundle[alternate]{} & \gate[wires=2]{SWAP} \qwbundle[alternate]{} & \qwbundle[alternate]{} & \\
\lstick[1]{$\ket{0}^{\otimes n}$} & \qwbundle[alternate]{} & \gate{U_{\phi(\bm{x^\prime})}} \qwbundle[alternate]{} & \qwbundle[alternate]{} & \qwbundle[alternate]{} & \\
\end{tikzcd}
\end{adjustbox}
\caption{The SWAP test can be used to estimate the implicit kernel implemented by an encoder circuit. Measurements of the $Z$ Pauli observable on the ancilla qubit yield the absolute value squared of the inner product between $U_{\phi(\bm{x})}\ket{0}$ and $U_{\phi(\bm{x^\prime})}\ket{0}$, respectively encoding data points $\bm{x}$ and $\bm{x}^\prime$. The SWAP test finds several applications in machine learning and is a ubiquitous routine in quantum computing in general.}
\label{f:swap_test}
\end{figure*}

The inner product of two data points in this space defines a similarity function, or kernel, ${ k(\bm{x},\bm{x}^\prime)=\abs{\bra{0} U^\dag_{\phi(\bm{x^\prime})} U^{\phantom{\dagger}}_{\phi(\bm{x})} \ket{0}}^2 }$. This quantity can be evaluated using the SWAP test shown in Fig.~\ref{f:swap_test}, and readily used in kernel-based models such as the support vector machine, the Gaussian process, and the principal component analysis.

Let us now discuss some examples. Stoudenmire and Schwab~\cite{stoudenmire2016supervised} encode data as products of local kernels, one for each component of the input vector, which results in a product quantum state (i.e., disentangled). This approach is often referred to as \textit{qubit encoding} and can produce highly non-linear kernels. As an example, for input vectors $\bm{x} \in [0,1]^n$ one can realize the feature map $\bm{x} \rightarrow \otimes_{i=1}^n \smqty( \cos(x_i \pi/2) \\ \sin(x_i \pi /2) )$ by applying suitable single-qubit rotations. Mitarai \textit{et al.}~\cite{mitarai2018quantum} use a similar approach, but encode each component of the data vector into multiple qubits. This redundancy populates the wave function with higher-order terms that can be exploited to fit non-linear functions of the data. Vidal and Theis~\cite{vidal2019input} investigate how this redundancy helps the task of data fitting. They found lower bounds of the redundancy that are logarithmic in the complexity of the function to be fit, using a linear-algebraic complexity measure. 

A different approach is taken by Wilson \textit{et al.}~\cite{wilson2018quantum}; the authors pre-process the input with a random linear map $\phi(\bm{x})=A\bm{x}+\bm{b}$, creating a quantum version of the \textit{random kitchen sink}~\cite{rahimi2008random}. They show that in the limit of many realizations of random linear maps, this approach implicitly implements a kernel. Interestingly, the form of the kernel depends on the layout of the encoder circuit, and not on all layouts are capable of implementing useful kernels. Another proposal that is based on random encoder circuits, but inspired by the convolutional filters used in neural networks, is the \textit{quanvolutional} network by \mbox{Henderson \textit{et al.}~\cite{henderson2019quanvolutional}}.

The examples discussed so far require low-depth encoder circuits and may therefore be robust depending on the noise characteristics and level. A different approach is the \textit{amplitude encoding}, a feature map encoding $2^n$-dimensional data vectors into the wave function of merely $n$ qubits. Assuming unit data vectors, the feature map ${ \bm{x} \rightarrow \ket{\bm{x}} }$ provides an exponential advantage in terms of memory and leads to a linear kernel. It is also known that by preparing copies of this feature map one can implement arbitrary polynomial kernels~\cite{rebentrost2014quantum}. Unfortunately, the depth of this encoder circuit is expected to scale exponentially with the number of qubits for generic inputs. Therefore, algorithms based on amplitude encoding could be impeded by our inability to coherently load data into quantum states.

On a different note, Havl{\'\i}{\v{c}}ek \textit{et al.}~\cite{havlivcek2019supervised} argue that a feature map can be constructed so that the kernel is hard to estimate using classical resources, and that this is a form of quantum supremacy. They consider, for example, ${ U_{\phi(\bm{x})} = \exp \small( i \sum_{j,k}^n \phi_{j,k}(\bm{x}) Z_j Z_k \small) H^{\otimes n} }$ where $Z_j$ is the Pauli-Z operator for the $j$-th qubit, $\phi_{j,k}$ are real functions, and H is the Hadamard gate. They conjecture that two layers of such an encoder circuit make the estimation of the kernel ${k(\bm{x},\bm{x^\prime}) = \lvert \bra{0} U_{\phi(\bm{x^\prime})}^\dag U_{\phi(\bm{x^\prime})}^\dag U^{\phantom{\dagger}}_{\phi(\bm{x})} U^{\phantom{\dagger}}_{\phi(\bm{x})} \ket{0} \rvert^2}$ classically intractable. This is due to its similarity to the circuits used in the hidden shift problem of Boolean bent functions, which are known to be classically hard to simulate~\cite{rotteler2010quantum}.

The design of feature maps inspired by quantum supremacy proposals is an interesting research direction. Whether this leads to an advantage in practical machine learning is an open question and should be tested empirically on existing computers. Ultimately, the form of the kernel and its parameters could be learned from data; this is a largely unexplored area in PQCs and has the potential to reduce the bias in kernel selection, and to automatically discover unknown feature maps that exhibit quantum supremacy.

\subsection{The variational circuit \texorpdfstring{$U_{\pmb{\theta}}$}{}}
\label{s:variational_circuit}

Similar to the universal approximation theorem in neural networks~\cite{hornik1989multilayer}, there always exists a quantum circuit that can represent a target function within an arbitrary small error. The caveat is that such a circuit may be exponentially deep and therefore impractical. Lin \textit{et al.}~\cite{lin2017does} argue that since real datasets arise from physical systems, they exhibit symmetry and locality; this suggests that it is possible to use `cheap' models, rather than exponentially costly ones, and still obtain a satisfactory result. With this in mind, the variational circuit aims to implement a function that can approximate the task at hand while remaining scalable in the number of parameters and depth. 

In practice, the circuit design follows a fixed structure of gates. Despite the dimension of the vector space growing exponentially with the number of qubits, the fixed structure reduces the model complexity, resulting in the number of free parameters to scale as a polynomial of the qubit count.

\begin{figure*}
\centering
\begin{tabular}{c c c}
(a) & \hspace{40pt}\vspace{5pt} & (b)\\
\begin{tikzcd}[row sep={24pt,between origins}]
\lstick[]{} & \gate{R_{P_1}} & \ctrl{1} & \qw & \qw\\
\lstick[]{} & \gate{R_{P_2}} & \control{} & \ctrl{1} & \qw\\
\lstick[]{} & \gate{R_{P_3}} & \ctrl{1} & \control{} & \qw\\
\lstick[]{} & \gate{R_{P_4}} & \control{} & \qw & \qw\\
\end{tikzcd}
& &
\begin{tikzcd}[row sep={24pt,between origins}]
\lstick[]{} & \gate{R_X} & \gate{R_Z} & \gate{XX} & \gate{XX} & \gate{XX} & \qw & \qw & \qw & \qw\\
\lstick[]{} & \gate{R_X} & \gate{R_Z} & \gate{XX}\vqw{-1} & \qw & \qw & \gate{XX} & \gate{XX} & \qw & \qw\\
\lstick[]{} & \gate{R_X} & \gate{R_Z} & \qw & \gate{XX}\vqw{-2} & \qw & \gate{XX}\vqw{-1} & \qw & \gate{XX} & \qw\\
\lstick[]{} & \gate{R_X} & \gate{R_Z} & \qw & \qw & \gate{XX}\vqw{-3} & \qw & \gate{XX}\vqw{-2} & \gate{XX}\vqw{-1} & \qw\\
\end{tikzcd}
\end{tabular}
\caption{Examples of hardware-efficient layers that can be used for encoder and variational circuits. Hardware-efficient constructions use entangling interactions that are naturally available on hardware and do not require compilation. Layers are repeated a number of times which is compatible with the hardware coherence time. (a) The construction in Ref.~\cite{mcclean2018barren} uses single-qubit rotations $R_{P_j} = \exp(-\frac{i}{2} \theta_j P_j)$ about randomly sampled directions $P_j \in \{X,Y,Z\}$, and a ladder of control-$Z$ entangling gates. Both the gate set and the connectivity are naturally implemented by many superconducting computers. (b) The construction in Ref.~\cite{benedetti2019generative} uses single-qubit rotations about $X$ and $Y$, and a fully-connected pattern of $XX$ entangling gates. Both the gate set and the connectivity are naturally implemented by trapped ions computers.}
\label{f:example_ansatz} 
\end{figure*}
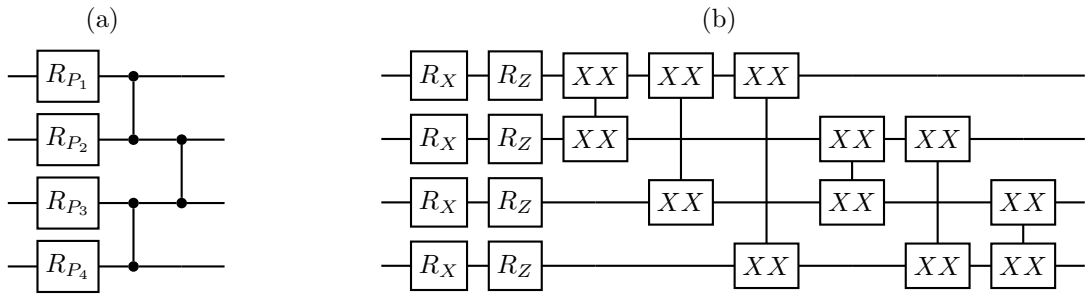

The first strategy to circuit design aims to comply with the fact that NISQ hardware has few qubits and usually operates on a sparse qubit-to-qubit connectivity graph with rather simple gates. \textit{Hardware-efficient} circuits alternate layers of native entangling gates and single-qubit rotations~\cite{kandala2017hardware}. Examples of these layers are shown in Fig.~\ref{f:example_ansatz}, where (a) and (b) are designed around the connectivity and gate set of superconducting and trapped ion computers, respectively. Heuristics can be used to strategically reduce the number of costly entangling gates. For example, Liu and Wang~\cite{liu2018differentiable} use the Chow-Liu tree graph~\cite{chow1968approximating} to setup the entangling layers. First, the mutual information between all pairs of variables is estimated form the dataset. Then, entangling gates are placed between qubits so that most of the mutual information is represented.

Another principled approach to circuit design is inspired by quantum many-body physics. \textit{Tensor networks} are methods to efficiently represent quantum states in terms of smaller interconnected tensors. In particular, these are often used to describe states whose entanglement is constrained by local interactions. By looking only at a smaller portion of the vector space, the computational cost is then reduced and becomes a polynomial function of the system size. This enables the numerical treatment of systems through layers of abstraction, reminiscent of deep neural networks. Indeed, some of the most studied tensor networks such as the matrix product state, the tree tensor network (TTN), and the multi-scale entanglement renormalization ansatz (MERA) have been tested for classification and generative modeling \cite{liu2017machine, Huggins2019Towards,grant2018hierarchical}.

\begin{figure*}
\includegraphics[width=.8\textwidth]{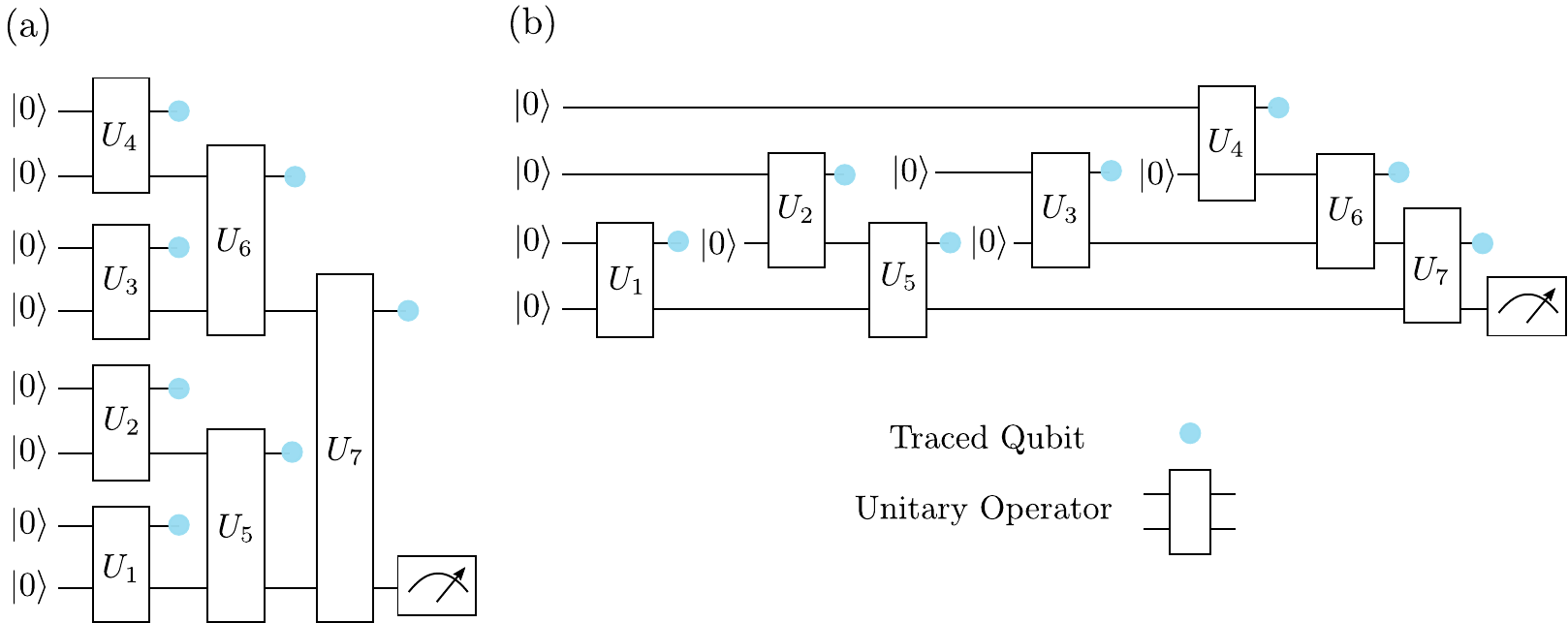}
\caption{Discriminative binary tree tensor network and its qubit-efficient version -- adapted from Ref.~\cite{Huggins2019Towards}. (a) The binary TTN implements a coarse graining procedure by tracing over half of the qubits after the application of each unitary. (b) A qubit-efficient version re-initializes the discarded qubits to be used in parallel operations. This scheme implements the same operation in (a) but requires fewer qubits on the device. It may however result in a deeper circuit.}
\label{f:ttn} 
\end{figure*}

Figure~\ref{f:ttn}~(a) shows an example of a TTN for supervised learning. After the application of each unitary, half of the qubits are traced out, while the other half continues to the next layer. Huggins \textit{et al.}~\cite{Huggins2019Towards} suggest a \textit{qubit-efficient} version where the traced qubits are reinitialized and used as the inputs of another unitary, as shown in Fig.~\ref{f:ttn}~(b). Qubit-efficient schemes could significantly reduce the required number of qubits, a favorable condition to some NISQ hardware. 

Neural networks and deep learning have proven to be very successful and therefore offer a further source of inspiration for circuit design. Both variational circuits and neural networks can be thought of as layers of connected computational units controlled by adjustable parameters. This has led some authors to refer to variational circuits as `quantum neural networks'. Here we shall briefly discuss the key differences that make this approach to circuit design rather difficult. 

First, quantum circuit operations are unitary and therefore linear; this is in contrast with the non-linear activation functions used in neural networks, which are key to their success and universality~\cite{hornik1991approximation}. There are several ways to construct non-linear operations in quantum circuits, both coherently (i.e., exploiting entanglement) or non-coherently (e.g., exploiting the natural coupling of the system to the environment). These can in turn be used to implement classical artificial neurons in quantum circuits~\cite{cao2017quantum,torrontegui2018universal,tacchino2018artificial}. 

The second key difference is that it is impossible to access the quantum state at intermediate points during computation. Although measurement of ancillary quantum variables can be used to extract limited information, any attempt to observe the full state of the system would disrupt its quantum character. This implies that executing the variational circuit cannot be seen as performing the forward pass of a neural network. Moreover, it is difficult to conceive a circuit learning algorithm that truly resembles \textit{backpropagation}, as it would rely on storing the intermediate state of the network during computation~\cite{hinton1986learning}. Backpropagation is the gold standard algorithm for neural networks and can be described as a computationally efficient organization of the chain rule that allows gradient descent to work on large-scale models.

The questions of how to generalize a quantum artificial neuron and design a quantum backpropagation algorithm have been open for quite some time~\cite{schuld2014quest}. Some recent work goes towards this direction. Verdon \textit{et al.}~\cite{verdon2018universal} quantize the parameters of the variational circuit which are then prepared in superposition in a dedicated register. This enables a backpropagation-like algorithm which exploits quantum effects such as phase kickback and tunneling. Beer \textit{et al.}~\cite{beer2019efficient} use separate qubit registers for input and output, and define the quantum neuron as a completely positive map between the two. The resulting network is universal for quantum computation and can be trained by an efficient process resembling backpropagation.

\subsection{Circuit learning}
\label{s:circuit_optimization}

Just like classical models, PQC models are trained to perform data-driven tasks. The task of learning an arbitrary function from data is mathematically expressed as the minimization of a loss function $L(\bm{\theta})$, also known as the objective function, with respect to the parameter vector $\bm{\theta}$. We discuss two types of algorithms, namely gradient-based and gradient-free, that can be applied to optimize the parameters of a variational circuit $U_{\bm{\theta}}$.

One instance of gradient-based algorithms is the iterative method called gradient descent. Here the parameters are updated towards the direction of steepest descent of the loss function 
\begin{equation}
\bm{\theta} \leftarrow \bm{\theta} - \eta \nabla_{\bm{\theta}} L ,
\label{e:gradient_descent}
\end{equation}
where $\nabla_{\bm{\theta}} L$ is the gradient vector and $\eta$ is the learning rate -- a hyperparameter controlling the magnitude of the update. This procedure is iterated and, assuming suitable conditions, converges to a local minimum of the loss function. 

The required partial derivatives can be calculated numerically using a finite difference scheme
\begin{equation}
\frac{\partial L}{\partial \theta_j} \approx \frac{L(\bm{\theta} + \Delta \bm{e}_j) - L(\bm{\theta} - \Delta \bm{e}_j)}{2 \Delta} ,
\end{equation}
where $\Delta$ is a (small) hyperparameter and $\bm{e}_j$ is the Cartesian unit vector in the $j$ direction. Note that in order to estimate the gradient vector $\nabla_{\bm{\theta}} L$, this approach evaluates the loss function twice for each parameter.

Alternatively, Spall's simultaneous perturbation stochastic approximation (SPSA)~\cite{spall1997one, spall2000adaptive} computes an approximate gradient vector with just two evaluations of the loss function as 
\begin{equation}
\frac{\partial L}{\partial \theta_j} \approx \frac{L(\bm{\theta}+c\, \bm{\Delta}) - L(\bm{\theta} - c\, \bm{\Delta})}{2\,c\,\Delta_j} ,
\end{equation}
where $\bm{\Delta}$ is a random perturbation vector and $c$ is a (small) hyperparameter.

There are cases when finite difference methods are ill-conditioned and unstable due to truncation and round-off errors. This is one of the reasons why machine learning relies on the analytical gradient when possible, and it is often calculated with automatic differentiation schemes~\cite{baydin2018automatic}. The analytical gradient can also be estimated for variational circuits, although the equations depend on the choice of parameterization for the gates. For our discussion, we consider circuits $U_{J:1}=U_J \cdots U_1$, where trainable gates are of the from $U_j = \exp( -\frac{i}{2} \theta_j P_j)$, and where $P_j \in \{ I, Z, X, Y \}^{\otimes n}$ is a tensor product of $n$ Pauli matrices. Arguably, this is the most common parameterization found in the literature.

Using this, Li \textit{et al.}~\cite{li2017hybrid} propose a way to efficiently compute analytical gradients in the context of quantum optimal control. Mitarai \textit{et al.}~\cite{mitarai2018quantum} bring this method into the context of supervised learning. Recall that the model's output is a function of expectation values $\expval{M_k}_{\bm{\theta}}$. Using the chain rule we can write the derivative $\frac{\partial L}{\partial \theta_j}$ as a function of the derivatives of the expectation values $\frac{\partial \expval{M_k}_{\bm{\theta}}}{\partial \theta_j}$. Each of these quantities can be estimated on quantum hardware using the  \textit{parameter shift rule}
\begin{equation}
\frac{\partial \expval{M_k}_{\bm{\theta}}}{\partial \theta_j} = \frac{\expval{M_{k}}_{\bm{\theta} + \frac{\pi}{2} \bm{e}_j} - \expval{M_{k}}_{\bm{\theta} - \frac{\pi}{2} \bm{e}_j}}{2} ,
\label{e:param_shift}
\end{equation}
where subscripts ${\bm{\theta} \pm \frac{\pi}{2} \bm{e}_j}$ indicate the shifted parameter vector to use for the evaluation (see Schuld \textit{et al.}~\cite{schuld2018evaluating} for a detailed derivation). Note that this estimation can be performed by executing two circuits. 

\begin{figure*}
\centering
\begin{adjustbox}{width=.9\textwidth}
\begin{tikzcd}
\lstick[1]{$\ket{0}$} & \gate{H} & \qw & \ctrl{1} & \qw & \ctrl{1} & \gate{H} & \meter{} & \rstick{$\expval{Z} = \frac{\partial}{\partial \theta_j} \expval{M_k}_{\bm{\theta}}$}\\
\lstick[1]{$\ket{0}^{\otimes n}$} & \qwbundle[alternate]{} & \gate{U_{1:j}} \qwbundle[alternate]{} & \gate{-i P_j} \qwbundle[alternate]{} & \gate{U_{j+1:J}} \qwbundle[alternate]{} & \gate{M_k} \qwbundle[alternate]{} & \qwbundle[alternate]{} & \\
\end{tikzcd}
\end{adjustbox}
\caption{The Hadamard test can be used to estimate the partial derivative of an expectation $\expval{M_k}_{\bm{\theta}}$ with respect to the parameter $\theta_j$. Here we show a simple case where gates are of the form ${ U_j = \exp \left (-\frac{i}{2} \theta_j P_j \right ) }$ and where both $P_j$ and $M_k$ are tensor products of Pauli matrices. It can be shown that measurements of the $Z$ Pauli observable on the ancilla qubit yield Eq.~\eqref{e:indirect}, the desired partial derivative. Hadamard tests can be designed to estimate higher order derivatives and to work with different measurements and gate parameterizations.}
\label{f:hadamard_test}
\end{figure*}
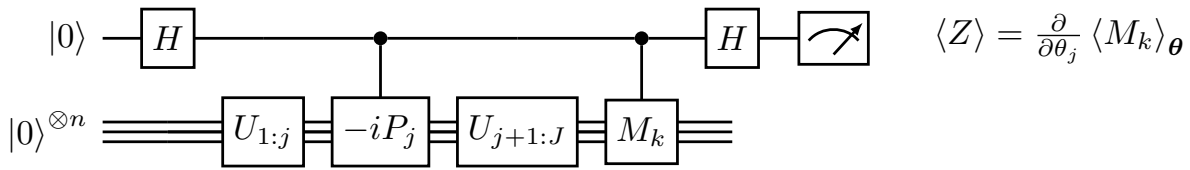

An alternative method can estimate the partial derivative with a single circuit, but at the cost of adding an ancilla qubit. A simple derivation using the gate parameterization introduced above (e.g., see Farhi and Neven~\cite{farhi2018classification}) shows that the partial derivative can be written as
\begin{equation}
\frac{\partial \expval{M_k}_{\bm{\theta}}}{\partial \theta_j} = \Im \left( \tr ( M^{\phantom{\dagger}}_k U^{\phantom{\dagger}}_{J : j+1 } P^{\phantom{\dagger}}_j U^{\phantom{\dagger}}_{j : 1 } \dyad{0}{0} U_{J : 1}^\dag ) \right) .
\label{e:indirect}
\end{equation}
This can be thought of as an \textit{indirect measurement} and can be evaluated using the Hadamard test shown in Fig.~\ref{f:hadamard_test}. This method can be generalized to compute higher order derivatives, as presented for example by Dallaire-Demers and Killoran~\cite{dallaire2018quantum}, and with alternative gate parameterizations, as done for example by Schuld \textit{et al.}~\cite{schuld2018circuit}.

We shall note that despite the apparent simplicity of the circuit in Fig.~\ref{f:hadamard_test}, the actual implementation of Hadamard tests may be challenging due to non-trivial controlled gates. Coherence must be guaranteed in order for quantum interference to produce the desired result. Mitarai and Fujii~\cite{mitarai2018methodology} propose a method for replacing a class of indirect measurements with direct ones. Instead of an interference circuit one can execute, in some cases, multiple simpler circuits that are suitable for implementations on NISQ computers. The `parameters shift rule' in Eq.~\eqref{e:param_shift} is nothing but the direct version of the measurement in Eq.~\eqref{e:indirect}. 

Compared to finite difference and SPSA, the analytical gradient has the advantage of providing an unbiased estimator. Additionally, Harrow and Napp~\cite{harrow2019low} find evidence that circuit learning using the analytical gradient outperforms any finite difference method. This is done by showing that for $n$ qubits and precision $\epsilon$, the query cost of an oracle for convex optimization in the vicinity of the optimum scales as $\mathcal{O} (\frac{n^2}{\epsilon} )$ for the analytical gradient, whereas finite difference needs at least $\Omega (\frac{n^3}{\epsilon^2} )$ calls to the oracle. In practice though, it is found that SPSA performs well in small-scale noisy experimental settings (e.g., see Kandala \textit{et al.}~\cite{kandala2017hardware} and Havl{\'\i}{\v{c}}ek \textit{et al.}~\cite{havlivcek2019supervised}). 

Particular attention should be given to the problems of exploding and vanishing gradients which are well-known to the machine learning community. Classical models, in particular recurrent neural networks, are often constrained to perform unitary operations so that their gradients cannot explode (see Wisdom \textit{et al.}~\cite{wisdom2016full} for an example). Quantum circuits implementing unitary operations naturally avoid the exploding gradient problem. On the other hand, McClean \textit{et al.}~\cite{mcclean2018barren} show that random circuits of reasonable depth lead to an optimization landscape with exponentially large plateaus of vanishing gradients with an exponentially decaying variance. This can be understood as a consequence of Levy's lemma~\cite{ledoux2001concentration} which states that a random variable that depends on many independent variables is essentially constant. The learning algorithm is thus unable to estimate the gradient and may perform a random walk in parameter space. While this limits the effectiveness of variational circuits initialized at random, the use of highly structured circuits could alleviate the problem (e.g., see Grant \textit{et al.}~\cite{grant2019initialization} for a structured initialization strategy).

We shall stress here that in hybrid systems parameter updates are performed classically. This implies that some of the most successful deep learning methods can be readily used for circuit learning. Indeed, heuristics such as stochastic gradient descent~\cite{robbins1951stochastic}, resilient backpropagation~\cite{riedmiller1993direct}, and adaptive momentum estimation~\cite{kingma2014adam}, have already been applied with success. These were designed to deal with issues of practical importance such as large datasets, large noise in gradient estimates, and the need to find adaptive learning rates in Eq.~\eqref{e:gradient_descent}. In practice, these choices can reduce the time for successful training from days to hours. 

There are cases where gradient-based optimization may be challenging. For example, in a noisy experimental setting the loss function may be highly non-smooth and not suitable for gradient descent. As another example, the objective function may be itself unknown and therefore should be treated as a black-box. In these cases, circuit learning can be carried out by gradient-free methods. A well-known method of this type is particle swarm optimization~\cite{eberhart1999human}. Here the system is initialized with a number of random solutions called particles, each one moving through solution space
with a certain velocity. The trajectory of each particle is adjusted according to its own experience and that of other particles so that they converge to a local minima. Another popular method is Bayesian optimization~\cite{frazier2018tutorial}. It uses evaluations of the objective function to construct a model of the function itself. Subsequent evaluations can be chosen either to improve the model or to find a minima. 

Zhu \textit{et al.}~\cite{zhu2018training} compare Bayesian and particle swarm optimization for training a generative model on a trapped ion quantum computer. While Bayesian optimization outperforms particle swarm in their setting, they found that the large number of parameters challenges both optimizers. They show that an ideal simulated system is not significantly faster than the experimental system, indicating that the actual bottleneck is the classical optimizer. Leyton-Ortega \textit{et al.}~\cite{leyton2019robust} train a generative model on a superconducting quantum computer and compare the gradient-free methods of zeroth-order optimization package~\cite{liu2017zoopt} and stochastic hill-climbing, with gradient descent. They find that on average zeroth-order optimization achieves the lowest loss on their hardware. They argue that the main optimization challenge is to overcome the variance of the loss function which is due to random parameter initialization, hardware noise, and finite number of measurements. 

Genetic algorithms~\cite{sastry2005genetic} are another large class of gradient-free optimization algorithms. At each step, candidate solutions are evolved using biology-inspired operations such as recombination, mutation, and natural selection. When used for circuit learning, genetic algorithms define a set of allowed gates and the maximum number to be employed. Lamata \textit{et al.}~\cite{lamata2018quantum} suggest the use of genetic algorithms to train a PQC model for compression using a universal set of single- and two-qubit gates. Ding \textit{et al.}~\cite{ding2019experimental} validate the idea experimentally by deploying a pre-trained PQC model on a superconducting computer and find that using a subsequent genetic algorithm improves its fidelity.

To conclude, we note that optimization algorithms should be tailored for PQC models if we want to achieve better scalability. Very recent work has been approaching circuit learning from this perspective (e.g., see Ostaszewski \textit{et al.}~\cite{ostaszewski2019quantum} and Nakanishi \textit{et al.}~\cite{nakanishi2019sequential}).

\section{Applications}
\label{s:applications}

In this Section we look at machine learning applications using PQC models where the goal is to obtain an advantage over classical models. For supervised learning with classical data we give a general overview of how PQC model can be applied to classification and regression. For unsupervised learning with classical data we focus on generative modeling since this comprises most of the literature. 

PQC models can also handle inputs and outputs that are inherently quantum mechanical, i.e., already in superposition. These are often referred to as \textit{quantum data}~\cite{aimeur2006machine}. Quantum input data could originate remotely, for example, from other quantum computers transmitting over a quantum Internet~\cite{kimble2008quantum}. Otherwise, if a preparation recipe is available, one could prepare the input data locally using a suitable encoder circuit. Assuming this data preparation is efficient, one can extend supervised and unsupervised learning to quantum states and quantum information.

\begin{figure*}
\includegraphics[width=.75\textwidth]{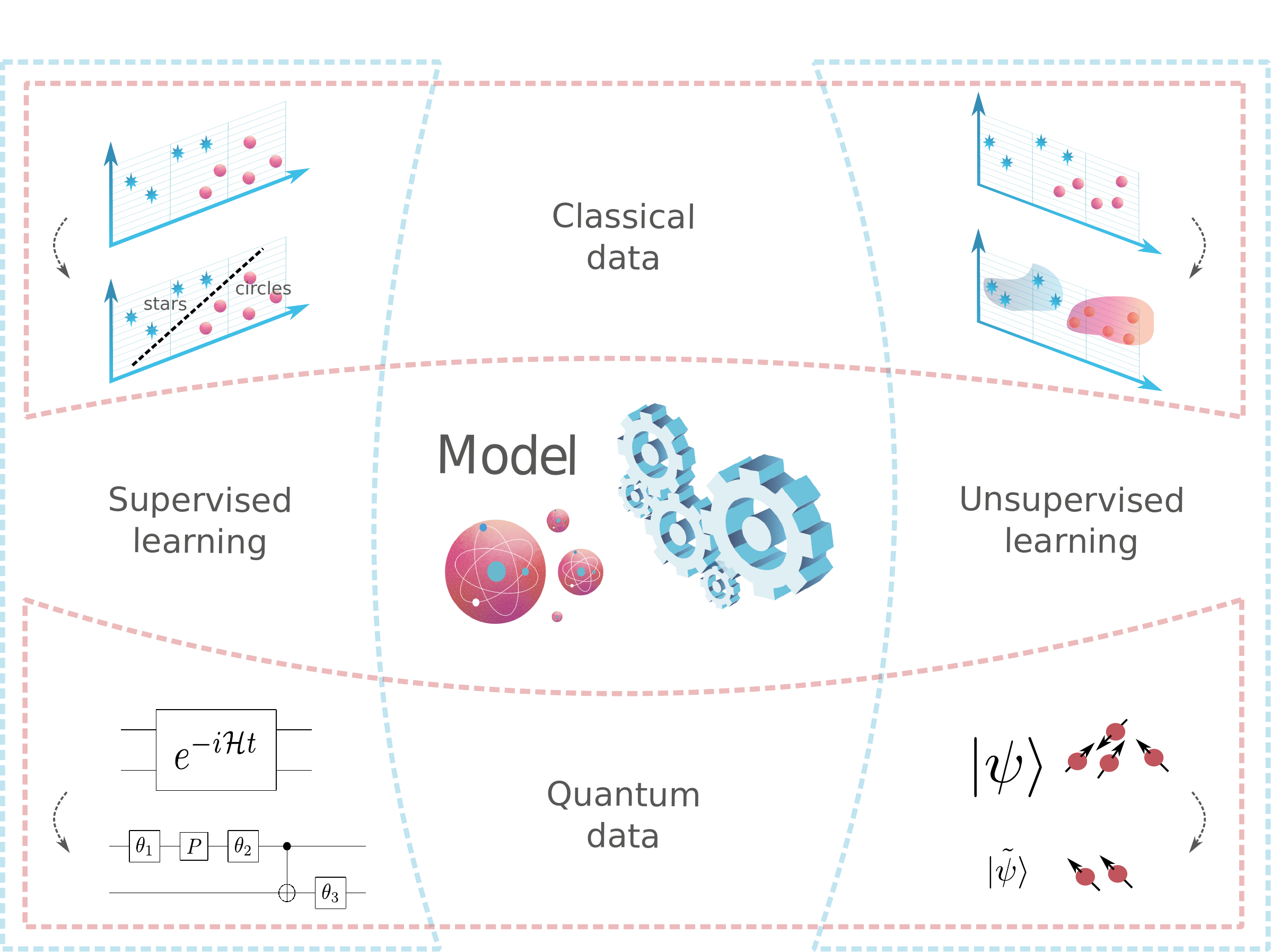}
\caption{Parameterized quantum circuit models can be trained for a variety of machine learning tasks, such as supervised and unsupervised learning, on both classical and quantum data. This figure shows examples from each category. In the top-left panel, the model learns to recognize patterns to classify the classical data. In the top-right panel, the model learns the probability distribution of the training data and can generate new synthetic data accordingly. For supervised learning of quantum data, bottom-left panel, the model assists the compilation of a high-level algorithm to low-level gates. Finally, for unsupervised learning of quantum data, bottom-right panel, the model performs lossy compression of a quantum state.}
\label{f:application} 
\end{figure*}

Figure~\ref{f:application} shows examples for all these cases. Intuitively each application is a specification of the components outlined in Fig.~\ref{f:PQC}, which the Reader is encouraged to refer to throughout the Section for clarity.

In many practical decision-making scenarios there is no available data concerning the best course of action. In this case, the model needs to interact with its environment to obtain information and learn how to perform a task from its own experience. This is known as reinforcement learning. An example would be a video game character that learns a successful strategy by repeatedly playing the game, analyzing results, and improving. Although quantum generalizations and algorithms for reinforcement learning have been proposed, to the best of our knowledge, none of them are based on hybrid systems and PQC models.

\subsection{Supervised learning}
\label{s:supervised}

Let us first consider supervised learning tasks, e.g., classification and regression, on classical data. Given a dataset ${\mathcal{D} = \{ (\bm{x}^{(i)}, \bm{y}^{(i)}) \}_{i=1}^N}$, of $N$ samples, the goal is to learn a model function $f: \mathcal{X}\rightarrow\mathcal{Y}$ that maps each $\bm{x} \in \mathcal{X}$ to its corresponding target $\bm{y} \in \mathcal{Y}$. A standard approach is to minimize a suitable regularized loss function, \mbox{that is,} 
\begin{equation}
\bm{\theta}^* = \arg\min_{\bm{\theta}} \frac{1}{N} \sum_{i=1}^N L \left ( f(\bm{x}^{(i)}, \bm{\theta}), \bm{y}^{(i)} \right ) + R(\bm{\theta}) ,
\label{eq:loss}
\end{equation}
where $\bm{\theta}$ is the set of parameters defining the model function, $L$ quantifies the error of a forecast, and $R$ is a regularization function penalizing undesired values for the parameters. The latter is used to prevent overfitting; indeed, if the training set is not sufficiently large, the model could simply memorize the training data and not generalize to unseen data.

In the PQC framework, once the encoder circuit $U_{\phi(\bm{x})}$ is set up, there are two main options for the remaining part of the circuit: the quantum kernel estimator (QKE), and the variational quantum model (VQM). We now briefly discuss both, and refer the Reader to Schuld and Killoran~\cite{Schuld2019quantum} for a more in-depth theoretical exposition.

The QKE does not use a variational circuit $U_{\bm{\theta}}$ to process the data; instead, it uses the SWAP test (e.g., see Fig.~\ref{f:swap_test}) to evaluate the possibly intractable kernel $k(\bm{x}^{(i)},\bm{x}^{(j)})$. Then, resorting to the representer theorem~\cite{Scholkopf2001Generalized}, the model function is expressed as an expansion over kernel functions $f(\bm{x}, \bm{w})=\sum^N_{i=1} w_i k(\bm{x},\bm{x}^{(i)}) $. The learning task is to find parameters $\bm{w}$ so that the model outputs correct forecasts. Note that these parameters define the classical post-processing function, as opposed to an operation of the PQC. A potential caveat is that QKE relies on a coherent SWAP test which may be non-trivial to implement on NISQ computers.

The VQM, on the other hand, uses a variational circuit $U_{\bm{\theta}}$ to process data directly in the feature space. A set of expectation values $\{\expval{M_k}_{\bm{x}, \bm{\theta}} \}_{k=1}^K$ is estimated and post-processed to obtain the model output (see Fig.~\ref{f:PQC}). In contrast to QKE, VQM parameters define the operations carried out by the quantum computer and require a circuit learning algorithm of the kind discussed in Section~\ref{s:circuit_optimization}.

Havl{\'\i}{\v{c}}ek \textit{et al.}~\cite{havlivcek2019supervised} experimentally demonstrate QKE and VQM classifiers on two superconducting qubits of the IBM Q5 Yorktown. Their QKE estimates a classically intractable feature map (see Section~\ref{s:encoder_circuit} for details) which is then fed into a support vector machine to find the separating hyper-plane. Their VQM uses a hardware-efficient circuit instead. By employing a suitable error mitigation protocol, they find an increase in classification success with increasing circuit depth. In the future, it would be interesting to systematically compare these proposals against established classical models by evaluating accuracy and training efficiency, for example.

We now focus our discussion on VQM proposals. Farhi and Neven~\cite{farhi2018classification} propose a VQM binary classifier for bitstrings. The encoder circuit simply maps bitstrings to the computational basis states by applying identity and NOT gates at almost no cost. The variational circuit acts on the input register and one ancilla qubit which is measured to yield a class forecast. With $n$-bit data strings as the input, there are $2^{2^n}$ possible binary functions that could generate the class labels. The authors show that for any of the possible label functions there exists a variational circuit that achieves zero classification error. For some of these functions, the circuit is exponentially deep and therefore impractical. This result parallels the well known universal approximation theorem~\cite{hornik1989multilayer} which states that neural networks with an exponentially large hidden layer of non-linear neurons are able to represent any Boolean function.

Mitarai \textit{et al.}~\cite{mitarai2018quantum} propose VQMs for classification and regression of real-valued data using a highly non-linear qubit encoding. The variational circuit must then entangle the qubits such that a local observable can extract the relevant non-linear features. As discussed in Section~\ref{s:variational_circuit} one possible way to strategically construct highly entangling variational circuits is inspired by tensor networks. Grant \textit{et al.}~\cite{grant2018hierarchical} use TTN and MERA variational circuits to perform binary classification on canonical datasets such as Iris and MNIST. In their simulations, MERA always outperforms TTN. One of their simplest models is efficiently trained classically and then deployed on the IBM Q5 Tenerife quantum computer with significant resilience to noise. 

Stoudenmire \textit{et al.}~\cite{liu2017machine} train a TTN to perform pairwise classification of the MNIST image data. In their simulations, they use entanglement entropy to quantify the amount of information in a detail of the image that is gained by observing the context. This is an example of how quantum properties can be used to characterize the complexity of classical data, which is a developing area of research. 

Schuld \textit{et al.}~\cite{schuld2018circuit} propose a VQM classifier assuming amplitude-encoded input data. Since this encoder circuit may be very expensive, the authors aim to keep the variational circuit low-depth and highly expressive at the same time. This is achieved through a systematic use of entangling gates, and by keeping the number of parameters polynomial in the number of qubits. Their simulations on benchmark datasets show performance comparable to that of off-the-shelf classical models while using significantly fewer parameters.

To date, all supervised learning experiments involved scaled-down, often trivial, datasets due to the limitation of available quantum hardware, and demonstrations at a more realistic scale are desirable. As a last comment, we note that a largely undeveloped area is that of regularization techniques specifically designed for PQC models which is, in our opinion, an interesting area for \mbox{future research}.

\subsection{Generative modeling}
\label{s:generative}

We now discuss generative modeling, an unsupervised learning task where the goal is to model an unknown probability distribution and generate synthetic data accordingly. Generative models have been successfully applied in computer vision, speech synthesis, inference of missing text, de-noising of images, chemical design, and many other automated tasks. It is believed that they will play a key role in the development of general artificial intelligence; a model that can generate realistic synthetic samples is likely to `understand' its environment. 

Concretely, the task is to learn a model distribution $q_{\bm{\theta}}$ that is close to a target distribution $p$. The closeness is defined in terms of a divergence $D$ on the statistical manifold, and learning consists of minimizing this divergence; that is,
\begin{equation}
\bm{\theta}^* = \arg\min_{\bm{\theta}} D ( p , q_{\bm{\theta}} ) .
\end{equation}
Since the target probability distribution is unknown, it is approximated using a dataset $\mathcal{D} = \{\bm{v}^{(i)}\}_{i=1}^N$ which we have access to and which is distributed according to the target distribution. As an example, $\bm{v}^{(i)}$ could be natural images extracted from the Internet.

The probabilistic nature of quantum mechanics suggests that a model distribution can be encoded in the wave function of a quantum system~\cite{cheng2017information,han2017unsupervised}. Let us see how a simple adaptation of the model shown in Fig.~\ref{f:PQC} gives a generative model for $n$-dimensional binary data $\bm{v}^{(i)} \in \{0,1\}^n$. First, we set the encoder circuit to the identity $U_{\phi(\bm{x})} = I$ since in this problem there is no input data. Second, we apply a variational circuit $U_{\bm{\theta}}$ to the initial state $\ket{0}^{\otimes n}$. Finally, we perform a measurement in the computational basis, i.e., we measure the set of operators $\{ \expval{M_{\bm{v}}}_{\bm{\theta}} \}_{\bm{v}}$ where $M_{\bm{v}} = \dyad{\bm{v}}{\bm{v}}$ are projectors for the bitstrings. The resulting generative model, known as the quantum circuit Born machine (QCBM)~\cite{benedetti2019generative,liu2018differentiable}, implements the probability distribution
\begin{equation}
q_{\bm{\theta}}(\bm{v}) = \tr ( M_{\bm{v}} U^{\phantom{\dagger}}_{\bm{\theta}} \dyad{0}{0} U_{\bm{\theta}}^\dag ) .
\end{equation}
Since the target data is binary, no post-processing is needed and each measurement outcome $\bm{v} \sim q_{\bm{\theta}}$ is an operational output. If the target data were instead real-valued, we could interpret bitstrings as discretized outputs and use a post-processing function to recover real-values.

As one does not have access to the wave function, characterizing the distribution $q_{\bm{\theta}}$ may be intractable for all but the smallest circuits. For this reason, QCBMs belong to the class of \textit{implicit models}, models where it is easy to obtain a sample $\bm{v} \sim q_{\bm{\theta}}$, but hard to estimate the likelihood $q_{\bm{\theta}}(\bm{v})$. Machine learning researchers have become increasingly interested in implicit models because of their generality, expressive power, and success in practice~\cite{goodfellow2016nips}. Interestingly, Du \textit{et al.}~\cite{du2018expressive} show that QCBMs have strictly more expressive power than classical models such as deep Boltzmann machines, when only a polynomial number of parameters are allowed. Coyle \textit{et al.}~\cite{coyle2019born} show that some QCBMs cannot be efficiently simulated by classical means in the worst case, and that this holds for all the circuit families encountered during training.

\begin{figure*}
\includegraphics[width=.94\textwidth]{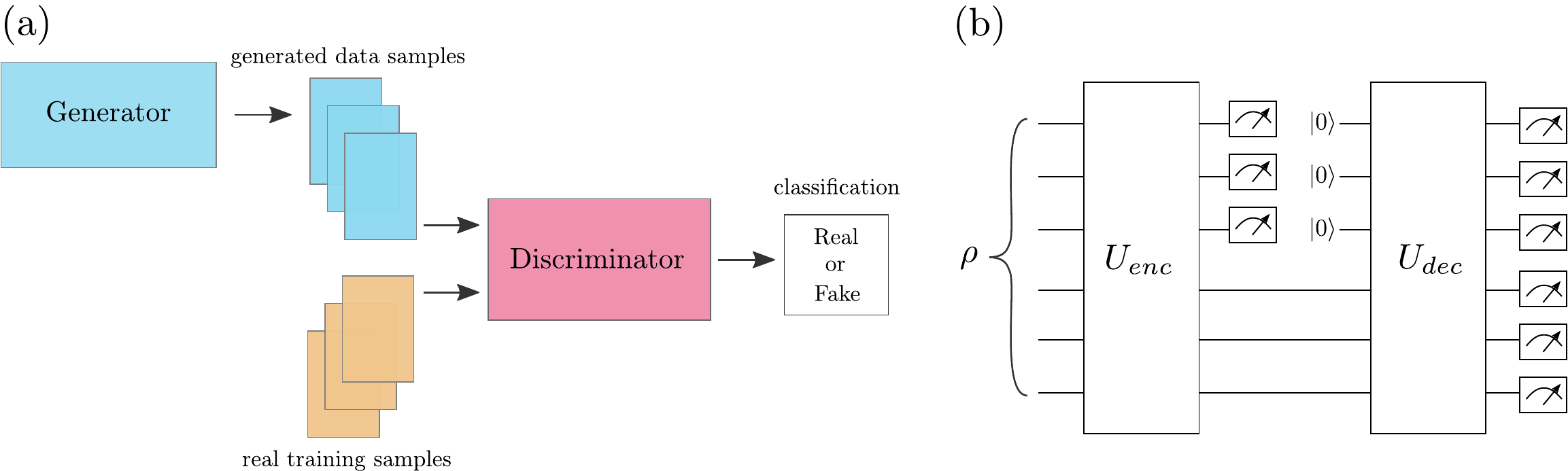}
\caption{Illustration of quantum generative models. (a) In the quantum generative adversarial network the generator creates synthetic samples and the discriminator tries to distinguish between the generated and the real samples. The network is trained until the generated samples are indistinguishable from the training samples. In this method the target data, the generator, and the discriminator can all be made quantum or classical. (b) The quantum autoencoder reduces the dimensionality of quantum data by applying an encoder circuit $U_{enc}$, tracing over a number of qubits and finally reconstructing the state with a decoder circuit $U_{\textit{dec}}$. Panels (a) and (b) are adapted from Refs.~\cite{zoufal2019quantum} and~\cite{romero2017quantum}, respectively.}
\label{f:qlearning} 
\end{figure*}

Benedetti \textit{et al.}~\cite{benedetti2019generative} build low-depth QCBMs using variational circuits suitable for trapped ion computers (see Fig.~\ref{f:example_ansatz} (b) for an example). They use particle swarms to minimize an approximation to the Kullback-Leibler divergence~\cite{kullback1951information} ${ D(p, q_{\bm{\theta}}) = \sum_{\bm{v}} p(\bm{v}) \ln \frac{ p(\bm{v}) } { q_{\bm{\theta}} (\bm{v}) } }$. In their simulations they successfully train models for the canonical Bars-and-Stripes dataset and for Boltzmann distributions, and use them to design a performance indicator for hybrid quantum-classical systems. Zhu \textit{et al.}~\cite{zhu2018training} implement this scheme on four qubits of an actual trapped ion computer and experimentally demonstrate convergence of the model to the target distribution. 

Liu and Wang~\cite{liu2018differentiable} propose the use of gradient descent to minimize the maximum mean discrepancy~\cite{gretton2007kernel} ${ D ( p , q_{\bm{\theta}} ) = \lVert \sum_{\bm{v}} p(\bm{v}) \phi(\bm{v}) - \sum_{\bm{v}} q_{\bm{\theta}}(\bm{v}) \phi(\bm{v}) \lVert^2 }$, where $\phi$ is a classical feature map, and the expectations are estimated from samples. Their approach allows for gradient estimates with discrete target data, which is often not possible in classical implicit models. In their simulations they successfully train QCBMs for the Bars-and-Stripes dataset and for discretized Gaussian distributions. Hamilton \textit{et al.}~\cite{hamilton2018generative} implement this schema on the IBM Q20 Tokyo computer, and examine how statistical and hardware noise affect convergence. They find that the generative performance of state-of-the-art hardware is usually significantly worse than that of the numerical simulations. Leyton-Ortega \textit{et al.}~\cite{leyton2019robust} perform a complementary experimental study on the Rigetti 16Q-Aspen computer. They argue that due to the many components involved in hybrid quantum-classical systems (e.g., choice for the entangling layers, optimizers, post-processing, etc.), the performance ultimately depends on the ability to correctly set hyperparameters; that is, research on automated hyperparameter setting will be key to the success of QCBMs.

Another challenge in QCBMs is the choice of a suitable loss functions.  Non-differentiable loss functions are often hard to optimize; one can use gradient-free methods, but these are likely to struggle as the number of parameters becomes large. Differentiable loss functions are often hard to design; recall that since QCBM are implicit models, one does not have access to the likelihood $q_{\bm{\theta}}(\bm{v})$. \textit{Adversarial} methods developed in deep learning can potentially overcome these limitations. Figure~\ref{f:qlearning} (a) shows the intuition; the adversarial method introduces a discriminative model whose task is to distinguish between true data coming from the dataset and synthetic data coming from the generative model. This creates a `game' where the two players, i.e., the models, compete. The advantage is that both models are trained at the same time, with the discriminator providing a differentiable loss function for the generator. 

Lloyd and Weedbrook~\cite{lloyd2018quantum} put forward the quantum generative adversarial network (QGAN) and theoretically examine variants where target data, generator and discriminator are either classical or quantum. We discuss the case of quantum data in the next Section while here we focus on classical data. Both Situ \textit{et al.}~\cite{situ2018adversarial} and Zeng \textit{et al.}~\cite{zeng2018learning} couple a PQC generator to a neural network discriminator and successfully reproduce the statistics of some discrete target distributions. Romero and Aspuru-Guzik~\cite{romero2019variational} extend this to continuous target distributions using a suitable post-processing function. Zoufal \textit{et al.}~\cite{zoufal2019quantum} propose a QGAN to approximately perform amplitude encoding. While the best known generic method has exponential complexity, their circuit uses a polynomial number of gates. If both the cost of training and the required precision are kept low, this method has the potential to facilitate algorithms that require amplitude encoding. 

One key aspect of generative models is their ability to perform \textit{inference}. That is, when some of the observable variables are `clamped' to known values, one can infer the expectation value of all other variables by sampling from the conditional probability. For example, inpainting, the process of reconstructing lost portions of images and videos, can be done by inferring missing values from a suitable generative model. Low \textit{et al.}~\cite{low2014quantum} use Grover's algorithm to perform inference on quantum circuits and obtain a quadratic speedup over na\"{i}ve methods, although the overall complexity remains exponential. Zeng \textit{et al.}~\cite{zeng2018learning} propose to equip QCBMs with this method, although this requires amplitude amplification and estimation methods that may be beyond NISQ hardware capabilities. It is an open question how to perform inference on QCBMs in the near term.

\subsection{Quantum learning tasks}
\label{s:quantum_tasks}

We finally consider learning tasks that are inherently quantum mechanical. As discussed in the Introduction, early hybrid approaches~\cite{bang2008quantum, gammelmark2009quantum} were proposed to assist the implementation of quantum algorithms (e.g., Deutsch's, Grover's, and Shor's) from datasets of input-output pairs. \textit{Quantum algorithm learning} has been recently rediscovered by the community. 

Morales \textit{et al.}~\cite{morales2018variational} train PQC models for the diffusion and oracle operators in Grover's algorithm. Noting that Grover's algorithm is optimal up to a constant, the authors show that the approach can find new improved operators for the specific case of three and four qubits. Wan \textit{et al.}~\cite{wan2018learning} train a PQC model to solve the hidden subgroup problem studied by Simon~\cite{simon1997power}. In their simulations, they recover the original Simon's algorithm with equal performance. Anschuetz \textit{et al.}~\cite{anschuetz2019variational} use known techniques to map integer factoring to an Ising Hamiltonian, then train a PQC model to find the ground state hence finding the factors. Cincio \textit{et al.}~\cite{cincio2018learning} train circuits to implement the SWAP test (see Fig.~\ref{f:swap_test}) and find solutions with a smaller number of gates than the known circuits. 

These methods promise to assist the implementation of algorithms on near-term computers. Experimental studies will be needed to assess their scaling under realistic NISQ constraints and noise. Theoretical studies will be needed to understand their \textit{sample complexity}, that is, the number of training samples required in order to successfully learn the target algorithm. Even in small-scale computers, we shall avoid exponential sampling complexity if we want these methods to be practical.

In the context of \textit{quantum state classification}, Grant \textit{et al.}~\cite{grant2018hierarchical} simulate the training of a TTN variational circuit for the classification of pure states that have different levels of entanglement. They found that, if the unitary operations in the TTN are too simple, classification accuracy on their synthetic dataset is no better than random class assignments. When using more complex operations involving ancilla qubits the TTN is able to classify quantum states with some accuracy. Chen \textit{et al.}~\cite{chen2018universal} simulate the training of PQC models to classify quantum states as pure or mixed, including a third possible output associated with an inconclusive result. Their circuits rely on layers of gates that are conditioned on measurement outcomes, with the purpose of introducing non-linear behaviour similar to that of neural networks.

State tomography is another ubiquitous task aiming at predicting the outcome probabilities of any measurement performed on an unknown state. To completely model the unknown state, one would require a number of measurements growing exponentially with the number of qubits. However, this can be formulated as a \textit{quantum state learning} problem with the hope of minimizing the number of required measurements. Aaronson~\cite{aaronson2007learnability} studies the sampling complexity of this problem under Valiant's probably approximately correct (PAC) learning model~\cite{valiant1984theory}. They find that for practical purposes one needs a number of measurements scaling linearly with the number of qubits. Rocchetto \textit{et al.}~\cite{rocchetto2019experimental} experimentally verify the linear scaling on a custom photonic computer and extrapolate the value of the scaling constant. In terms of methodology, Lee \textit{et al.}~\cite{lee2018learning} propose to train a variational circuit $U_{\bm{\theta}}$ that transforms the unknown state $\ket{\psi}$ to a known fiducial state $\ket{f}$. The unknown state can be reproduced by evaluating the adjoint circuit on the fiducial state, that is, $\ket{\psi}\approx U_{\bm{\theta}}^{\dagger}\ket{f}$. A related learning tasks is that of \textit{quantum state diagonalization} for mixed states. LaRose \textit{et al.}~\cite{larose2018variational} propose to train a variational circuit $U_{\bm{\theta}}$ such that the density matrix ${ \Tilde{\rho}= U^{\phantom{\dagger}}_{\bm{\theta}} \rho U^\dag_{\bm{\theta}} }$ is diagonalized, hence representing a classical probability distribution.

In the previous Section and in Fig.~\ref{f:qlearning}~(a) we introduced QGANs for classical data. We now discuss the case where all components are quantum mechanical, hence enabling the generative modeling of quantum data. The discriminator, now taking target and synthetic quantum states in input, aims at modeling the measurement for optimal distinguishability, also known as the Helstrom measurement~\cite{helstrom1969quantum}. In turn, the generator tries to make the task of distinguishing more difficult by minimizing its distance from the target state~\cite{lloyd2018quantum,benedetti2019adversarial}. In practice, this game can be implemented by coupling two PQC models and optimizing them in tandem. For example, Dallaire-Demers and Killoran~\cite{dallaire2018quantum} propose a QGAN that generates states conditioned on labels. This may find application in chemistry where the label is `clamped' to a desired physical property and the model generates new molecular states accordingly. Benedetti \textit{et al.}~\cite{benedetti2019adversarial} propose a QGAN that generates approximations of pure states. They numerically show how the depths of generator and discriminator impact the quality of approximation. They also design a heuristic for stopping training, which is a non-trivial problem even in classical adversarial methods. Hu \textit{et al.}~\cite{hu2019quantum} experimentally demonstrate adversarial learning on a custom superconducting qubit. 

Finally, PQC models can be used to attack well-known problems in quantum information from a novel machine learning perspective. Let us see some examples within the context of compression, error correction and compilation.

Romero \textit{et al.}~\cite{romero2017quantum} propose a quantum autoencoder (QAE) to reduce the amount of resources needed to store quantum data. As shown in Fig.~\ref{f:qlearning}~(b) an encoder circuit $U_{enc}$ is applied to the quantum data stored in $n$ qubits. After tracing out $n-k$ qubits, a decoder circuit $U_{dec}$ is used to reconstruct the initial state. The circuits are trained to maximize the expected fidelity between inputs and outputs, effectively performing a lossy compression of an $n$-qubit state into a $k$-qubit state.

Fault-tolerant quantum computers require error correction schemes that can deal with noisy and faulty operations. Leading proposals such as the color code and the surface code devote a large number of physical qubits to implement error-corrected logical qubits (see Gottesman~\cite{gottesman2010introduction} for an introduction to quantum error correction). Johnson \textit{et al.}~\cite{johnson2017qvector} suggest that a reduced overhead could be achieved in NISQ devices by training encoding and recovery circuits to optimize the average code fidelity. 

The implementation of a quantum algorithm is also limited by the available gate set and qubit-to-qubit connectivity of the underlying hardware. This is where quantum compilers come into play, by abstracting the user from the low-level details. Khatari \textit{et al.}~\cite{khatri2018quantum} propose to train a hardware-efficient variational circuit $U_{\bm{\theta}}$ to approximately execute the same action as a target unitary $U=e^{-i\mathcal{H}t}$.

\onecolumngrid

\begin{table*}[h]
\begin{tabularx}{\textwidth}{L{1.1}|L{.7}|L{.6}|L{.8}|L{.4}|L{1.25}}
\textbf{Reference} & \textbf{Task} & \textbf{Model} & \textbf{Learning} & \textbf{Qubits} & \textbf{Computer} \\ \hline\hline
Schuld \textit{et al.}~\cite{Schuld2017implementing} & Classification & QKE & N/A & 4 & IBM Q5 Yorktown (S) \\ \hline
Grant \textit{et al.}~\cite{grant2018hierarchical} & Classification & VQM & N/A & 4 & IBM Q5 Tenerife (S) \\ \hline
Havl{\'\i}{\v{c}}ek \textit{et al.}~\cite{havlivcek2019supervised} & Classification & QKE, VQM & Gradient-based & 2 & IBM Q5 Yorktown (S) \\ \hline
Tacchino \textit{et al.}~\cite{tacchino2018artificial} & Classification & Perceptron & Gradient-based & 3 & IBM Q5 Tenerife (S) \\ \hline
Benedetti \textit{et al.}~\cite{benedetti2019generative} & Generative & QCBM & N/A & 4 & Custom (T) \\ \hline
Hamilton \textit{et al.}~\cite{hamilton2018generative} & Generative & QCBM & Gradient-based & 4 & IBM Q20 Tokyo (S) \\ \hline
Zhu \textit{et al.}~\cite{zhu2018training} & Generative & QCBM & Gradient-free & 4 & Custom (T) \\ \hline
Leyton-Ortega \textit{et al.}~\cite{leyton2019robust} & Generative & QCBM & Gradient-based, gradient-free & 4 & Rigetti 16Q-Aspen (S) \\ \hline
Coyle \textit{et al.}~\cite{coyle2019born} & Generative & QCBM & Gradient-based & 4 & Rigetti 16Q-Aspen (S) \\ \hline
Hu \textit{et al.}~\cite{hu2019quantum} & State learning & QGAN & Gradient-based & 1 & Custom (S) \\ \hline
Zoufal \textit{et al.}~\cite{zoufal2019quantum} & State learning & QGAN & Gradient-based & 3 & IBM Q20 Poughkeepsie (S)\\ \hline
Rocchetto \textit{et al.}~\cite{rocchetto2019experimental} & State learning & PAC & N/A & 6 & Custom (P) \\ \hline
Otterbach \textit{et al.}~\cite{otterbach2017unsupervised} & Clustering & QAOA & Gradient-free & 19 & Rigetti 19Q-Acorn (S) \\ \hline
Ding \textit{et al.}~\cite{ding2019experimental} & Compression & QAE & Gradient-free & 3 & Rigetti 8Q-Agave (S) \\ \hline
Rist{\`e} \textit{et al.}~\cite{riste2017demonstration} & Learning parity with noise & Oracle & N/A & 5 & IBM Q5 Yorktown (S) \\
\end{tabularx}
\caption{Overview of parameterized quantum circuit models that have been demonstrated experimentally on superconducting (S), trapped ion (T), and photonic (P) hardware. N/A labels the cases where a learning algorithm was either not required or not used, e.g., when learning is simulated classically and the model is deployed on quantum hardware.}
\label{tab:experimental}
\end{table*}

\begin{table*}[h]
\begin{tabularx}{\textwidth}{L{1.2}|L{.7}|L{.75}|L{.8}|L{.6}|L{1}}
\textbf{Reference} & \textbf{Name} & \textbf{Developer} & \textbf{PQC models} & \textbf{Language} & \textbf{Backend} \\ \hline\hline
Aleksandrowicz \textit{et al.} \cite{qiskit} & Qiskit Aqua & IBM Research & VQE, QAOA, VQM, QKE & Python & Superconducting, Simulator \,\\ \hline
Bergholm \textit{et al.} \cite{bergholm2018pennylane} & Pennylane & Xanadu & VQE, VQM, QGAN & Python & Superconducting, Simulator\\ \hline
Luo \textit{et al.} \cite{Luo2018yao} & Yao & QuantumBFS & VQE, QAOA, QCBM, QGAN & Julia & Simulator \\
\end{tabularx}
\caption{Open-source software for developing machine learning models based on parameterized quantum circuits and, in some cases, for experimenting on existing quantum computers.}
\label{tab:software}
\end{table*}

\FloatBarrier
\twocolumngrid

\section{Outlook}
\label{s:outlook}

In this Review we discussed parameterized quantum circuits (PQCs), a novel framework at the intersection of quantum computing and machine learning. This approach has not been restricted to theory and simulation but involved a series of experimental demonstrations on scaled-down problems being performed in the past two years. In Table~\ref{tab:experimental} we summarize the relevant demonstrations, and the Reader interested in experimental setups is invited to delve into the references therein.

The software development has also been moving at a fast pace (see Fingerhuth \textit{et al.}~\cite{fingerhuth2018open} for a Review of general quantum computing software). There now exist several platforms for hybrid quantum-classical computation which are specifically dedicated to machine learning and provide PQC models, automatic differentiation techniques, and interfaces to both simulators and existing quantum computers. We shall stress here the importance of open-source software and the key role of numerical analysis. While traditional quantum algorithms have been subject to much analytical study of their performance, algorithms for PQC models often relies on heavy numerical study. This is due to the large number of components of the hybrid system, each one affecting the overall performance in a complex way. Open-source software enables experimentation at a much higher rate than previously possible, a scenario reminiscent of the deep learning developments a decade ago. It is therefore recommended to use available libraries when possible, enabling comparison of algorithms on an equal footing and to facilitate the replicability of the results. We summarize the relevant open-source software in Table~\ref{tab:software}, without claiming to be comprehensive.

Researchers have also begun to explore connections between quantum supremacy proposals and quantum algorithms for optimization~\cite{farhi2016quantum}, getting us closer to practical utility if some key requirements can be met~\cite{guerreschi2018qaoa, zhou2018quantum, crooks2018performance}. It is natural to explore similar connections between quantum supremacy and machine learning~\cite{coyle2019born,tangpanitanon2019quantum}.

We have seen that PQCs can implement classically intractable feature maps and kernel functions. Further studies will be needed to assess whether these can improve the performance of established kernel-based models such as the support vector machine, the Gaussian process and the principal component analysis. We also know that sampling from the probability distribution generated by instantaneous quantum polynomial-time circuits is classically intractable in the average case. A natural application for them is in generative modeling where the task itself requires sampling from complex probability distributions. But does classical intractability of these circuits imply an advantage in practice? One possible pitfall is that as the circuits become more expressive, the optimization landscape might also become harder to explore. As previously mentioned, demonstrations on real-world datasets of meaningful scale could answer these questions and should therefore be prioritised.

PQC models can also help in the study of quantum mechanical systems. For systems that exhibit quantum supremacy, a classical model cannot learn to reproduce the statistics unless it uses exponentially scaling resources. Provided that we can efficiently load or prepare quantum data in a qubit register, PQC models will deliver a clear advantage over classical methods for quantum learning tasks.

From the machine learning practitioner's point of view, there are several desirable properties that are naturally captured by PQC models. For example, recurrent neural networks may suffer from the exploding gradient problem. This can be prevented by constraining the operations to be unitary and much work has been done to efficiently parameterize the unitary group~\cite{jing2017tunable,hyland2017learning}. PQC models have the advantage of naturally implementing unitary operations on an exponentially large vector space. As another example, state-of-the-art classical generative models may not allow gradient-based training when the data is discrete~\cite{goodfellow2016nips}. In PQC models discrete data arises from measurements on the qubits and, as we have seen, this does not preclude the computation of gradients. We believe that this is only the "tip of the iceberg" and that there are a number of research opportunities in this field. Largely unexplored aspects of PQC models include Vapnik-Chervonenkis dimensions, regularization techniques, Bayesian inference, and applications to reinforcement learning.

Finally, hybrid systems based on PQCs provide a framework for the incremental development of algorithms. In the near term, hybrid algorithms will rely heavily on classical resources. As quantum hardware improves, classical resources shall gradually be replaced by quantum resources and generic methods. For example, Wang \textit{\textit{et al.}}~\cite{wang2018generalised} propose a method that interpolates between the near-term variational quantum eigensolver and the long-term quantum phase estimation. Similarly, destructive SWAP and Hadamard tests~\cite{garcia2013swap,mitarai2018methodology} could be gradually replaced by non-destructive variants. Hardware-efficient circuits shall be replaced by new parameterizations driven by the theory of tensor networks. Quantum compilers~\cite{cowtan2019qubit,iten2019introduction} will enable the implementation of these higher level constructions on existing devices.

In passing, we envisage that a closer integration between the quantum and the classical components is desirable. This will entail a new generation of hardware facilities, such as hybrid data centers, the improvement of the software interfaces for cloud access to these computational resources, and the development of software frameworks that are \textit{native} of hybrid systems. We believe that the accomplishment of these goals will firstly, facilitate the general research efforts, secondly, it will enable more extensive demonstrations of hybrid algorithms' potential on real-world application, and ultimately pave the way for the implementation in production environments.

The ideas and examples presented in this Review show the remarkable flexibility of the hybrid framework and its potential to use existing quantum hardware to its full extent. If PQC models can be shown to scale well to realistic machine learning tasks, they may become an integral part of automated forecasting and decision-making systems.

\section{Acknowledgements}
The authors would like to thank Tiya-Renee Jones for her help with Figures~\ref{fig:1} and~\ref{f:application}, Ilyas Khan for his support, and Miles Stoudenmire and Leonard Wossnig for useful feedback on an early version of this manuscript. M.B. is supported by the UK Engineering and Physical Sciences Research Council (EPSRC).


\begin{thebibliography}{119}%
\makeatletter
\providecommand \@ifxundefined [1]{%
 \@ifx{#1\undefined}
}%
\providecommand \@ifnum [1]{%
 \ifnum #1\expandafter \@firstoftwo
 \else \expandafter \@secondoftwo
 \fi
}%
\providecommand \@ifx [1]{%
 \ifx #1\expandafter \@firstoftwo
 \else \expandafter \@secondoftwo
 \fi
}%
\providecommand \natexlab [1]{#1}%
\providecommand \enquote  [1]{``#1''}%
\providecommand \bibnamefont  [1]{#1}%
\providecommand \bibfnamefont [1]{#1}%
\providecommand \citenamefont [1]{#1}%
\providecommand \href@noop [0]{\@secondoftwo}%
\providecommand \href [0]{\begingroup \@sanitize@url \@href}%
\providecommand \@href[1]{\@@startlink{#1}\@@href}%
\providecommand \@@href[1]{\endgroup#1\@@endlink}%
\providecommand \@sanitize@url [0]{\catcode `\\12\catcode `\$12\catcode
  `\&12\catcode `\#12\catcode `\^12\catcode `\_12\catcode `\%12\relax}%
\providecommand \@@startlink[1]{}%
\providecommand \@@endlink[0]{}%
\providecommand \url  [0]{\begingroup\@sanitize@url \@url }%
\providecommand \@url [1]{\endgroup\@href {#1}{\urlprefix }}%
\providecommand \urlprefix  [0]{URL }%
\providecommand \Eprint [0]{\href }%
\providecommand \doibase [0]{http://dx.doi.org/}%
\providecommand \selectlanguage [0]{\@gobble}%
\providecommand \bibinfo  [0]{\@secondoftwo}%
\providecommand \bibfield  [0]{\@secondoftwo}%
\providecommand \translation [1]{[#1]}%
\providecommand \BibitemOpen [0]{}%
\providecommand \bibitemStop [0]{}%
\providecommand \bibitemNoStop [0]{.\EOS\space}%
\providecommand \EOS [0]{\spacefactor3000\relax}%
\providecommand \BibitemShut  [1]{\csname bibitem#1\endcsname}%
\let\auto@bib@innerbib\@empty
\bibitem [{\citenamefont {Preskill}(2018)}]{preskill2018quantum}%
  \BibitemOpen
  \bibfield  {author} {\bibinfo {author} {\bibfnamefont {John}\ \bibnamefont
  {Preskill}},\ }\bibfield  {title} {\enquote {\bibinfo {title} {Quantum
  computing in the {NISQ} era and beyond},}\ }\href {\doibase
  10.22331/q-2018-08-06-79} {\bibfield  {journal} {\bibinfo  {journal}
  {Quantum}\ }\textbf {\bibinfo {volume} {2}},\ \bibinfo {pages} {79} (\bibinfo
  {year} {2018})}\BibitemShut {NoStop}%
\bibitem [{\citenamefont {Mohseni}\ \emph {et~al.}(2017)\citenamefont
  {Mohseni}, \citenamefont {Read}, \citenamefont {Neven}, \citenamefont
  {Boixo}, \citenamefont {Denchev}, \citenamefont {Babbush}, \citenamefont
  {Fowler}, \citenamefont {Smelyanskiy},\ and\ \citenamefont
  {Martinis}}]{mohseni2017commercialize}%
  \BibitemOpen
  \bibfield  {author} {\bibinfo {author} {\bibfnamefont {Masoud}\ \bibnamefont
  {Mohseni}}, \bibinfo {author} {\bibfnamefont {Peter}\ \bibnamefont {Read}},
  \bibinfo {author} {\bibfnamefont {Hartmut}\ \bibnamefont {Neven}}, \bibinfo
  {author} {\bibfnamefont {Sergio}\ \bibnamefont {Boixo}}, \bibinfo {author}
  {\bibfnamefont {Vasil}\ \bibnamefont {Denchev}}, \bibinfo {author}
  {\bibfnamefont {Ryan}\ \bibnamefont {Babbush}}, \bibinfo {author}
  {\bibfnamefont {Austin}\ \bibnamefont {Fowler}}, \bibinfo {author}
  {\bibfnamefont {Vadim}\ \bibnamefont {Smelyanskiy}}, \ and\ \bibinfo {author}
  {\bibfnamefont {John}\ \bibnamefont {Martinis}},\ }\bibfield  {title}
  {\enquote {\bibinfo {title} {Commercialize quantum technologies in five
  years},}\ }\href {\doibase 10.1038/543171a} {\bibfield  {journal} {\bibinfo
  {journal} {Nature}\ }\textbf {\bibinfo {volume} {543}},\ \bibinfo {pages}
  {171--174} (\bibinfo {year} {2017})}\BibitemShut {NoStop}%
\bibitem [{\citenamefont {Lund}\ \emph {et~al.}(2017)\citenamefont {Lund},
  \citenamefont {Bremner},\ and\ \citenamefont {Ralph}}]{lund2017quantum}%
  \BibitemOpen
  \bibfield  {author} {\bibinfo {author} {\bibfnamefont {AP}~\bibnamefont
  {Lund}}, \bibinfo {author} {\bibfnamefont {Michael~J}\ \bibnamefont
  {Bremner}}, \ and\ \bibinfo {author} {\bibfnamefont {TC}~\bibnamefont
  {Ralph}},\ }\bibfield  {title} {\enquote {\bibinfo {title} {Quantum sampling
  problems, bosonsampling and quantum supremacy},}\ }\href {\doibase
  10.1038/s41534-017-0018-2} {\bibfield  {journal} {\bibinfo  {journal} {npj
  Quantum Information}\ }\textbf {\bibinfo {volume} {3}},\ \bibinfo {pages}
  {15} (\bibinfo {year} {2017})}\BibitemShut {NoStop}%
\bibitem [{\citenamefont {Harrow}\ and\ \citenamefont
  {Montanaro}(2017)}]{harrow2017quantum}%
  \BibitemOpen
  \bibfield  {author} {\bibinfo {author} {\bibfnamefont {Aram~W}\ \bibnamefont
  {Harrow}}\ and\ \bibinfo {author} {\bibfnamefont {Ashley}\ \bibnamefont
  {Montanaro}},\ }\bibfield  {title} {\enquote {\bibinfo {title} {Quantum
  computational supremacy},}\ }\href {\doibase 10.1038/nature23458} {\bibfield
  {journal} {\bibinfo  {journal} {Nature}\ }\textbf {\bibinfo {volume} {549}},\
  \bibinfo {pages} {203} (\bibinfo {year} {2017})}\BibitemShut {NoStop}%
\bibitem [{\citenamefont {Peruzzo}\ \emph {et~al.}(2014)\citenamefont
  {Peruzzo}, \citenamefont {McClean}, \citenamefont {Shadbolt}, \citenamefont
  {Yung}, \citenamefont {Zhou}, \citenamefont {Love}, \citenamefont
  {Aspuru-Guzik},\ and\ \citenamefont {O’brien}}]{peruzzo2014variational}%
  \BibitemOpen
  \bibfield  {author} {\bibinfo {author} {\bibfnamefont {Alberto}\ \bibnamefont
  {Peruzzo}}, \bibinfo {author} {\bibfnamefont {Jarrod}\ \bibnamefont
  {McClean}}, \bibinfo {author} {\bibfnamefont {Peter}\ \bibnamefont
  {Shadbolt}}, \bibinfo {author} {\bibfnamefont {Man-Hong}\ \bibnamefont
  {Yung}}, \bibinfo {author} {\bibfnamefont {Xiao-Qi}\ \bibnamefont {Zhou}},
  \bibinfo {author} {\bibfnamefont {Peter~J}\ \bibnamefont {Love}}, \bibinfo
  {author} {\bibfnamefont {Al{\'a}n}\ \bibnamefont {Aspuru-Guzik}}, \ and\
  \bibinfo {author} {\bibfnamefont {Jeremy~L}\ \bibnamefont {O’brien}},\
  }\bibfield  {title} {\enquote {\bibinfo {title} {A variational eigenvalue
  solver on a photonic quantum processor},}\ }\href {\doibase
  10.1038/ncomms5213} {\bibfield  {journal} {\bibinfo  {journal} {Nature
  communications}\ }\textbf {\bibinfo {volume} {5}},\ \bibinfo {pages} {4213}
  (\bibinfo {year} {2014})}\BibitemShut {NoStop}%
\bibitem [{\citenamefont {O'Malley}\ \emph {et~al.}(2016)\citenamefont
  {O'Malley}, \citenamefont {Babbush}, \citenamefont {Kivlichan}, \citenamefont
  {Romero}, \citenamefont {McClean}, \citenamefont {Barends}, \citenamefont
  {Kelly}, \citenamefont {Roushan}, \citenamefont {Tranter}, \citenamefont
  {Ding} \emph {et~al.}}]{o2016scalable}%
  \BibitemOpen
  \bibfield  {author} {\bibinfo {author} {\bibfnamefont {PJJ}\ \bibnamefont
  {O'Malley}}, \bibinfo {author} {\bibfnamefont {Ryan}\ \bibnamefont
  {Babbush}}, \bibinfo {author} {\bibfnamefont {ID}~\bibnamefont {Kivlichan}},
  \bibinfo {author} {\bibfnamefont {Jonathan}\ \bibnamefont {Romero}}, \bibinfo
  {author} {\bibfnamefont {JR}~\bibnamefont {McClean}}, \bibinfo {author}
  {\bibfnamefont {Rami}\ \bibnamefont {Barends}}, \bibinfo {author}
  {\bibfnamefont {Julian}\ \bibnamefont {Kelly}}, \bibinfo {author}
  {\bibfnamefont {Pedram}\ \bibnamefont {Roushan}}, \bibinfo {author}
  {\bibfnamefont {Andrew}\ \bibnamefont {Tranter}}, \bibinfo {author}
  {\bibfnamefont {Nan}\ \bibnamefont {Ding}},  \emph {et~al.},\ }\bibfield
  {title} {\enquote {\bibinfo {title} {Scalable quantum simulation of molecular
  energies},}\ }\href {\doibase 10.1103/physrevx.6.031007} {\bibfield
  {journal} {\bibinfo  {journal} {Physical Review X}\ }\textbf {\bibinfo
  {volume} {6}},\ \bibinfo {pages} {031007} (\bibinfo {year}
  {2016})}\BibitemShut {NoStop}%
\bibitem [{\citenamefont {Kandala}\ \emph {et~al.}(2017)\citenamefont
  {Kandala}, \citenamefont {Mezzacapo}, \citenamefont {Temme}, \citenamefont
  {Takita}, \citenamefont {Brink}, \citenamefont {Chow},\ and\ \citenamefont
  {Gambetta}}]{kandala2017hardware}%
  \BibitemOpen
  \bibfield  {author} {\bibinfo {author} {\bibfnamefont {Abhinav}\ \bibnamefont
  {Kandala}}, \bibinfo {author} {\bibfnamefont {Antonio}\ \bibnamefont
  {Mezzacapo}}, \bibinfo {author} {\bibfnamefont {Kristan}\ \bibnamefont
  {Temme}}, \bibinfo {author} {\bibfnamefont {Maika}\ \bibnamefont {Takita}},
  \bibinfo {author} {\bibfnamefont {Markus}\ \bibnamefont {Brink}}, \bibinfo
  {author} {\bibfnamefont {Jerry~M}\ \bibnamefont {Chow}}, \ and\ \bibinfo
  {author} {\bibfnamefont {Jay~M}\ \bibnamefont {Gambetta}},\ }\bibfield
  {title} {\enquote {\bibinfo {title} {Hardware-efficient variational quantum
  eigensolver for small molecules and quantum magnets},}\ }\href {\doibase
  10.1038/nature23879} {\bibfield  {journal} {\bibinfo  {journal} {Nature}\
  }\textbf {\bibinfo {volume} {549}},\ \bibinfo {pages} {242} (\bibinfo {year}
  {2017})}\BibitemShut {NoStop}%
\bibitem [{\citenamefont {Farhi}\ \emph {et~al.}(2014)\citenamefont {Farhi},
  \citenamefont {Goldstone},\ and\ \citenamefont {Gutmann}}]{farhi2014quantum}%
  \BibitemOpen
  \bibfield  {author} {\bibinfo {author} {\bibfnamefont {Edward}\ \bibnamefont
  {Farhi}}, \bibinfo {author} {\bibfnamefont {Jeffrey}\ \bibnamefont
  {Goldstone}}, \ and\ \bibinfo {author} {\bibfnamefont {Sam}\ \bibnamefont
  {Gutmann}},\ }\bibfield  {title} {\enquote {\bibinfo {title} {A quantum
  approximate optimization algorithm},}\ }\href@noop {} {\bibfield  {journal}
  {\bibinfo  {journal} {arXiv preprint arXiv:1411.4028}\ } (\bibinfo {year}
  {2014})}\BibitemShut {NoStop}%
\bibitem [{\citenamefont {Moll}\ \emph {et~al.}(2018)\citenamefont {Moll},
  \citenamefont {Barkoutsos}, \citenamefont {Bishop}, \citenamefont {Chow},
  \citenamefont {Cross}, \citenamefont {Egger}, \citenamefont {Filipp},
  \citenamefont {Fuhrer}, \citenamefont {Gambetta}, \citenamefont {Ganzhorn},
  \citenamefont {Kandala}, \citenamefont {Mezzacapo}, \citenamefont {Müller},
  \citenamefont {Riess}, \citenamefont {Salis}, \citenamefont {Smolin},
  \citenamefont {Tavernelli},\ and\ \citenamefont {Temme}}]{Moll2017}%
  \BibitemOpen
  \bibfield  {author} {\bibinfo {author} {\bibfnamefont {Nikolaj}\ \bibnamefont
  {Moll}}, \bibinfo {author} {\bibfnamefont {Panagiotis}\ \bibnamefont
  {Barkoutsos}}, \bibinfo {author} {\bibfnamefont {Lev~S}\ \bibnamefont
  {Bishop}}, \bibinfo {author} {\bibfnamefont {Jerry~M}\ \bibnamefont {Chow}},
  \bibinfo {author} {\bibfnamefont {Andrew}\ \bibnamefont {Cross}}, \bibinfo
  {author} {\bibfnamefont {Daniel~J}\ \bibnamefont {Egger}}, \bibinfo {author}
  {\bibfnamefont {Stefan}\ \bibnamefont {Filipp}}, \bibinfo {author}
  {\bibfnamefont {Andreas}\ \bibnamefont {Fuhrer}}, \bibinfo {author}
  {\bibfnamefont {Jay~M}\ \bibnamefont {Gambetta}}, \bibinfo {author}
  {\bibfnamefont {Marc}\ \bibnamefont {Ganzhorn}}, \bibinfo {author}
  {\bibfnamefont {Abhinav}\ \bibnamefont {Kandala}}, \bibinfo {author}
  {\bibfnamefont {Antonio}\ \bibnamefont {Mezzacapo}}, \bibinfo {author}
  {\bibfnamefont {Peter}\ \bibnamefont {Müller}}, \bibinfo {author}
  {\bibfnamefont {Walter}\ \bibnamefont {Riess}}, \bibinfo {author}
  {\bibfnamefont {Gian}\ \bibnamefont {Salis}}, \bibinfo {author}
  {\bibfnamefont {John}\ \bibnamefont {Smolin}}, \bibinfo {author}
  {\bibfnamefont {Ivano}\ \bibnamefont {Tavernelli}}, \ and\ \bibinfo {author}
  {\bibfnamefont {Kristan}\ \bibnamefont {Temme}},\ }\bibfield  {title}
  {\enquote {\bibinfo {title} {Quantum optimization using variational
  algorithms on near-term quantum devices},}\ }\href
  {http://stacks.iop.org/2058-9565/3/i=3/a=030503} {\bibfield  {journal}
  {\bibinfo  {journal} {Quantum Science and Technology}\ }\textbf {\bibinfo
  {volume} {3}},\ \bibinfo {pages} {030503} (\bibinfo {year}
  {2018})}\BibitemShut {NoStop}%
\bibitem [{\citenamefont {Otterbach}\ \emph {et~al.}(2017)\citenamefont
  {Otterbach}, \citenamefont {Manenti}, \citenamefont {Alidoust}, \citenamefont
  {Bestwick}, \citenamefont {Block}, \citenamefont {Bloom}, \citenamefont
  {Caldwell}, \citenamefont {Didier}, \citenamefont {Fried}, \citenamefont
  {Hong} \emph {et~al.}}]{otterbach2017unsupervised}%
  \BibitemOpen
  \bibfield  {author} {\bibinfo {author} {\bibfnamefont {JS}~\bibnamefont
  {Otterbach}}, \bibinfo {author} {\bibfnamefont {R}~\bibnamefont {Manenti}},
  \bibinfo {author} {\bibfnamefont {N}~\bibnamefont {Alidoust}}, \bibinfo
  {author} {\bibfnamefont {A}~\bibnamefont {Bestwick}}, \bibinfo {author}
  {\bibfnamefont {M}~\bibnamefont {Block}}, \bibinfo {author} {\bibfnamefont
  {B}~\bibnamefont {Bloom}}, \bibinfo {author} {\bibfnamefont {S}~\bibnamefont
  {Caldwell}}, \bibinfo {author} {\bibfnamefont {N}~\bibnamefont {Didier}},
  \bibinfo {author} {\bibfnamefont {E~Schuyler}\ \bibnamefont {Fried}},
  \bibinfo {author} {\bibfnamefont {S}~\bibnamefont {Hong}},  \emph {et~al.},\
  }\bibfield  {title} {\enquote {\bibinfo {title} {Unsupervised machine
  learning on a hybrid quantum computer},}\ }\href@noop {} {\bibfield
  {journal} {\bibinfo  {journal} {arXiv preprint arXiv:1712.05771}\ } (\bibinfo
  {year} {2017})}\BibitemShut {NoStop}%
\bibitem [{\citenamefont {Bang}\ \emph {et~al.}(2008)\citenamefont {Bang},
  \citenamefont {Lim}, \citenamefont {Kim},\ and\ \citenamefont
  {Lee}}]{bang2008quantum}%
  \BibitemOpen
  \bibfield  {author} {\bibinfo {author} {\bibfnamefont {Jeongho}\ \bibnamefont
  {Bang}}, \bibinfo {author} {\bibfnamefont {James}\ \bibnamefont {Lim}},
  \bibinfo {author} {\bibfnamefont {MS}~\bibnamefont {Kim}}, \ and\ \bibinfo
  {author} {\bibfnamefont {Jinhyoung}\ \bibnamefont {Lee}},\ }\bibfield
  {title} {\enquote {\bibinfo {title} {Quantum learning machine},}\ }\href@noop
  {} {\bibfield  {journal} {\bibinfo  {journal} {arXiv preprint
  arXiv:0803.2976}\ } (\bibinfo {year} {2008})}\BibitemShut {NoStop}%
\bibitem [{\citenamefont {Gammelmark}\ and\ \citenamefont
  {M{\o}lmer}(2009)}]{gammelmark2009quantum}%
  \BibitemOpen
  \bibfield  {author} {\bibinfo {author} {\bibfnamefont {S{\o}ren}\
  \bibnamefont {Gammelmark}}\ and\ \bibinfo {author} {\bibfnamefont {Klaus}\
  \bibnamefont {M{\o}lmer}},\ }\bibfield  {title} {\enquote {\bibinfo {title}
  {Quantum learning by measurement and feedback},}\ }\href {\doibase
  10.1088/1367-2630/11/3/033017} {\bibfield  {journal} {\bibinfo  {journal}
  {New Journal of Physics}\ }\textbf {\bibinfo {volume} {11}},\ \bibinfo
  {pages} {033017} (\bibinfo {year} {2009})}\BibitemShut {NoStop}%
\bibitem [{\citenamefont {Mehta}\ \emph {et~al.}(2019)\citenamefont {Mehta},
  \citenamefont {Bukov}, \citenamefont {Wang}, \citenamefont {Day},
  \citenamefont {Richardson}, \citenamefont {Fisher},\ and\ \citenamefont
  {Schwab}}]{mehta2019high}%
  \BibitemOpen
  \bibfield  {author} {\bibinfo {author} {\bibfnamefont {Pankaj}\ \bibnamefont
  {Mehta}}, \bibinfo {author} {\bibfnamefont {Marin}\ \bibnamefont {Bukov}},
  \bibinfo {author} {\bibfnamefont {Ching-Hao}\ \bibnamefont {Wang}}, \bibinfo
  {author} {\bibfnamefont {Alexandre~GR}\ \bibnamefont {Day}}, \bibinfo
  {author} {\bibfnamefont {Clint}\ \bibnamefont {Richardson}}, \bibinfo
  {author} {\bibfnamefont {Charles~K}\ \bibnamefont {Fisher}}, \ and\ \bibinfo
  {author} {\bibfnamefont {David~J}\ \bibnamefont {Schwab}},\ }\bibfield
  {title} {\enquote {\bibinfo {title} {A high-bias, low-variance introduction
  to machine learning for physicists},}\ }\href {\doibase
  https://doi.org/10.1016/j.physrep.2019.03.001} {\bibfield  {journal}
  {\bibinfo  {journal} {Physics Reports}\ } (\bibinfo {year} {2019}),\
  https://doi.org/10.1016/j.physrep.2019.03.001}\BibitemShut {NoStop}%
\bibitem [{\citenamefont {Nielsen}\ and\ \citenamefont
  {Chuang}(2011)}]{nielsen2002quantum}%
  \BibitemOpen
  \bibfield  {author} {\bibinfo {author} {\bibfnamefont {Michael~A.}\
  \bibnamefont {Nielsen}}\ and\ \bibinfo {author} {\bibfnamefont {Isaac~L.}\
  \bibnamefont {Chuang}},\ }\href@noop {} {\enquote {\bibinfo {title} {Quantum
  computation and quantum information: 10th anniversary edition},}\ } (\bibinfo
  {year} {2011})\BibitemShut {NoStop}%
\bibitem [{\citenamefont {Stoudenmire}\ and\ \citenamefont
  {Schwab}(2016)}]{stoudenmire2016supervised}%
  \BibitemOpen
  \bibfield  {author} {\bibinfo {author} {\bibfnamefont {Edwin}\ \bibnamefont
  {Stoudenmire}}\ and\ \bibinfo {author} {\bibfnamefont {David~J}\ \bibnamefont
  {Schwab}},\ }\bibfield  {title} {\enquote {\bibinfo {title} {Supervised
  learning with tensor networks},}\ }in\ \href
  {http://papers.nips.cc/paper/6211-supervised-learning-with-tensor-networks.pdf}
  {\emph {\bibinfo {booktitle} {Advances in Neural Information Processing
  Systems 29}}},\ \bibinfo {editor} {edited by\ \bibinfo {editor}
  {\bibfnamefont {D.~D.}\ \bibnamefont {Lee}}, \bibinfo {editor} {\bibfnamefont
  {M.}~\bibnamefont {Sugiyama}}, \bibinfo {editor} {\bibfnamefont {U.~V.}\
  \bibnamefont {Luxburg}}, \bibinfo {editor} {\bibfnamefont {I.}~\bibnamefont
  {Guyon}}, \ and\ \bibinfo {editor} {\bibfnamefont {R.}~\bibnamefont
  {Garnett}}}\ (\bibinfo  {publisher} {Curran Associates, Inc.},\ \bibinfo
  {year} {2016})\ pp.\ \bibinfo {pages} {4799--4807}\BibitemShut {NoStop}%
\bibitem [{\citenamefont {Mitarai}\ \emph {et~al.}(2018)\citenamefont
  {Mitarai}, \citenamefont {Negoro}, \citenamefont {Kitagawa},\ and\
  \citenamefont {Fujii}}]{mitarai2018quantum}%
  \BibitemOpen
  \bibfield  {author} {\bibinfo {author} {\bibfnamefont {Kosuke}\ \bibnamefont
  {Mitarai}}, \bibinfo {author} {\bibfnamefont {Makoto}\ \bibnamefont
  {Negoro}}, \bibinfo {author} {\bibfnamefont {Masahiro}\ \bibnamefont
  {Kitagawa}}, \ and\ \bibinfo {author} {\bibfnamefont {Keisuke}\ \bibnamefont
  {Fujii}},\ }\bibfield  {title} {\enquote {\bibinfo {title} {Quantum circuit
  learning},}\ }\href {\doibase 10.1103/PhysRevA.98.032309} {\bibfield
  {journal} {\bibinfo  {journal} {Physical Review A}\ }\textbf {\bibinfo
  {volume} {98}},\ \bibinfo {pages} {032309} (\bibinfo {year}
  {2018})}\BibitemShut {NoStop}%
\bibitem [{\citenamefont {Vidal}\ and\ \citenamefont
  {Theis}(2019)}]{vidal2019input}%
  \BibitemOpen
  \bibfield  {author} {\bibinfo {author} {\bibfnamefont {Javier~Gil}\
  \bibnamefont {Vidal}}\ and\ \bibinfo {author} {\bibfnamefont {Dirk~Oliver}\
  \bibnamefont {Theis}},\ }\bibfield  {title} {\enquote {\bibinfo {title}
  {Input redundancy for parameterized quantum circuits},}\ }\href@noop {}
  {\bibfield  {journal} {\bibinfo  {journal} {arXiv preprint arXiv:1901.11434}\
  } (\bibinfo {year} {2019})}\BibitemShut {NoStop}%
\bibitem [{\citenamefont {Wilson}\ \emph {et~al.}(2018)\citenamefont {Wilson},
  \citenamefont {Otterbach}, \citenamefont {Tezak}, \citenamefont {Smith},
  \citenamefont {Crooks},\ and\ \citenamefont {da~Silva}}]{wilson2018quantum}%
  \BibitemOpen
  \bibfield  {author} {\bibinfo {author} {\bibfnamefont {CM}~\bibnamefont
  {Wilson}}, \bibinfo {author} {\bibfnamefont {JS}~\bibnamefont {Otterbach}},
  \bibinfo {author} {\bibfnamefont {Nikolas}\ \bibnamefont {Tezak}}, \bibinfo
  {author} {\bibfnamefont {RS}~\bibnamefont {Smith}}, \bibinfo {author}
  {\bibfnamefont {GE}~\bibnamefont {Crooks}}, \ and\ \bibinfo {author}
  {\bibfnamefont {MP}~\bibnamefont {da~Silva}},\ }\bibfield  {title} {\enquote
  {\bibinfo {title} {Quantum kitchen sinks: An algorithm for machine learning
  on near-term quantum computers},}\ }\href@noop {} {\bibfield  {journal}
  {\bibinfo  {journal} {arXiv preprint arXiv:1806.08321}\ } (\bibinfo {year}
  {2018})}\BibitemShut {NoStop}%
\bibitem [{\citenamefont {Rahimi}\ and\ \citenamefont
  {Recht}(2008)}]{rahimi2008random}%
  \BibitemOpen
  \bibfield  {author} {\bibinfo {author} {\bibfnamefont {Ali}\ \bibnamefont
  {Rahimi}}\ and\ \bibinfo {author} {\bibfnamefont {Benjamin}\ \bibnamefont
  {Recht}},\ }\bibfield  {title} {\enquote {\bibinfo {title} {Random features
  for large-scale kernel machines},}\ }in\ \href
  {http://papers.nips.cc/paper/3182-random-features-for-large-scale-kernel-machines.pdf}
  {\emph {\bibinfo {booktitle} {Advances in neural information processing
  systems}}}\ (\bibinfo {year} {2008})\ pp.\ \bibinfo {pages}
  {1177--1184}\BibitemShut {NoStop}%
\bibitem [{\citenamefont {Henderson}\ \emph {et~al.}(2019)\citenamefont
  {Henderson}, \citenamefont {Shakya}, \citenamefont {Pradhan},\ and\
  \citenamefont {Cook}}]{henderson2019quanvolutional}%
  \BibitemOpen
  \bibfield  {author} {\bibinfo {author} {\bibfnamefont {Maxwell}\ \bibnamefont
  {Henderson}}, \bibinfo {author} {\bibfnamefont {Samriddhi}\ \bibnamefont
  {Shakya}}, \bibinfo {author} {\bibfnamefont {Shashindra}\ \bibnamefont
  {Pradhan}}, \ and\ \bibinfo {author} {\bibfnamefont {Tristan}\ \bibnamefont
  {Cook}},\ }\bibfield  {title} {\enquote {\bibinfo {title} {Quanvolutional
  neural networks: Powering image recognition with quantum circuits},}\
  }\href@noop {} {\bibfield  {journal} {\bibinfo  {journal} {arXiv preprint
  arXiv:1904.04767}\ } (\bibinfo {year} {2019})}\BibitemShut {NoStop}%
\bibitem [{\citenamefont {Rebentrost}\ \emph {et~al.}(2014)\citenamefont
  {Rebentrost}, \citenamefont {Mohseni},\ and\ \citenamefont
  {Lloyd}}]{rebentrost2014quantum}%
  \BibitemOpen
  \bibfield  {author} {\bibinfo {author} {\bibfnamefont {Patrick}\ \bibnamefont
  {Rebentrost}}, \bibinfo {author} {\bibfnamefont {Masoud}\ \bibnamefont
  {Mohseni}}, \ and\ \bibinfo {author} {\bibfnamefont {Seth}\ \bibnamefont
  {Lloyd}},\ }\bibfield  {title} {\enquote {\bibinfo {title} {Quantum support
  vector machine for big data classification},}\ }\href {\doibase
  10.1103/PhysRevLett.113.130503} {\bibfield  {journal} {\bibinfo  {journal}
  {Phys. Rev. Lett.}\ }\textbf {\bibinfo {volume} {113}},\ \bibinfo {pages}
  {130503} (\bibinfo {year} {2014})}\BibitemShut {NoStop}%
\bibitem [{\citenamefont {Havl{\'\i}{\v{c}}ek}\ \emph
  {et~al.}(2019)\citenamefont {Havl{\'\i}{\v{c}}ek}, \citenamefont
  {C{\'o}rcoles}, \citenamefont {Temme}, \citenamefont {Harrow}, \citenamefont
  {Kandala}, \citenamefont {Chow},\ and\ \citenamefont
  {Gambetta}}]{havlivcek2019supervised}%
  \BibitemOpen
  \bibfield  {author} {\bibinfo {author} {\bibfnamefont {Vojt{\v{e}}ch}\
  \bibnamefont {Havl{\'\i}{\v{c}}ek}}, \bibinfo {author} {\bibfnamefont
  {Antonio~D}\ \bibnamefont {C{\'o}rcoles}}, \bibinfo {author} {\bibfnamefont
  {Kristan}\ \bibnamefont {Temme}}, \bibinfo {author} {\bibfnamefont {Aram~W}\
  \bibnamefont {Harrow}}, \bibinfo {author} {\bibfnamefont {Abhinav}\
  \bibnamefont {Kandala}}, \bibinfo {author} {\bibfnamefont {Jerry~M}\
  \bibnamefont {Chow}}, \ and\ \bibinfo {author} {\bibfnamefont {Jay~M}\
  \bibnamefont {Gambetta}},\ }\bibfield  {title} {\enquote {\bibinfo {title}
  {Supervised learning with quantum-enhanced feature spaces},}\ }\href
  {\doibase 10.1038/s41586-019-0980-2} {\bibfield  {journal} {\bibinfo
  {journal} {Nature}\ }\textbf {\bibinfo {volume} {567}},\ \bibinfo {pages}
  {209} (\bibinfo {year} {2019})}\BibitemShut {NoStop}%
\bibitem [{\citenamefont {R{\"o}tteler}(2010)}]{rotteler2010quantum}%
  \BibitemOpen
  \bibfield  {author} {\bibinfo {author} {\bibfnamefont {Martin}\ \bibnamefont
  {R{\"o}tteler}},\ }\bibfield  {title} {\enquote {\bibinfo {title} {Quantum
  algorithms for highly non-linear boolean functions},}\ }in\ \href
  {http://dl.acm.org/citation.cfm?id=1873601.1873638} {\emph {\bibinfo
  {booktitle} {Proceedings of the twenty-first annual ACM-SIAM symposium on
  Discrete algorithms}}}\ (\bibinfo {organization} {Society for Industrial and
  Applied Mathematics},\ \bibinfo {year} {2010})\ pp.\ \bibinfo {pages}
  {448--457}\BibitemShut {NoStop}%
\bibitem [{\citenamefont {Hornik}\ \emph {et~al.}(1989)\citenamefont {Hornik},
  \citenamefont {Stinchcombe},\ and\ \citenamefont
  {White}}]{hornik1989multilayer}%
  \BibitemOpen
  \bibfield  {author} {\bibinfo {author} {\bibfnamefont {Kurt}\ \bibnamefont
  {Hornik}}, \bibinfo {author} {\bibfnamefont {Maxwell}\ \bibnamefont
  {Stinchcombe}}, \ and\ \bibinfo {author} {\bibfnamefont {Halbert}\
  \bibnamefont {White}},\ }\bibfield  {title} {\enquote {\bibinfo {title}
  {Multilayer feedforward networks are universal approximators},}\ }\href
  {\doibase https://doi.org/10.1016/0893-6080(89)90020-8} {\bibfield  {journal}
  {\bibinfo  {journal} {Neural networks}\ }\textbf {\bibinfo {volume} {2}},\
  \bibinfo {pages} {359--366} (\bibinfo {year} {1989})}\BibitemShut {NoStop}%
\bibitem [{\citenamefont {Lin}\ \emph {et~al.}(2017)\citenamefont {Lin},
  \citenamefont {Tegmark},\ and\ \citenamefont {Rolnick}}]{lin2017does}%
  \BibitemOpen
  \bibfield  {author} {\bibinfo {author} {\bibfnamefont {Henry~W}\ \bibnamefont
  {Lin}}, \bibinfo {author} {\bibfnamefont {Max}\ \bibnamefont {Tegmark}}, \
  and\ \bibinfo {author} {\bibfnamefont {David}\ \bibnamefont {Rolnick}},\
  }\bibfield  {title} {\enquote {\bibinfo {title} {Why does deep and cheap
  learning work so well?}}\ }\href {\doibase 10.1007/s10955-017-1836-5}
  {\bibfield  {journal} {\bibinfo  {journal} {Journal of Statistical Physics}\
  }\textbf {\bibinfo {volume} {168}},\ \bibinfo {pages} {1223--1247} (\bibinfo
  {year} {2017})}\BibitemShut {NoStop}%
\bibitem [{\citenamefont {Liu}\ and\ \citenamefont
  {Wang}(2018)}]{liu2018differentiable}%
  \BibitemOpen
  \bibfield  {author} {\bibinfo {author} {\bibfnamefont {Jin-Guo}\ \bibnamefont
  {Liu}}\ and\ \bibinfo {author} {\bibfnamefont {Lei}\ \bibnamefont {Wang}},\
  }\bibfield  {title} {\enquote {\bibinfo {title} {Differentiable learning of
  quantum circuit born machines},}\ }\href {\doibase
  10.1103/PhysRevA.98.062324} {\bibfield  {journal} {\bibinfo  {journal} {Phys.
  Rev. A}\ }\textbf {\bibinfo {volume} {98}},\ \bibinfo {pages} {062324}
  (\bibinfo {year} {2018})}\BibitemShut {NoStop}%
\bibitem [{\citenamefont {Chow}\ and\ \citenamefont
  {Liu}(1968)}]{chow1968approximating}%
  \BibitemOpen
  \bibfield  {author} {\bibinfo {author} {\bibfnamefont {C}~\bibnamefont
  {Chow}}\ and\ \bibinfo {author} {\bibfnamefont {Cong}\ \bibnamefont {Liu}},\
  }\bibfield  {title} {\enquote {\bibinfo {title} {Approximating discrete
  probability distributions with dependence trees},}\ }\href {\doibase
  10.1109/TIT.1968.1054142} {\bibfield  {journal} {\bibinfo  {journal} {IEEE
  transactions on Information Theory}\ }\textbf {\bibinfo {volume} {14}},\
  \bibinfo {pages} {462--467} (\bibinfo {year} {1968})}\BibitemShut {NoStop}%
\bibitem [{\citenamefont {Liu}\ \emph {et~al.}(2017{\natexlab{a}})\citenamefont
  {Liu}, \citenamefont {Ran}, \citenamefont {Wittek}, \citenamefont {Peng},
  \citenamefont {Garc{\'\i}a}, \citenamefont {Su},\ and\ \citenamefont
  {Lewenstein}}]{liu2017machine}%
  \BibitemOpen
  \bibfield  {author} {\bibinfo {author} {\bibfnamefont {Ding}\ \bibnamefont
  {Liu}}, \bibinfo {author} {\bibfnamefont {Shi-Ju}\ \bibnamefont {Ran}},
  \bibinfo {author} {\bibfnamefont {Peter}\ \bibnamefont {Wittek}}, \bibinfo
  {author} {\bibfnamefont {Cheng}\ \bibnamefont {Peng}}, \bibinfo {author}
  {\bibfnamefont {Raul~Bl{\'a}zquez}\ \bibnamefont {Garc{\'\i}a}}, \bibinfo
  {author} {\bibfnamefont {Gang}\ \bibnamefont {Su}}, \ and\ \bibinfo {author}
  {\bibfnamefont {Maciej}\ \bibnamefont {Lewenstein}},\ }\bibfield  {title}
  {\enquote {\bibinfo {title} {Machine learning by unitary tensor network of
  hierarchical tree structure},}\ }\href@noop {} {\bibfield  {journal}
  {\bibinfo  {journal} {arXiv preprint arXiv:1710.04833}\ } (\bibinfo {year}
  {2017}{\natexlab{a}})}\BibitemShut {NoStop}%
\bibitem [{\citenamefont {Huggins}\ \emph {et~al.}(2019)\citenamefont
  {Huggins}, \citenamefont {Patil}, \citenamefont {Mitchell}, \citenamefont
  {Whaley},\ and\ \citenamefont {Stoudenmire}}]{Huggins2019Towards}%
  \BibitemOpen
  \bibfield  {author} {\bibinfo {author} {\bibfnamefont {William}\ \bibnamefont
  {Huggins}}, \bibinfo {author} {\bibfnamefont {Piyush}\ \bibnamefont {Patil}},
  \bibinfo {author} {\bibfnamefont {Bradley}\ \bibnamefont {Mitchell}},
  \bibinfo {author} {\bibfnamefont {K~Birgitta}\ \bibnamefont {Whaley}}, \ and\
  \bibinfo {author} {\bibfnamefont {E~Miles}\ \bibnamefont {Stoudenmire}},\
  }\bibfield  {title} {\enquote {\bibinfo {title} {Towards quantum machine
  learning with tensor networks},}\ }\href {\doibase 10.1088/2058-9565/aaea94}
  {\bibfield  {journal} {\bibinfo  {journal} {Quantum Science and Technology}\
  }\textbf {\bibinfo {volume} {4}},\ \bibinfo {pages} {024001} (\bibinfo {year}
  {2019})}\BibitemShut {NoStop}%
\bibitem [{\citenamefont {Grant}\ \emph {et~al.}(2018)\citenamefont {Grant},
  \citenamefont {Benedetti}, \citenamefont {Cao}, \citenamefont {Hallam},
  \citenamefont {Lockhart}, \citenamefont {Stojevic}, \citenamefont {Green},\
  and\ \citenamefont {Severini}}]{grant2018hierarchical}%
  \BibitemOpen
  \bibfield  {author} {\bibinfo {author} {\bibfnamefont {Edward}\ \bibnamefont
  {Grant}}, \bibinfo {author} {\bibfnamefont {Marcello}\ \bibnamefont
  {Benedetti}}, \bibinfo {author} {\bibfnamefont {Shuxiang}\ \bibnamefont
  {Cao}}, \bibinfo {author} {\bibfnamefont {Andrew}\ \bibnamefont {Hallam}},
  \bibinfo {author} {\bibfnamefont {Joshua}\ \bibnamefont {Lockhart}}, \bibinfo
  {author} {\bibfnamefont {Vid}\ \bibnamefont {Stojevic}}, \bibinfo {author}
  {\bibfnamefont {Andrew~G.}\ \bibnamefont {Green}}, \ and\ \bibinfo {author}
  {\bibfnamefont {Simone}\ \bibnamefont {Severini}},\ }\bibfield  {title}
  {\enquote {\bibinfo {title} {Hierarchical quantum classifiers},}\ }\href
  {\doibase 10.1038/s41534-018-0116-9} {\bibfield  {journal} {\bibinfo
  {journal} {npj Quantum Information}\ }\textbf {\bibinfo {volume} {4}},\
  \bibinfo {pages} {65} (\bibinfo {year} {2018})}\BibitemShut {NoStop}%
\bibitem [{\citenamefont {McClean}\ \emph {et~al.}(2018)\citenamefont
  {McClean}, \citenamefont {Boixo}, \citenamefont {Smelyanskiy}, \citenamefont
  {Babbush},\ and\ \citenamefont {Neven}}]{mcclean2018barren}%
  \BibitemOpen
  \bibfield  {author} {\bibinfo {author} {\bibfnamefont {Jarrod~R}\
  \bibnamefont {McClean}}, \bibinfo {author} {\bibfnamefont {Sergio}\
  \bibnamefont {Boixo}}, \bibinfo {author} {\bibfnamefont {Vadim~N}\
  \bibnamefont {Smelyanskiy}}, \bibinfo {author} {\bibfnamefont {Ryan}\
  \bibnamefont {Babbush}}, \ and\ \bibinfo {author} {\bibfnamefont {Hartmut}\
  \bibnamefont {Neven}},\ }\bibfield  {title} {\enquote {\bibinfo {title}
  {Barren plateaus in quantum neural network training landscapes},}\ }\href
  {\doibase 10.1038/s41467-018-07090-4} {\bibfield  {journal} {\bibinfo
  {journal} {Nature communications}\ }\textbf {\bibinfo {volume} {9}},\
  \bibinfo {pages} {4812} (\bibinfo {year} {2018})}\BibitemShut {NoStop}%
\bibitem [{\citenamefont {Benedetti}\ \emph
  {et~al.}(2019{\natexlab{a}})\citenamefont {Benedetti}, \citenamefont
  {Garcia-Pintos}, \citenamefont {Perdomo}, \citenamefont {Leyton-Ortega},
  \citenamefont {Nam},\ and\ \citenamefont
  {Perdomo-Ortiz}}]{benedetti2019generative}%
  \BibitemOpen
  \bibfield  {author} {\bibinfo {author} {\bibfnamefont {Marcello}\
  \bibnamefont {Benedetti}}, \bibinfo {author} {\bibfnamefont {Delfina}\
  \bibnamefont {Garcia-Pintos}}, \bibinfo {author} {\bibfnamefont {Oscar}\
  \bibnamefont {Perdomo}}, \bibinfo {author} {\bibfnamefont {Vicente}\
  \bibnamefont {Leyton-Ortega}}, \bibinfo {author} {\bibfnamefont {Yunseong}\
  \bibnamefont {Nam}}, \ and\ \bibinfo {author} {\bibfnamefont {Alejandro}\
  \bibnamefont {Perdomo-Ortiz}},\ }\bibfield  {title} {\enquote {\bibinfo
  {title} {A generative modeling approach for benchmarking and training shallow
  quantum circuits},}\ }\href {\doibase 10.1038/s41534-019-0157-8} {\bibfield
  {journal} {\bibinfo  {journal} {npj Quantum Information}\ }\textbf {\bibinfo
  {volume} {5}},\ \bibinfo {pages} {45} (\bibinfo {year}
  {2019}{\natexlab{a}})}\BibitemShut {NoStop}%
\bibitem [{\citenamefont {Hornik}(1991)}]{hornik1991approximation}%
  \BibitemOpen
  \bibfield  {author} {\bibinfo {author} {\bibfnamefont {Kurt}\ \bibnamefont
  {Hornik}},\ }\bibfield  {title} {\enquote {\bibinfo {title} {Approximation
  capabilities of multilayer feedforward networks},}\ }\href {\doibase
  10.1016/0893-6080(91)90009-T} {\bibfield  {journal} {\bibinfo  {journal}
  {Neural networks}\ }\textbf {\bibinfo {volume} {4}},\ \bibinfo {pages}
  {251--257} (\bibinfo {year} {1991})}\BibitemShut {NoStop}%
\bibitem [{\citenamefont {Cao}\ \emph {et~al.}(2017)\citenamefont {Cao},
  \citenamefont {Guerreschi},\ and\ \citenamefont
  {Aspuru-Guzik}}]{cao2017quantum}%
  \BibitemOpen
  \bibfield  {author} {\bibinfo {author} {\bibfnamefont {Yudong}\ \bibnamefont
  {Cao}}, \bibinfo {author} {\bibfnamefont {Gian~Giacomo}\ \bibnamefont
  {Guerreschi}}, \ and\ \bibinfo {author} {\bibfnamefont {Al{\'a}n}\
  \bibnamefont {Aspuru-Guzik}},\ }\bibfield  {title} {\enquote {\bibinfo
  {title} {Quantum neuron: an elementary building block for machine learning on
  quantum computers},}\ }\href@noop {} {\bibfield  {journal} {\bibinfo
  {journal} {arXiv preprint arXiv:1711.11240}\ } (\bibinfo {year}
  {2017})}\BibitemShut {NoStop}%
\bibitem [{\citenamefont {Torrontegui}\ and\ \citenamefont
  {Garc{\'{\i}}a-Ripoll}(2019)}]{torrontegui2018universal}%
  \BibitemOpen
  \bibfield  {author} {\bibinfo {author} {\bibfnamefont {E.}~\bibnamefont
  {Torrontegui}}\ and\ \bibinfo {author} {\bibfnamefont {J.~J.}\ \bibnamefont
  {Garc{\'{\i}}a-Ripoll}},\ }\bibfield  {title} {\enquote {\bibinfo {title}
  {Unitary quantum perceptron as efficient universal approximator},}\ }\href
  {\doibase 10.1209/0295-5075/125/30004} {\bibfield  {journal} {\bibinfo
  {journal} {{EPL} (Europhysics Letters)}\ }\textbf {\bibinfo {volume} {125}},\
  \bibinfo {pages} {30004} (\bibinfo {year} {2019})}\BibitemShut {NoStop}%
\bibitem [{\citenamefont {Francesco~Tacchino}\ and\ \citenamefont
  {Bajoni}(2019)}]{tacchino2018artificial}%
  \BibitemOpen
  \bibfield  {author} {\bibinfo {author} {\bibfnamefont {Dario~Gerace}\
  \bibnamefont {Francesco~Tacchino}, \bibfnamefont {Chiara~Macchiavello}}\ and\
  \bibinfo {author} {\bibfnamefont {Daniele}\ \bibnamefont {Bajoni}},\
  }\bibfield  {title} {\enquote {\bibinfo {title} {An artificial neuron
  implemented on an actual quantum processor},}\ }\href {\doibase
  10.1038/s41534-019-0140-4} {\bibfield  {journal} {\bibinfo  {journal} {npj
  Quantum Information}\ }\textbf {\bibinfo {volume} {5}},\ \bibinfo {pages}
  {26} (\bibinfo {year} {2019})}\BibitemShut {NoStop}%
\bibitem [{\citenamefont {Hinton}\ \emph {et~al.}(1986)\citenamefont {Hinton},
  \citenamefont {Rumelhart},\ and\ \citenamefont
  {Williams}}]{hinton1986learning}%
  \BibitemOpen
  \bibfield  {author} {\bibinfo {author} {\bibfnamefont {Geoffrey~E}\
  \bibnamefont {Hinton}}, \bibinfo {author} {\bibfnamefont {DE}~\bibnamefont
  {Rumelhart}}, \ and\ \bibinfo {author} {\bibfnamefont {Ronald~J}\
  \bibnamefont {Williams}},\ }\bibfield  {title} {\enquote {\bibinfo {title}
  {Learning representations by back-propagating errors},}\ }\href {\doibase
  10.1038/323533a0} {\bibfield  {journal} {\bibinfo  {journal} {Nature}\
  }\textbf {\bibinfo {volume} {323}},\ \bibinfo {pages} {533--536} (\bibinfo
  {year} {1986})}\BibitemShut {NoStop}%
\bibitem [{\citenamefont {Schuld}\ \emph {et~al.}(2014)\citenamefont {Schuld},
  \citenamefont {Sinayskiy},\ and\ \citenamefont
  {Petruccione}}]{schuld2014quest}%
  \BibitemOpen
  \bibfield  {author} {\bibinfo {author} {\bibfnamefont {Maria}\ \bibnamefont
  {Schuld}}, \bibinfo {author} {\bibfnamefont {Ilya}\ \bibnamefont
  {Sinayskiy}}, \ and\ \bibinfo {author} {\bibfnamefont {Francesco}\
  \bibnamefont {Petruccione}},\ }\bibfield  {title} {\enquote {\bibinfo {title}
  {The quest for a quantum neural network},}\ }\href {\doibase
  10.1007/s11128-014-0809-8} {\bibfield  {journal} {\bibinfo  {journal}
  {Quantum Information Processing}\ }\textbf {\bibinfo {volume} {13}},\
  \bibinfo {pages} {2567--2586} (\bibinfo {year} {2014})}\BibitemShut {NoStop}%
\bibitem [{\citenamefont {Verdon}\ \emph {et~al.}(2018)\citenamefont {Verdon},
  \citenamefont {Pye},\ and\ \citenamefont {Broughton}}]{verdon2018universal}%
  \BibitemOpen
  \bibfield  {author} {\bibinfo {author} {\bibfnamefont {Guillaume}\
  \bibnamefont {Verdon}}, \bibinfo {author} {\bibfnamefont {Jason}\
  \bibnamefont {Pye}}, \ and\ \bibinfo {author} {\bibfnamefont {Michael}\
  \bibnamefont {Broughton}},\ }\bibfield  {title} {\enquote {\bibinfo {title}
  {A universal training algorithm for quantum deep learning},}\ }\href@noop {}
  {\bibfield  {journal} {\bibinfo  {journal} {arXiv preprint arXiv:1806.09729}\
  } (\bibinfo {year} {2018})}\BibitemShut {NoStop}%
\bibitem [{\citenamefont {Beer}\ \emph {et~al.}(2019)\citenamefont {Beer},
  \citenamefont {Bondarenko}, \citenamefont {Farrelly}, \citenamefont
  {Osborne}, \citenamefont {Salzmann},\ and\ \citenamefont
  {Wolf}}]{beer2019efficient}%
  \BibitemOpen
  \bibfield  {author} {\bibinfo {author} {\bibfnamefont {Kerstin}\ \bibnamefont
  {Beer}}, \bibinfo {author} {\bibfnamefont {Dmytro}\ \bibnamefont
  {Bondarenko}}, \bibinfo {author} {\bibfnamefont {Terry}\ \bibnamefont
  {Farrelly}}, \bibinfo {author} {\bibfnamefont {Tobias~J}\ \bibnamefont
  {Osborne}}, \bibinfo {author} {\bibfnamefont {Robert}\ \bibnamefont
  {Salzmann}}, \ and\ \bibinfo {author} {\bibfnamefont {Ramona}\ \bibnamefont
  {Wolf}},\ }\bibfield  {title} {\enquote {\bibinfo {title} {Efficient learning
  for deep quantum neural networks},}\ }\href@noop {} {\bibfield  {journal}
  {\bibinfo  {journal} {arXiv preprint arXiv:1902.10445}\ } (\bibinfo {year}
  {2019})}\BibitemShut {NoStop}%
\bibitem [{\citenamefont {Spall}(1997)}]{spall1997one}%
  \BibitemOpen
  \bibfield  {author} {\bibinfo {author} {\bibfnamefont {James~C}\ \bibnamefont
  {Spall}},\ }\bibfield  {title} {\enquote {\bibinfo {title} {A one-measurement
  form of simultaneous perturbation stochastic approximation},}\ }\href
  {\doibase 10.1016/S0005-1098(96)00149-5} {\bibfield  {journal} {\bibinfo
  {journal} {Automatica}\ }\textbf {\bibinfo {volume} {33}},\ \bibinfo {pages}
  {109--112} (\bibinfo {year} {1997})}\BibitemShut {NoStop}%
\bibitem [{\citenamefont {Spall}(2000)}]{spall2000adaptive}%
  \BibitemOpen
  \bibfield  {author} {\bibinfo {author} {\bibfnamefont {James~C}\ \bibnamefont
  {Spall}},\ }\bibfield  {title} {\enquote {\bibinfo {title} {Adaptive
  stochastic approximation by the simultaneous perturbation method},}\ }\href
  {\doibase 10.1109/CDC.1998.761833} {\bibfield  {journal} {\bibinfo  {journal}
  {IEEE transactions on automatic control}\ }\textbf {\bibinfo {volume} {45}},\
  \bibinfo {pages} {1839--1853} (\bibinfo {year} {2000})}\BibitemShut {NoStop}%
\bibitem [{\citenamefont {Baydin}\ \emph {et~al.}(2018)\citenamefont {Baydin},
  \citenamefont {Pearlmutter}, \citenamefont {Radul},\ and\ \citenamefont
  {Siskind}}]{baydin2018automatic}%
  \BibitemOpen
  \bibfield  {author} {\bibinfo {author} {\bibfnamefont {Atilim~Gunes}\
  \bibnamefont {Baydin}}, \bibinfo {author} {\bibfnamefont {Barak~A}\
  \bibnamefont {Pearlmutter}}, \bibinfo {author} {\bibfnamefont
  {Alexey~Andreyevich}\ \bibnamefont {Radul}}, \ and\ \bibinfo {author}
  {\bibfnamefont {Jeffrey~Mark}\ \bibnamefont {Siskind}},\ }\bibfield  {title}
  {\enquote {\bibinfo {title} {Automatic differentiation in machine learning: a
  survey},}\ }\href {http://dl.acm.org/citation.cfm?id=3122009.3242010}
  {\bibfield  {journal} {\bibinfo  {journal} {Journal of Marchine Learning
  Research}\ }\textbf {\bibinfo {volume} {18}},\ \bibinfo {pages} {1--43}
  (\bibinfo {year} {2018})}\BibitemShut {NoStop}%
\bibitem [{\citenamefont {Li}\ \emph {et~al.}(2017)\citenamefont {Li},
  \citenamefont {Yang}, \citenamefont {Peng},\ and\ \citenamefont
  {Sun}}]{li2017hybrid}%
  \BibitemOpen
  \bibfield  {author} {\bibinfo {author} {\bibfnamefont {Jun}\ \bibnamefont
  {Li}}, \bibinfo {author} {\bibfnamefont {Xiaodong}\ \bibnamefont {Yang}},
  \bibinfo {author} {\bibfnamefont {Xinhua}\ \bibnamefont {Peng}}, \ and\
  \bibinfo {author} {\bibfnamefont {Chang-Pu}\ \bibnamefont {Sun}},\ }\bibfield
   {title} {\enquote {\bibinfo {title} {Hybrid quantum-classical approach to
  quantum optimal control},}\ }\href {\doibase 10.1103/PhysRevLett.118.150503}
  {\bibfield  {journal} {\bibinfo  {journal} {Physical review letters}\
  }\textbf {\bibinfo {volume} {118}},\ \bibinfo {pages} {150503} (\bibinfo
  {year} {2017})}\BibitemShut {NoStop}%
\bibitem [{\citenamefont {Schuld}\ \emph
  {et~al.}(2018{\natexlab{a}})\citenamefont {Schuld}, \citenamefont {Bergholm},
  \citenamefont {Gogolin}, \citenamefont {Izaac},\ and\ \citenamefont
  {Killoran}}]{schuld2018evaluating}%
  \BibitemOpen
  \bibfield  {author} {\bibinfo {author} {\bibfnamefont {Maria}\ \bibnamefont
  {Schuld}}, \bibinfo {author} {\bibfnamefont {Ville}\ \bibnamefont
  {Bergholm}}, \bibinfo {author} {\bibfnamefont {Christian}\ \bibnamefont
  {Gogolin}}, \bibinfo {author} {\bibfnamefont {Josh}\ \bibnamefont {Izaac}}, \
  and\ \bibinfo {author} {\bibfnamefont {Nathan}\ \bibnamefont {Killoran}},\
  }\bibfield  {title} {\enquote {\bibinfo {title} {Evaluating analytic
  gradients on quantum hardware},}\ }\href@noop {} {\bibfield  {journal}
  {\bibinfo  {journal} {arXiv preprint arXiv:1811.11184}\ } (\bibinfo {year}
  {2018}{\natexlab{a}})}\BibitemShut {NoStop}%
\bibitem [{\citenamefont {Farhi}\ and\ \citenamefont
  {Neven}(2018)}]{farhi2018classification}%
  \BibitemOpen
  \bibfield  {author} {\bibinfo {author} {\bibfnamefont {Edward}\ \bibnamefont
  {Farhi}}\ and\ \bibinfo {author} {\bibfnamefont {Hartmut}\ \bibnamefont
  {Neven}},\ }\bibfield  {title} {\enquote {\bibinfo {title} {Classification
  with quantum neural networks on near term processors},}\ }\href@noop {}
  {\bibfield  {journal} {\bibinfo  {journal} {arXiv:1802.06002}\ } (\bibinfo
  {year} {2018})}\BibitemShut {NoStop}%
\bibitem [{\citenamefont {Dallaire-Demers}\ and\ \citenamefont
  {Killoran}(2018)}]{dallaire2018quantum}%
  \BibitemOpen
  \bibfield  {author} {\bibinfo {author} {\bibfnamefont {Pierre-Luc}\
  \bibnamefont {Dallaire-Demers}}\ and\ \bibinfo {author} {\bibfnamefont
  {Nathan}\ \bibnamefont {Killoran}},\ }\bibfield  {title} {\enquote {\bibinfo
  {title} {Quantum generative adversarial networks},}\ }\href {\doibase
  10.1103/PhysRevA.98.012324} {\bibfield  {journal} {\bibinfo  {journal}
  {Physical Review A}\ }\textbf {\bibinfo {volume} {98}},\ \bibinfo {pages}
  {012324} (\bibinfo {year} {2018})}\BibitemShut {NoStop}%
\bibitem [{\citenamefont {Schuld}\ \emph
  {et~al.}(2018{\natexlab{b}})\citenamefont {Schuld}, \citenamefont {Bocharov},
  \citenamefont {Svore},\ and\ \citenamefont {Wiebe}}]{schuld2018circuit}%
  \BibitemOpen
  \bibfield  {author} {\bibinfo {author} {\bibfnamefont {Maria}\ \bibnamefont
  {Schuld}}, \bibinfo {author} {\bibfnamefont {Alex}\ \bibnamefont {Bocharov}},
  \bibinfo {author} {\bibfnamefont {Krysta}\ \bibnamefont {Svore}}, \ and\
  \bibinfo {author} {\bibfnamefont {Nathan}\ \bibnamefont {Wiebe}},\ }\bibfield
   {title} {\enquote {\bibinfo {title} {Circuit-centric quantum classifiers},}\
  }\href@noop {} {\bibfield  {journal} {\bibinfo  {journal} {arXiv preprint
  arXiv:1804.00633}\ } (\bibinfo {year} {2018}{\natexlab{b}})}\BibitemShut
  {NoStop}%
\bibitem [{\citenamefont {Mitarai}\ and\ \citenamefont
  {Fujii}(2018)}]{mitarai2018methodology}%
  \BibitemOpen
  \bibfield  {author} {\bibinfo {author} {\bibfnamefont {Kosuke}\ \bibnamefont
  {Mitarai}}\ and\ \bibinfo {author} {\bibfnamefont {Keisuke}\ \bibnamefont
  {Fujii}},\ }\bibfield  {title} {\enquote {\bibinfo {title} {Methodology for
  replacing indirect measurements with direct measurements},}\ }\href@noop {}
  {\bibfield  {journal} {\bibinfo  {journal} {arXiv preprint arXiv:1901.00015}\
  } (\bibinfo {year} {2018})}\BibitemShut {NoStop}%
\bibitem [{\citenamefont {Harrow}\ and\ \citenamefont
  {Napp}(2019)}]{harrow2019low}%
  \BibitemOpen
  \bibfield  {author} {\bibinfo {author} {\bibfnamefont {Aram}\ \bibnamefont
  {Harrow}}\ and\ \bibinfo {author} {\bibfnamefont {John}\ \bibnamefont
  {Napp}},\ }\bibfield  {title} {\enquote {\bibinfo {title} {Low-depth gradient
  measurements can improve convergence in variational hybrid quantum-classical
  algorithms},}\ }\href@noop {} {\bibfield  {journal} {\bibinfo  {journal}
  {arXiv preprint arXiv:1901.05374}\ } (\bibinfo {year} {2019})}\BibitemShut
  {NoStop}%
\bibitem [{\citenamefont {Wisdom}\ \emph {et~al.}(2016)\citenamefont {Wisdom},
  \citenamefont {Powers}, \citenamefont {Hershey}, \citenamefont {Roux},\ and\
  \citenamefont {Atlas}}]{wisdom2016full}%
  \BibitemOpen
  \bibfield  {author} {\bibinfo {author} {\bibfnamefont {Scott}\ \bibnamefont
  {Wisdom}}, \bibinfo {author} {\bibfnamefont {Thomas}\ \bibnamefont {Powers}},
  \bibinfo {author} {\bibfnamefont {John~R.}\ \bibnamefont {Hershey}}, \bibinfo
  {author} {\bibfnamefont {Jonathan~Le}\ \bibnamefont {Roux}}, \ and\ \bibinfo
  {author} {\bibfnamefont {Les}\ \bibnamefont {Atlas}},\ }\bibfield  {title}
  {\enquote {\bibinfo {title} {Full-capacity unitary recurrent neural
  networks},}\ }in\ \href {http://dl.acm.org/citation.cfm?id=3157382.3157643}
  {\emph {\bibinfo {booktitle} {Proceedings of the 30th International
  Conference on Neural Information Processing Systems}}},\ \bibinfo {series and
  number} {NIPS'16}\ (\bibinfo  {publisher} {Curran Associates Inc.},\ \bibinfo
  {address} {USA},\ \bibinfo {year} {2016})\ pp.\ \bibinfo {pages}
  {4887--4895}\BibitemShut {NoStop}%
\bibitem [{\citenamefont {Ledoux}(2001)}]{ledoux2001concentration}%
  \BibitemOpen
  \bibfield  {author} {\bibinfo {author} {\bibfnamefont {M.}~\bibnamefont
  {Ledoux}},\ }\href {https://books.google.co.uk/books?id=mHS-oQEACAAJ} {\emph
  {\bibinfo {title} {The Concentration of Measure Phenomenon}}},\ Mathematical
  surveys and monographs\ (\bibinfo  {publisher} {American Mathematical
  Society},\ \bibinfo {year} {2001})\BibitemShut {NoStop}%
\bibitem [{\citenamefont {Grant}\ \emph {et~al.}(2019)\citenamefont {Grant},
  \citenamefont {Wossnig}, \citenamefont {Ostaszewski},\ and\ \citenamefont
  {Benedetti}}]{grant2019initialization}%
  \BibitemOpen
  \bibfield  {author} {\bibinfo {author} {\bibfnamefont {Edward}\ \bibnamefont
  {Grant}}, \bibinfo {author} {\bibfnamefont {Leonard}\ \bibnamefont
  {Wossnig}}, \bibinfo {author} {\bibfnamefont {Mateusz}\ \bibnamefont
  {Ostaszewski}}, \ and\ \bibinfo {author} {\bibfnamefont {Marcello}\
  \bibnamefont {Benedetti}},\ }\bibfield  {title} {\enquote {\bibinfo {title}
  {An initialization strategy for addressing barren plateaus in parametrized
  quantum circuits},}\ }\href@noop {} {\bibfield  {journal} {\bibinfo
  {journal} {arXiv preprint arXiv:1903.05076}\ } (\bibinfo {year}
  {2019})}\BibitemShut {NoStop}%
\bibitem [{\citenamefont {Robbins}\ and\ \citenamefont
  {Monro}(1951)}]{robbins1951stochastic}%
  \BibitemOpen
  \bibfield  {author} {\bibinfo {author} {\bibfnamefont {Herbert}\ \bibnamefont
  {Robbins}}\ and\ \bibinfo {author} {\bibfnamefont {Sutton}\ \bibnamefont
  {Monro}},\ }\bibfield  {title} {\enquote {\bibinfo {title} {A stochastic
  approximation method},}\ }\href {\doibase 10.1214/aoms/1177729586} {\bibfield
   {journal} {\bibinfo  {journal} {The annals of mathematical statistics}\ ,\
  \bibinfo {pages} {400--407}} (\bibinfo {year} {1951})}\BibitemShut {NoStop}%
\bibitem [{\citenamefont {Riedmiller}\ and\ \citenamefont
  {Braun}(1993)}]{riedmiller1993direct}%
  \BibitemOpen
  \bibfield  {author} {\bibinfo {author} {\bibfnamefont {Martin}\ \bibnamefont
  {Riedmiller}}\ and\ \bibinfo {author} {\bibfnamefont {Heinrich}\ \bibnamefont
  {Braun}},\ }\bibfield  {title} {\enquote {\bibinfo {title} {A direct adaptive
  method for faster backpropagation learning: The rprop algorithm},}\ }in\
  \href {\doibase 10.1109/ICNN.1993.298623} {\emph {\bibinfo {booktitle} {IEEE
  International Conference on Neural Networks}}}\ (\bibinfo {year} {1993})\
  pp.\ \bibinfo {pages} {586--591 vol.1}\BibitemShut {NoStop}%
\bibitem [{\citenamefont {Kingma}\ and\ \citenamefont
  {Ba}(2014)}]{kingma2014adam}%
  \BibitemOpen
  \bibfield  {author} {\bibinfo {author} {\bibfnamefont {Diederik~P}\
  \bibnamefont {Kingma}}\ and\ \bibinfo {author} {\bibfnamefont {Jimmy}\
  \bibnamefont {Ba}},\ }\bibfield  {title} {\enquote {\bibinfo {title} {Adam: A
  method for stochastic optimization},}\ }\href@noop {} {\bibfield  {journal}
  {\bibinfo  {journal} {arXiv preprint arXiv:1412.6980}\ } (\bibinfo {year}
  {2014})}\BibitemShut {NoStop}%
\bibitem [{\citenamefont {Eberhart}\ and\ \citenamefont
  {Hu}(1999)}]{eberhart1999human}%
  \BibitemOpen
  \bibfield  {author} {\bibinfo {author} {\bibfnamefont {Russell~C}\
  \bibnamefont {Eberhart}}\ and\ \bibinfo {author} {\bibfnamefont {Xiaohui}\
  \bibnamefont {Hu}},\ }\bibfield  {title} {\enquote {\bibinfo {title} {Human
  tremor analysis using particle swarm optimization},}\ }in\ \href {\doibase
  10.1109/CEC.1999.785508} {\emph {\bibinfo {booktitle} {Proceedings of the
  1999 Congress on Evolutionary Computation-CEC99 (Cat. No. 99TH8406)}}},\
  Vol.~\bibinfo {volume} {3}\ (\bibinfo {organization} {IEEE},\ \bibinfo {year}
  {1999})\ pp.\ \bibinfo {pages} {1927--1930}\BibitemShut {NoStop}%
\bibitem [{\citenamefont {Frazier}(2018)}]{frazier2018tutorial}%
  \BibitemOpen
  \bibfield  {author} {\bibinfo {author} {\bibfnamefont {Peter~I}\ \bibnamefont
  {Frazier}},\ }\bibfield  {title} {\enquote {\bibinfo {title} {A tutorial on
  bayesian optimization},}\ }\href@noop {} {\bibfield  {journal} {\bibinfo
  {journal} {arXiv preprint arXiv:1807.02811}\ } (\bibinfo {year}
  {2018})}\BibitemShut {NoStop}%
\bibitem [{\citenamefont {Zhu}\ \emph {et~al.}(2018)\citenamefont {Zhu},
  \citenamefont {Linke}, \citenamefont {Benedetti}, \citenamefont {Landsman},
  \citenamefont {Nguyen}, \citenamefont {Alderete}, \citenamefont
  {Perdomo-Ortiz}, \citenamefont {Korda}, \citenamefont {Garfoot},
  \citenamefont {Brecque} \emph {et~al.}}]{zhu2018training}%
  \BibitemOpen
  \bibfield  {author} {\bibinfo {author} {\bibfnamefont {D}~\bibnamefont
  {Zhu}}, \bibinfo {author} {\bibfnamefont {NM}~\bibnamefont {Linke}}, \bibinfo
  {author} {\bibfnamefont {M}~\bibnamefont {Benedetti}}, \bibinfo {author}
  {\bibfnamefont {KA}~\bibnamefont {Landsman}}, \bibinfo {author}
  {\bibfnamefont {NH}~\bibnamefont {Nguyen}}, \bibinfo {author} {\bibfnamefont
  {CH}~\bibnamefont {Alderete}}, \bibinfo {author} {\bibfnamefont
  {A}~\bibnamefont {Perdomo-Ortiz}}, \bibinfo {author} {\bibfnamefont
  {N}~\bibnamefont {Korda}}, \bibinfo {author} {\bibfnamefont {A}~\bibnamefont
  {Garfoot}}, \bibinfo {author} {\bibfnamefont {C}~\bibnamefont {Brecque}},
  \emph {et~al.},\ }\bibfield  {title} {\enquote {\bibinfo {title} {Training of
  quantum circuits on a hybrid quantum computer},}\ }\href@noop {} {\bibfield
  {journal} {\bibinfo  {journal} {arXiv preprint arXiv:1812.08862}\ } (\bibinfo
  {year} {2018})}\BibitemShut {NoStop}%
\bibitem [{\citenamefont {Leyton-Ortega}\ \emph {et~al.}(2019)\citenamefont
  {Leyton-Ortega}, \citenamefont {Perdomo-Ortiz},\ and\ \citenamefont
  {Perdomo}}]{leyton2019robust}%
  \BibitemOpen
  \bibfield  {author} {\bibinfo {author} {\bibfnamefont {Vicente}\ \bibnamefont
  {Leyton-Ortega}}, \bibinfo {author} {\bibfnamefont {Alejandro}\ \bibnamefont
  {Perdomo-Ortiz}}, \ and\ \bibinfo {author} {\bibfnamefont {Oscar}\
  \bibnamefont {Perdomo}},\ }\bibfield  {title} {\enquote {\bibinfo {title}
  {Robust implementation of generative modeling with parametrized quantum
  circuits},}\ }\href@noop {} {\bibfield  {journal} {\bibinfo  {journal} {arXiv
  preprint arXiv:1901.08047}\ } (\bibinfo {year} {2019})}\BibitemShut {NoStop}%
\bibitem [{\citenamefont {Liu}\ \emph {et~al.}(2017{\natexlab{b}})\citenamefont
  {Liu}, \citenamefont {Hu}, \citenamefont {Qian}, \citenamefont {Yu},\ and\
  \citenamefont {Qian}}]{liu2017zoopt}%
  \BibitemOpen
  \bibfield  {author} {\bibinfo {author} {\bibfnamefont {Yu-Ren}\ \bibnamefont
  {Liu}}, \bibinfo {author} {\bibfnamefont {Yi-Qi}\ \bibnamefont {Hu}},
  \bibinfo {author} {\bibfnamefont {Hong}\ \bibnamefont {Qian}}, \bibinfo
  {author} {\bibfnamefont {Yang}\ \bibnamefont {Yu}}, \ and\ \bibinfo {author}
  {\bibfnamefont {Chao}\ \bibnamefont {Qian}},\ }\bibfield  {title} {\enquote
  {\bibinfo {title} {Zoopt: Toolbox for derivative-free optimization},}\
  }\href@noop {} {\bibfield  {journal} {\bibinfo  {journal} {arXiv preprint
  arXiv:1801.00329}\ } (\bibinfo {year} {2017}{\natexlab{b}})}\BibitemShut
  {NoStop}%
\bibitem [{\citenamefont {Sastry}\ \emph {et~al.}(2005)\citenamefont {Sastry},
  \citenamefont {Goldberg},\ and\ \citenamefont {Kendall}}]{sastry2005genetic}%
  \BibitemOpen
  \bibfield  {author} {\bibinfo {author} {\bibfnamefont {Kumara}\ \bibnamefont
  {Sastry}}, \bibinfo {author} {\bibfnamefont {David}\ \bibnamefont
  {Goldberg}}, \ and\ \bibinfo {author} {\bibfnamefont {Graham}\ \bibnamefont
  {Kendall}},\ }\bibfield  {title} {\enquote {\bibinfo {title} {Genetic
  algorithms},}\ }in\ \href {\doibase 10.1007/0-387-28356-0_4} {\emph {\bibinfo
  {booktitle} {Search methodologies}}}\ (\bibinfo  {publisher} {Springer},\
  \bibinfo {year} {2005})\ pp.\ \bibinfo {pages} {97--125}\BibitemShut
  {NoStop}%
\bibitem [{\citenamefont {Lamata}\ \emph {et~al.}(2018)\citenamefont {Lamata},
  \citenamefont {Alvarez-Rodriguez}, \citenamefont {Mart{\'\i}n-Guerrero},
  \citenamefont {Sanz},\ and\ \citenamefont {Solano}}]{lamata2018quantum}%
  \BibitemOpen
  \bibfield  {author} {\bibinfo {author} {\bibfnamefont {Lucas}\ \bibnamefont
  {Lamata}}, \bibinfo {author} {\bibfnamefont {Unai}\ \bibnamefont
  {Alvarez-Rodriguez}}, \bibinfo {author} {\bibfnamefont {Jos{\'e}~David}\
  \bibnamefont {Mart{\'\i}n-Guerrero}}, \bibinfo {author} {\bibfnamefont
  {Mikel}\ \bibnamefont {Sanz}}, \ and\ \bibinfo {author} {\bibfnamefont
  {Enrique}\ \bibnamefont {Solano}},\ }\bibfield  {title} {\enquote {\bibinfo
  {title} {Quantum autoencoders via quantum adders with genetic algorithms},}\
  }\href {\doibase 10.1088/2058-9565/aae22b} {\bibfield  {journal} {\bibinfo
  {journal} {Quantum Science and Technology}\ }\textbf {\bibinfo {volume}
  {4}},\ \bibinfo {pages} {014007} (\bibinfo {year} {2018})}\BibitemShut
  {NoStop}%
\bibitem [{\citenamefont {Ding}\ \emph {et~al.}(2019)\citenamefont {Ding},
  \citenamefont {Lamata}, \citenamefont {Sanz}, \citenamefont {Chen},\ and\
  \citenamefont {Solano}}]{ding2019experimental}%
  \BibitemOpen
  \bibfield  {author} {\bibinfo {author} {\bibfnamefont {Yongcheng}\
  \bibnamefont {Ding}}, \bibinfo {author} {\bibfnamefont {Lucas}\ \bibnamefont
  {Lamata}}, \bibinfo {author} {\bibfnamefont {Mikel}\ \bibnamefont {Sanz}},
  \bibinfo {author} {\bibfnamefont {Xi}~\bibnamefont {Chen}}, \ and\ \bibinfo
  {author} {\bibfnamefont {Enrique}\ \bibnamefont {Solano}},\ }\bibfield
  {title} {\enquote {\bibinfo {title} {Experimental implementation of a quantum
  autoencoder via quantum adders},}\ }\href {\doibase 10.1002/qute.201800065}
  {\bibfield  {journal} {\bibinfo  {journal} {Advanced Quantum Technologies}\
  ,\ \bibinfo {pages} {1800065}} (\bibinfo {year} {2019})}\BibitemShut
  {NoStop}%
\bibitem [{\citenamefont {Ostaszewski}\ \emph {et~al.}(2019)\citenamefont
  {Ostaszewski}, \citenamefont {Grant},\ and\ \citenamefont
  {Benedetti}}]{ostaszewski2019quantum}%
  \BibitemOpen
  \bibfield  {author} {\bibinfo {author} {\bibfnamefont {Mateusz}\ \bibnamefont
  {Ostaszewski}}, \bibinfo {author} {\bibfnamefont {Edward}\ \bibnamefont
  {Grant}}, \ and\ \bibinfo {author} {\bibfnamefont {Marcello}\ \bibnamefont
  {Benedetti}},\ }\bibfield  {title} {\enquote {\bibinfo {title} {Quantum
  circuit structure learning},}\ }\href@noop {} {\bibfield  {journal} {\bibinfo
   {journal} {arXiv preprint arXiv:1905.09692}\ } (\bibinfo {year}
  {2019})}\BibitemShut {NoStop}%
\bibitem [{\citenamefont {Nakanishi}\ \emph {et~al.}(2019)\citenamefont
  {Nakanishi}, \citenamefont {Fujii},\ and\ \citenamefont
  {Todo}}]{nakanishi2019sequential}%
  \BibitemOpen
  \bibfield  {author} {\bibinfo {author} {\bibfnamefont {Ken~M}\ \bibnamefont
  {Nakanishi}}, \bibinfo {author} {\bibfnamefont {Keisuke}\ \bibnamefont
  {Fujii}}, \ and\ \bibinfo {author} {\bibfnamefont {Synge}\ \bibnamefont
  {Todo}},\ }\bibfield  {title} {\enquote {\bibinfo {title} {Sequential minimal
  optimization for quantum-classical hybrid algorithms},}\ }\href@noop {}
  {\bibfield  {journal} {\bibinfo  {journal} {arXiv preprint arXiv:1903.12166}\
  } (\bibinfo {year} {2019})}\BibitemShut {NoStop}%
\bibitem [{\citenamefont {A{\"\i}meur}\ \emph {et~al.}(2006)\citenamefont
  {A{\"\i}meur}, \citenamefont {Brassard},\ and\ \citenamefont
  {Gambs}}]{aimeur2006machine}%
  \BibitemOpen
  \bibfield  {author} {\bibinfo {author} {\bibfnamefont {Esma}\ \bibnamefont
  {A{\"\i}meur}}, \bibinfo {author} {\bibfnamefont {Gilles}\ \bibnamefont
  {Brassard}}, \ and\ \bibinfo {author} {\bibfnamefont {S{\'e}bastien}\
  \bibnamefont {Gambs}},\ }\bibfield  {title} {\enquote {\bibinfo {title}
  {Machine learning in a quantum world},}\ }in\ \href@noop {} {\emph {\bibinfo
  {booktitle} {Conference of the Canadian Society for Computational Studies of
  Intelligence}}}\ (\bibinfo {organization} {Springer},\ \bibinfo {year}
  {2006})\ pp.\ \bibinfo {pages} {431--442}\BibitemShut {NoStop}%
\bibitem [{\citenamefont {Kimble}(2008)}]{kimble2008quantum}%
  \BibitemOpen
  \bibfield  {author} {\bibinfo {author} {\bibfnamefont {H~Jeff}\ \bibnamefont
  {Kimble}},\ }\bibfield  {title} {\enquote {\bibinfo {title} {The quantum
  internet},}\ }\href {\doibase 10.1038/nature07127} {\bibfield  {journal}
  {\bibinfo  {journal} {Nature}\ }\textbf {\bibinfo {volume} {453}},\ \bibinfo
  {pages} {1023} (\bibinfo {year} {2008})}\BibitemShut {NoStop}%
\bibitem [{\citenamefont {Schuld}\ and\ \citenamefont
  {Killoran}(2019)}]{Schuld2019quantum}%
  \BibitemOpen
  \bibfield  {author} {\bibinfo {author} {\bibfnamefont {Maria}\ \bibnamefont
  {Schuld}}\ and\ \bibinfo {author} {\bibfnamefont {Nathan}\ \bibnamefont
  {Killoran}},\ }\bibfield  {title} {\enquote {\bibinfo {title} {Quantum
  machine learning in feature hilbert spaces},}\ }\href {\doibase
  10.1103/PhysRevLett.122.040504} {\bibfield  {journal} {\bibinfo  {journal}
  {Phys. Rev. Lett.}\ }\textbf {\bibinfo {volume} {122}},\ \bibinfo {pages}
  {040504} (\bibinfo {year} {2019})}\BibitemShut {NoStop}%
\bibitem [{\citenamefont {Sch{\"o}lkopf}\ \emph {et~al.}(2001)\citenamefont
  {Sch{\"o}lkopf}, \citenamefont {Herbrich},\ and\ \citenamefont
  {Smola}}]{Scholkopf2001Generalized}%
  \BibitemOpen
  \bibfield  {author} {\bibinfo {author} {\bibfnamefont {Bernhard}\
  \bibnamefont {Sch{\"o}lkopf}}, \bibinfo {author} {\bibfnamefont {Ralf}\
  \bibnamefont {Herbrich}}, \ and\ \bibinfo {author} {\bibfnamefont {Alex~J.}\
  \bibnamefont {Smola}},\ }\bibfield  {title} {\enquote {\bibinfo {title} {A
  generalized representer theorem},}\ }in\ \href@noop {} {\emph {\bibinfo
  {booktitle} {Computational Learning Theory}}},\ \bibinfo {editor} {edited by\
  \bibinfo {editor} {\bibfnamefont {David}\ \bibnamefont {Helmbold}}\ and\
  \bibinfo {editor} {\bibfnamefont {Bob}\ \bibnamefont {Williamson}}}\
  (\bibinfo  {publisher} {Springer Berlin Heidelberg},\ \bibinfo {address}
  {Berlin, Heidelberg},\ \bibinfo {year} {2001})\ pp.\ \bibinfo {pages}
  {416--426}\BibitemShut {NoStop}%
\bibitem [{\citenamefont {Cheng}\ \emph {et~al.}(2018)\citenamefont {Cheng},
  \citenamefont {Chen},\ and\ \citenamefont {Wang}}]{cheng2017information}%
  \BibitemOpen
  \bibfield  {author} {\bibinfo {author} {\bibfnamefont {Song}\ \bibnamefont
  {Cheng}}, \bibinfo {author} {\bibfnamefont {Jing}\ \bibnamefont {Chen}}, \
  and\ \bibinfo {author} {\bibfnamefont {Lei}\ \bibnamefont {Wang}},\
  }\bibfield  {title} {\enquote {\bibinfo {title} {Information perspective to
  probabilistic modeling: Boltzmann machines versus born machines},}\ }\href
  {\doibase 10.3390/e20080583} {\bibfield  {journal} {\bibinfo  {journal}
  {Entropy}\ }\textbf {\bibinfo {volume} {20}},\ \bibinfo {pages} {583}
  (\bibinfo {year} {2018})}\BibitemShut {NoStop}%
\bibitem [{\citenamefont {Han}\ \emph {et~al.}(2018)\citenamefont {Han},
  \citenamefont {Wang}, \citenamefont {Fan}, \citenamefont {Wang},\ and\
  \citenamefont {Zhang}}]{han2017unsupervised}%
  \BibitemOpen
  \bibfield  {author} {\bibinfo {author} {\bibfnamefont {Zhao-Yu}\ \bibnamefont
  {Han}}, \bibinfo {author} {\bibfnamefont {Jun}\ \bibnamefont {Wang}},
  \bibinfo {author} {\bibfnamefont {Heng}\ \bibnamefont {Fan}}, \bibinfo
  {author} {\bibfnamefont {Lei}\ \bibnamefont {Wang}}, \ and\ \bibinfo {author}
  {\bibfnamefont {Pan}\ \bibnamefont {Zhang}},\ }\bibfield  {title} {\enquote
  {\bibinfo {title} {Unsupervised generative modeling using matrix product
  states},}\ }\href {\doibase 10.1103/physrevx.8.031012} {\bibfield  {journal}
  {\bibinfo  {journal} {Physical Review X}\ }\textbf {\bibinfo {volume} {8}}
  (\bibinfo {year} {2018}),\ 10.1103/physrevx.8.031012}\BibitemShut {NoStop}%
\bibitem [{\citenamefont {Goodfellow}(2016)}]{goodfellow2016nips}%
  \BibitemOpen
  \bibfield  {author} {\bibinfo {author} {\bibfnamefont {Ian}\ \bibnamefont
  {Goodfellow}},\ }\bibfield  {title} {\enquote {\bibinfo {title} {Nips 2016
  tutorial: Generative adversarial networks},}\ }\href@noop {} {\bibfield
  {journal} {\bibinfo  {journal} {arXiv preprint arXiv:1701.00160}\ } (\bibinfo
  {year} {2016})}\BibitemShut {NoStop}%
\bibitem [{\citenamefont {Du}\ \emph {et~al.}(2018)\citenamefont {Du},
  \citenamefont {Hsieh}, \citenamefont {Liu},\ and\ \citenamefont
  {Tao}}]{du2018expressive}%
  \BibitemOpen
  \bibfield  {author} {\bibinfo {author} {\bibfnamefont {Yuxuan}\ \bibnamefont
  {Du}}, \bibinfo {author} {\bibfnamefont {Min-Hsiu}\ \bibnamefont {Hsieh}},
  \bibinfo {author} {\bibfnamefont {Tongliang}\ \bibnamefont {Liu}}, \ and\
  \bibinfo {author} {\bibfnamefont {Dacheng}\ \bibnamefont {Tao}},\ }\bibfield
  {title} {\enquote {\bibinfo {title} {The expressive power of parameterized
  quantum circuits},}\ }\href@noop {} {\bibfield  {journal} {\bibinfo
  {journal} {arXiv preprint arXiv:1810.11922}\ } (\bibinfo {year}
  {2018})}\BibitemShut {NoStop}%
\bibitem [{\citenamefont {Coyle}\ \emph {et~al.}(2019)\citenamefont {Coyle},
  \citenamefont {Mills}, \citenamefont {Danos},\ and\ \citenamefont
  {Kashefi}}]{coyle2019born}%
  \BibitemOpen
  \bibfield  {author} {\bibinfo {author} {\bibfnamefont {Brian}\ \bibnamefont
  {Coyle}}, \bibinfo {author} {\bibfnamefont {Daniel}\ \bibnamefont {Mills}},
  \bibinfo {author} {\bibfnamefont {Vincent}\ \bibnamefont {Danos}}, \ and\
  \bibinfo {author} {\bibfnamefont {Elham}\ \bibnamefont {Kashefi}},\
  }\bibfield  {title} {\enquote {\bibinfo {title} {The born supremacy: Quantum
  advantage and training of an ising born machine},}\ }\href@noop {} {\bibfield
   {journal} {\bibinfo  {journal} {arXiv preprint arXiv:1904.02214}\ }
  (\bibinfo {year} {2019})}\BibitemShut {NoStop}%
\bibitem [{\citenamefont {Zoufal}\ \emph {et~al.}(2019)\citenamefont {Zoufal},
  \citenamefont {Lucchi},\ and\ \citenamefont {Woerner}}]{zoufal2019quantum}%
  \BibitemOpen
  \bibfield  {author} {\bibinfo {author} {\bibfnamefont {Christa}\ \bibnamefont
  {Zoufal}}, \bibinfo {author} {\bibfnamefont {Aur{\'e}lien}\ \bibnamefont
  {Lucchi}}, \ and\ \bibinfo {author} {\bibfnamefont {Stefan}\ \bibnamefont
  {Woerner}},\ }\bibfield  {title} {\enquote {\bibinfo {title} {Quantum
  generative adversarial networks for learning and loading random
  distributions},}\ }\href@noop {} {\bibfield  {journal} {\bibinfo  {journal}
  {arXiv preprint arXiv:1904.00043}\ } (\bibinfo {year} {2019})}\BibitemShut
  {NoStop}%
\bibitem [{\citenamefont {Romero}\ \emph {et~al.}(2017)\citenamefont {Romero},
  \citenamefont {Olson},\ and\ \citenamefont
  {Aspuru-Guzik}}]{romero2017quantum}%
  \BibitemOpen
  \bibfield  {author} {\bibinfo {author} {\bibfnamefont {Jonathan}\
  \bibnamefont {Romero}}, \bibinfo {author} {\bibfnamefont {Jonathan~P}\
  \bibnamefont {Olson}}, \ and\ \bibinfo {author} {\bibfnamefont {Alan}\
  \bibnamefont {Aspuru-Guzik}},\ }\bibfield  {title} {\enquote {\bibinfo
  {title} {Quantum autoencoders for efficient compression of quantum data},}\
  }\href {https://doi.org/10.1088/2058-9565/aa8072} {\bibfield  {journal}
  {\bibinfo  {journal} {Quantum Science and Technology}\ }\textbf {\bibinfo
  {volume} {2}},\ \bibinfo {pages} {045001} (\bibinfo {year}
  {2017})}\BibitemShut {NoStop}%
\bibitem [{\citenamefont {Kullback}\ and\ \citenamefont
  {Leibler}(1951)}]{kullback1951information}%
  \BibitemOpen
  \bibfield  {author} {\bibinfo {author} {\bibfnamefont {Solomon}\ \bibnamefont
  {Kullback}}\ and\ \bibinfo {author} {\bibfnamefont {Richard~A}\ \bibnamefont
  {Leibler}},\ }\bibfield  {title} {\enquote {\bibinfo {title} {On information
  and sufficiency},}\ }\href {\doibase 10.1214/aoms/1177729694} {\bibfield
  {journal} {\bibinfo  {journal} {The annals of mathematical statistics}\
  }\textbf {\bibinfo {volume} {22}},\ \bibinfo {pages} {79--86} (\bibinfo
  {year} {1951})}\BibitemShut {NoStop}%
\bibitem [{\citenamefont {Gretton}\ \emph {et~al.}(2007)\citenamefont
  {Gretton}, \citenamefont {Borgwardt}, \citenamefont {Rasch}, \citenamefont
  {Scholk\"{o}pf},\ and\ \citenamefont {Smola}}]{gretton2007kernel}%
  \BibitemOpen
  \bibfield  {author} {\bibinfo {author} {\bibfnamefont {Arthur}\ \bibnamefont
  {Gretton}}, \bibinfo {author} {\bibfnamefont {Karsten~M.}\ \bibnamefont
  {Borgwardt}}, \bibinfo {author} {\bibfnamefont {Malte}\ \bibnamefont
  {Rasch}}, \bibinfo {author} {\bibfnamefont {Bernhard}\ \bibnamefont
  {Scholk\"{o}pf}}, \ and\ \bibinfo {author} {\bibfnamefont {Alexander~J.}\
  \bibnamefont {Smola}},\ }\bibfield  {title} {\enquote {\bibinfo {title} {A
  kernel approach to comparing distributions},}\ }in\ \href
  {http://dl.acm.org/citation.cfm?id=1619797.1619910} {\emph {\bibinfo
  {booktitle} {Proceedings of the 22Nd National Conference on Artificial
  Intelligence - Volume 2}}},\ \bibinfo {series and number} {AAAI'07}\
  (\bibinfo  {publisher} {AAAI Press},\ \bibinfo {year} {2007})\ pp.\ \bibinfo
  {pages} {1637--1641}\BibitemShut {NoStop}%
\bibitem [{\citenamefont {Hamilton}\ \emph {et~al.}(2018)\citenamefont
  {Hamilton}, \citenamefont {Dumitrescu},\ and\ \citenamefont
  {Pooser}}]{hamilton2018generative}%
  \BibitemOpen
  \bibfield  {author} {\bibinfo {author} {\bibfnamefont {Kathleen~E}\
  \bibnamefont {Hamilton}}, \bibinfo {author} {\bibfnamefont {Eugene~F}\
  \bibnamefont {Dumitrescu}}, \ and\ \bibinfo {author} {\bibfnamefont
  {Raphael~C}\ \bibnamefont {Pooser}},\ }\bibfield  {title} {\enquote {\bibinfo
  {title} {Generative model benchmarks for superconducting qubits},}\
  }\href@noop {} {\bibfield  {journal} {\bibinfo  {journal} {arXiv preprint
  arXiv:1811.09905}\ } (\bibinfo {year} {2018})}\BibitemShut {NoStop}%
\bibitem [{\citenamefont {Lloyd}\ and\ \citenamefont
  {Weedbrook}(2018)}]{lloyd2018quantum}%
  \BibitemOpen
  \bibfield  {author} {\bibinfo {author} {\bibfnamefont {Seth}\ \bibnamefont
  {Lloyd}}\ and\ \bibinfo {author} {\bibfnamefont {Christian}\ \bibnamefont
  {Weedbrook}},\ }\bibfield  {title} {\enquote {\bibinfo {title} {Quantum
  generative adversarial learning},}\ }\href {\doibase
  10.1103/PhysRevLett.121.040502} {\bibfield  {journal} {\bibinfo  {journal}
  {Physical review letters}\ }\textbf {\bibinfo {volume} {121}},\ \bibinfo
  {pages} {040502} (\bibinfo {year} {2018})}\BibitemShut {NoStop}%
\bibitem [{\citenamefont {Situ}\ \emph {et~al.}(2018)\citenamefont {Situ},
  \citenamefont {He}, \citenamefont {Li},\ and\ \citenamefont
  {Zheng}}]{situ2018adversarial}%
  \BibitemOpen
  \bibfield  {author} {\bibinfo {author} {\bibfnamefont {Haozhen}\ \bibnamefont
  {Situ}}, \bibinfo {author} {\bibfnamefont {Zhimin}\ \bibnamefont {He}},
  \bibinfo {author} {\bibfnamefont {Lvzhou}\ \bibnamefont {Li}}, \ and\
  \bibinfo {author} {\bibfnamefont {Shenggen}\ \bibnamefont {Zheng}},\
  }\bibfield  {title} {\enquote {\bibinfo {title} {Quantum generative
  adversarial network for generating discrete data},}\ }\href@noop {}
  {\bibfield  {journal} {\bibinfo  {journal} {arXiv preprint arXiv:1807.01235}\
  } (\bibinfo {year} {2018})}\BibitemShut {NoStop}%
\bibitem [{\citenamefont {Zeng}\ \emph {et~al.}(2019)\citenamefont {Zeng},
  \citenamefont {Wu}, \citenamefont {Liu}, \citenamefont {Wang},\ and\
  \citenamefont {Hu}}]{zeng2018learning}%
  \BibitemOpen
  \bibfield  {author} {\bibinfo {author} {\bibfnamefont {Jinfeng}\ \bibnamefont
  {Zeng}}, \bibinfo {author} {\bibfnamefont {Yufeng}\ \bibnamefont {Wu}},
  \bibinfo {author} {\bibfnamefont {Jin-Guo}\ \bibnamefont {Liu}}, \bibinfo
  {author} {\bibfnamefont {Lei}\ \bibnamefont {Wang}}, \ and\ \bibinfo {author}
  {\bibfnamefont {Jiangping}\ \bibnamefont {Hu}},\ }\bibfield  {title}
  {\enquote {\bibinfo {title} {Learning and inference on generative adversarial
  quantum circuits},}\ }\href {\doibase 10.1103/PhysRevA.99.052306} {\bibfield
  {journal} {\bibinfo  {journal} {Phys. Rev. A}\ }\textbf {\bibinfo {volume}
  {99}},\ \bibinfo {pages} {052306} (\bibinfo {year} {2019})}\BibitemShut
  {NoStop}%
\bibitem [{\citenamefont {Romero}\ and\ \citenamefont
  {Aspuru-Guzik}(2019)}]{romero2019variational}%
  \BibitemOpen
  \bibfield  {author} {\bibinfo {author} {\bibfnamefont {Jonathan}\
  \bibnamefont {Romero}}\ and\ \bibinfo {author} {\bibfnamefont {Alan}\
  \bibnamefont {Aspuru-Guzik}},\ }\bibfield  {title} {\enquote {\bibinfo
  {title} {Variational quantum generators: Generative adversarial quantum
  machine learning for continuous distributions},}\ }\href@noop {} {\bibfield
  {journal} {\bibinfo  {journal} {arXiv preprint arXiv:1901.00848}\ } (\bibinfo
  {year} {2019})}\BibitemShut {NoStop}%
\bibitem [{\citenamefont {Low}\ \emph {et~al.}(2014)\citenamefont {Low},
  \citenamefont {Yoder},\ and\ \citenamefont {Chuang}}]{low2014quantum}%
  \BibitemOpen
  \bibfield  {author} {\bibinfo {author} {\bibfnamefont {Guang~Hao}\
  \bibnamefont {Low}}, \bibinfo {author} {\bibfnamefont {Theodore~J}\
  \bibnamefont {Yoder}}, \ and\ \bibinfo {author} {\bibfnamefont {Isaac~L}\
  \bibnamefont {Chuang}},\ }\bibfield  {title} {\enquote {\bibinfo {title}
  {Quantum inference on bayesian networks},}\ }\href {\doibase
  10.1103/PhysRevA.89.062315} {\bibfield  {journal} {\bibinfo  {journal}
  {Physical Review A}\ }\textbf {\bibinfo {volume} {89}},\ \bibinfo {pages}
  {062315} (\bibinfo {year} {2014})}\BibitemShut {NoStop}%
\bibitem [{\citenamefont {Morales}\ \emph {et~al.}(2018)\citenamefont
  {Morales}, \citenamefont {Tlyachev},\ and\ \citenamefont
  {Biamonte}}]{morales2018variational}%
  \BibitemOpen
  \bibfield  {author} {\bibinfo {author} {\bibfnamefont {Mauro~ES}\
  \bibnamefont {Morales}}, \bibinfo {author} {\bibfnamefont {Timur}\
  \bibnamefont {Tlyachev}}, \ and\ \bibinfo {author} {\bibfnamefont {Jacob}\
  \bibnamefont {Biamonte}},\ }\bibfield  {title} {\enquote {\bibinfo {title}
  {Variational learning of grover's quantum search algorithm},}\ }\href
  {\doibase 10.1103/PhysRevA.98.062333} {\bibfield  {journal} {\bibinfo
  {journal} {Physical Review A}\ }\textbf {\bibinfo {volume} {98}},\ \bibinfo
  {pages} {062333} (\bibinfo {year} {2018})}\BibitemShut {NoStop}%
\bibitem [{\citenamefont {Wan}\ \emph {et~al.}(2018)\citenamefont {Wan},
  \citenamefont {Liu}, \citenamefont {Dahlsten},\ and\ \citenamefont
  {Kim}}]{wan2018learning}%
  \BibitemOpen
  \bibfield  {author} {\bibinfo {author} {\bibfnamefont {Kwok~Ho}\ \bibnamefont
  {Wan}}, \bibinfo {author} {\bibfnamefont {Feiyang}\ \bibnamefont {Liu}},
  \bibinfo {author} {\bibfnamefont {Oscar}\ \bibnamefont {Dahlsten}}, \ and\
  \bibinfo {author} {\bibfnamefont {MS}~\bibnamefont {Kim}},\ }\bibfield
  {title} {\enquote {\bibinfo {title} {Learning simon's quantum algorithm},}\
  }\href@noop {} {\bibfield  {journal} {\bibinfo  {journal} {arXiv preprint
  arXiv:1806.10448}\ } (\bibinfo {year} {2018})}\BibitemShut {NoStop}%
\bibitem [{\citenamefont {Simon}(1997)}]{simon1997power}%
  \BibitemOpen
  \bibfield  {author} {\bibinfo {author} {\bibfnamefont {Daniel~R}\
  \bibnamefont {Simon}},\ }\bibfield  {title} {\enquote {\bibinfo {title} {On
  the power of quantum computation},}\ }\href
  {https://doi.org/10.1137/S0097539796298637} {\bibfield  {journal} {\bibinfo
  {journal} {SIAM journal on computing}\ }\textbf {\bibinfo {volume} {26}},\
  \bibinfo {pages} {1474--1483} (\bibinfo {year} {1997})}\BibitemShut {NoStop}%
\bibitem [{\citenamefont {Anschuetz}\ \emph {et~al.}(2019)\citenamefont
  {Anschuetz}, \citenamefont {Olson}, \citenamefont {Aspuru-Guzik},\ and\
  \citenamefont {Cao}}]{anschuetz2019variational}%
  \BibitemOpen
  \bibfield  {author} {\bibinfo {author} {\bibfnamefont {Eric}\ \bibnamefont
  {Anschuetz}}, \bibinfo {author} {\bibfnamefont {Jonathan}\ \bibnamefont
  {Olson}}, \bibinfo {author} {\bibfnamefont {Al{\'a}n}\ \bibnamefont
  {Aspuru-Guzik}}, \ and\ \bibinfo {author} {\bibfnamefont {Yudong}\
  \bibnamefont {Cao}},\ }\bibfield  {title} {\enquote {\bibinfo {title}
  {Variational quantum factoring},}\ }in\ \href
  {https://link.springer.com/chapter/10.1007/978-3-030-14082-3_7} {\emph
  {\bibinfo {booktitle} {International Workshop on Quantum Technology and
  Optimization Problems}}}\ (\bibinfo {organization} {Springer},\ \bibinfo
  {year} {2019})\ pp.\ \bibinfo {pages} {74--85}\BibitemShut {NoStop}%
\bibitem [{\citenamefont {Cincio}\ \emph {et~al.}(2018)\citenamefont {Cincio},
  \citenamefont {Suba{\c{s}}{\i}}, \citenamefont {Sornborger},\ and\
  \citenamefont {Coles}}]{cincio2018learning}%
  \BibitemOpen
  \bibfield  {author} {\bibinfo {author} {\bibfnamefont {Lukasz}\ \bibnamefont
  {Cincio}}, \bibinfo {author} {\bibfnamefont {Yi{\u{g}}it}\ \bibnamefont
  {Suba{\c{s}}{\i}}}, \bibinfo {author} {\bibfnamefont {Andrew~T}\ \bibnamefont
  {Sornborger}}, \ and\ \bibinfo {author} {\bibfnamefont {Patrick~J}\
  \bibnamefont {Coles}},\ }\bibfield  {title} {\enquote {\bibinfo {title}
  {Learning the quantum algorithm for state overlap},}\ }\href
  {https://doi.org/10.1088/1367-2630/aae94a} {\bibfield  {journal} {\bibinfo
  {journal} {New Journal of Physics}\ }\textbf {\bibinfo {volume} {20}},\
  \bibinfo {pages} {113022} (\bibinfo {year} {2018})}\BibitemShut {NoStop}%
\bibitem [{\citenamefont {Chen}\ \emph {et~al.}(2018)\citenamefont {Chen},
  \citenamefont {Wossnig}, \citenamefont {Severini}, \citenamefont {Neven},\
  and\ \citenamefont {Mohseni}}]{chen2018universal}%
  \BibitemOpen
  \bibfield  {author} {\bibinfo {author} {\bibfnamefont {Hongxiang}\
  \bibnamefont {Chen}}, \bibinfo {author} {\bibfnamefont {Leonard}\
  \bibnamefont {Wossnig}}, \bibinfo {author} {\bibfnamefont {Simone}\
  \bibnamefont {Severini}}, \bibinfo {author} {\bibfnamefont {Hartmut}\
  \bibnamefont {Neven}}, \ and\ \bibinfo {author} {\bibfnamefont {Masoud}\
  \bibnamefont {Mohseni}},\ }\bibfield  {title} {\enquote {\bibinfo {title}
  {Universal discriminative quantum neural networks},}\ }\href@noop {}
  {\bibfield  {journal} {\bibinfo  {journal} {arXiv preprint arXiv:1805.08654}\
  } (\bibinfo {year} {2018})}\BibitemShut {NoStop}%
\bibitem [{\citenamefont {Aaronson}(2007)}]{aaronson2007learnability}%
  \BibitemOpen
  \bibfield  {author} {\bibinfo {author} {\bibfnamefont {Scott}\ \bibnamefont
  {Aaronson}},\ }\bibfield  {title} {\enquote {\bibinfo {title} {The
  learnability of quantum states},}\ }\href
  {https://doi.org/10.1098/rspa.2007.0113} {\bibfield  {journal} {\bibinfo
  {journal} {Proceedings of the Royal Society A: Mathematical, Physical and
  Engineering Sciences}\ }\textbf {\bibinfo {volume} {463}},\ \bibinfo {pages}
  {3089--3114} (\bibinfo {year} {2007})}\BibitemShut {NoStop}%
\bibitem [{\citenamefont {Valiant}(1984)}]{valiant1984theory}%
  \BibitemOpen
  \bibfield  {author} {\bibinfo {author} {\bibfnamefont {Leslie~G}\
  \bibnamefont {Valiant}},\ }\bibfield  {title} {\enquote {\bibinfo {title} {A
  theory of the learnable},}\ }in\ \href {\doibase 10.1145/800057.808710}
  {\emph {\bibinfo {booktitle} {Proceedings of the sixteenth annual ACM
  symposium on Theory of computing}}}\ (\bibinfo {organization} {ACM},\
  \bibinfo {year} {1984})\ pp.\ \bibinfo {pages} {436--445}\BibitemShut
  {NoStop}%
\bibitem [{\citenamefont {Rocchetto}\ \emph {et~al.}(2019)\citenamefont
  {Rocchetto}, \citenamefont {Aaronson}, \citenamefont {Severini},
  \citenamefont {Carvacho}, \citenamefont {Poderini}, \citenamefont {Agresti},
  \citenamefont {Bentivegna},\ and\ \citenamefont
  {Sciarrino}}]{rocchetto2019experimental}%
  \BibitemOpen
  \bibfield  {author} {\bibinfo {author} {\bibfnamefont {Andrea}\ \bibnamefont
  {Rocchetto}}, \bibinfo {author} {\bibfnamefont {Scott}\ \bibnamefont
  {Aaronson}}, \bibinfo {author} {\bibfnamefont {Simone}\ \bibnamefont
  {Severini}}, \bibinfo {author} {\bibfnamefont {Gonzalo}\ \bibnamefont
  {Carvacho}}, \bibinfo {author} {\bibfnamefont {Davide}\ \bibnamefont
  {Poderini}}, \bibinfo {author} {\bibfnamefont {Iris}\ \bibnamefont
  {Agresti}}, \bibinfo {author} {\bibfnamefont {Marco}\ \bibnamefont
  {Bentivegna}}, \ and\ \bibinfo {author} {\bibfnamefont {Fabio}\ \bibnamefont
  {Sciarrino}},\ }\bibfield  {title} {\enquote {\bibinfo {title} {Experimental
  learning of quantum states},}\ }\href {\doibase 10.1126/sciadv.aau1946}
  {\bibfield  {journal} {\bibinfo  {journal} {Science Advances}\ }\textbf
  {\bibinfo {volume} {5}},\ \bibinfo {pages} {eaau1946} (\bibinfo {year}
  {2019})}\BibitemShut {NoStop}%
\bibitem [{\citenamefont {Lee}\ \emph {et~al.}(2018)\citenamefont {Lee},
  \citenamefont {Lee},\ and\ \citenamefont {Bang}}]{lee2018learning}%
  \BibitemOpen
  \bibfield  {author} {\bibinfo {author} {\bibfnamefont {Sang~Min}\
  \bibnamefont {Lee}}, \bibinfo {author} {\bibfnamefont {Jinhyoung}\
  \bibnamefont {Lee}}, \ and\ \bibinfo {author} {\bibfnamefont {Jeongho}\
  \bibnamefont {Bang}},\ }\bibfield  {title} {\enquote {\bibinfo {title}
  {Learning unknown pure quantum states},}\ }\href {\doibase
  10.1103/PhysRevA.98.052302} {\bibfield  {journal} {\bibinfo  {journal}
  {Physical Review A}\ }\textbf {\bibinfo {volume} {98}},\ \bibinfo {pages}
  {052302} (\bibinfo {year} {2018})}\BibitemShut {NoStop}%
\bibitem [{\citenamefont {LaRose}\ \emph {et~al.}(2018)\citenamefont {LaRose},
  \citenamefont {Tikku}, \citenamefont {O'Neel-Judy}, \citenamefont {Cincio},\
  and\ \citenamefont {Coles}}]{larose2018variational}%
  \BibitemOpen
  \bibfield  {author} {\bibinfo {author} {\bibfnamefont {Ryan}\ \bibnamefont
  {LaRose}}, \bibinfo {author} {\bibfnamefont {Arkin}\ \bibnamefont {Tikku}},
  \bibinfo {author} {\bibfnamefont {{\'E}tude}\ \bibnamefont {O'Neel-Judy}},
  \bibinfo {author} {\bibfnamefont {Lukasz}\ \bibnamefont {Cincio}}, \ and\
  \bibinfo {author} {\bibfnamefont {Patrick~J}\ \bibnamefont {Coles}},\
  }\bibfield  {title} {\enquote {\bibinfo {title} {Variational quantum state
  diagonalization},}\ }\href@noop {} {\bibfield  {journal} {\bibinfo  {journal}
  {arXiv preprint arXiv:1810.10506}\ } (\bibinfo {year} {2018})}\BibitemShut
  {NoStop}%
\bibitem [{\citenamefont {Helstrom}(1969)}]{helstrom1969quantum}%
  \BibitemOpen
  \bibfield  {author} {\bibinfo {author} {\bibfnamefont {Carl~W}\ \bibnamefont
  {Helstrom}},\ }\bibfield  {title} {\enquote {\bibinfo {title} {Quantum
  detection and estimation theory},}\ }\href {\doibase 10.1007/BF01007479}
  {\bibfield  {journal} {\bibinfo  {journal} {Journal of Statistical Physics}\
  }\textbf {\bibinfo {volume} {1}},\ \bibinfo {pages} {231--252} (\bibinfo
  {year} {1969})}\BibitemShut {NoStop}%
\bibitem [{\citenamefont {Benedetti}\ \emph
  {et~al.}(2019{\natexlab{b}})\citenamefont {Benedetti}, \citenamefont {Grant},
  \citenamefont {Wossnig},\ and\ \citenamefont
  {Severini}}]{benedetti2019adversarial}%
  \BibitemOpen
  \bibfield  {author} {\bibinfo {author} {\bibfnamefont {Marcello}\
  \bibnamefont {Benedetti}}, \bibinfo {author} {\bibfnamefont {Edward}\
  \bibnamefont {Grant}}, \bibinfo {author} {\bibfnamefont {Leonard}\
  \bibnamefont {Wossnig}}, \ and\ \bibinfo {author} {\bibfnamefont {Simone}\
  \bibnamefont {Severini}},\ }\bibfield  {title} {\enquote {\bibinfo {title}
  {Adversarial quantum circuit learning for pure state approximation},}\ }\href
  {\doibase 10.1088/1367-2630/ab14b5} {\bibfield  {journal} {\bibinfo
  {journal} {New Journal of Physics}\ } (\bibinfo {year}
  {2019}{\natexlab{b}}),\ 10.1088/1367-2630/ab14b5}\BibitemShut {NoStop}%
\bibitem [{\citenamefont {Hu}\ \emph {et~al.}(2019)\citenamefont {Hu},
  \citenamefont {Wu}, \citenamefont {Cai}, \citenamefont {Ma}, \citenamefont
  {Mu}, \citenamefont {Xu}, \citenamefont {Wang}, \citenamefont {Song},
  \citenamefont {Deng}, \citenamefont {Zou} \emph {et~al.}}]{hu2019quantum}%
  \BibitemOpen
  \bibfield  {author} {\bibinfo {author} {\bibfnamefont {Ling}\ \bibnamefont
  {Hu}}, \bibinfo {author} {\bibfnamefont {Shu-Hao}\ \bibnamefont {Wu}},
  \bibinfo {author} {\bibfnamefont {Weizhou}\ \bibnamefont {Cai}}, \bibinfo
  {author} {\bibfnamefont {Yuwei}\ \bibnamefont {Ma}}, \bibinfo {author}
  {\bibfnamefont {Xianghao}\ \bibnamefont {Mu}}, \bibinfo {author}
  {\bibfnamefont {Yuan}\ \bibnamefont {Xu}}, \bibinfo {author} {\bibfnamefont
  {Haiyan}\ \bibnamefont {Wang}}, \bibinfo {author} {\bibfnamefont {Yipu}\
  \bibnamefont {Song}}, \bibinfo {author} {\bibfnamefont {Dong-Ling}\
  \bibnamefont {Deng}}, \bibinfo {author} {\bibfnamefont {Chang-Ling}\
  \bibnamefont {Zou}},  \emph {et~al.},\ }\bibfield  {title} {\enquote
  {\bibinfo {title} {Quantum generative adversarial learning in a
  superconducting quantum circuit},}\ }\href {\doibase 10.1126/sciadv.aav2761}
  {\bibfield  {journal} {\bibinfo  {journal} {Science advances}\ }\textbf
  {\bibinfo {volume} {5}},\ \bibinfo {pages} {eaav2761} (\bibinfo {year}
  {2019})}\BibitemShut {NoStop}%
\bibitem [{\citenamefont {Gottesman}(2010)}]{gottesman2010introduction}%
  \BibitemOpen
  \bibfield  {author} {\bibinfo {author} {\bibfnamefont {Daniel}\ \bibnamefont
  {Gottesman}},\ }\bibfield  {title} {\enquote {\bibinfo {title} {An
  introduction to quantum error correction and fault-tolerant quantum
  computation},}\ }in\ \href {\doibase 10.1090/psapm/068/2762145} {\emph
  {\bibinfo {booktitle} {Quantum information science and its contributions to
  mathematics, Proceedings of Symposia in Applied Mathematics}}},\
  Vol.~\bibinfo {volume} {68}\ (\bibinfo {year} {2010})\ pp.\ \bibinfo {pages}
  {13--58}\BibitemShut {NoStop}%
\bibitem [{\citenamefont {Johnson}\ \emph {et~al.}(2017)\citenamefont
  {Johnson}, \citenamefont {Romero}, \citenamefont {Olson}, \citenamefont
  {Cao},\ and\ \citenamefont {Aspuru-Guzik}}]{johnson2017qvector}%
  \BibitemOpen
  \bibfield  {author} {\bibinfo {author} {\bibfnamefont {Peter~D}\ \bibnamefont
  {Johnson}}, \bibinfo {author} {\bibfnamefont {Jonathan}\ \bibnamefont
  {Romero}}, \bibinfo {author} {\bibfnamefont {Jonathan}\ \bibnamefont
  {Olson}}, \bibinfo {author} {\bibfnamefont {Yudong}\ \bibnamefont {Cao}}, \
  and\ \bibinfo {author} {\bibfnamefont {Al{\'a}n}\ \bibnamefont
  {Aspuru-Guzik}},\ }\bibfield  {title} {\enquote {\bibinfo {title} {Qvector:
  an algorithm for device-tailored quantum error correction},}\ }\href@noop {}
  {\bibfield  {journal} {\bibinfo  {journal} {arXiv preprint arXiv:1711.02249}\
  } (\bibinfo {year} {2017})}\BibitemShut {NoStop}%
\bibitem [{\citenamefont {Khatri}\ \emph {et~al.}(2019)\citenamefont {Khatri},
  \citenamefont {LaRose}, \citenamefont {Poremba}, \citenamefont {Cincio},
  \citenamefont {Sornborger},\ and\ \citenamefont {Coles}}]{khatri2018quantum}%
  \BibitemOpen
  \bibfield  {author} {\bibinfo {author} {\bibfnamefont {Sumeet}\ \bibnamefont
  {Khatri}}, \bibinfo {author} {\bibfnamefont {Ryan}\ \bibnamefont {LaRose}},
  \bibinfo {author} {\bibfnamefont {Alexander}\ \bibnamefont {Poremba}},
  \bibinfo {author} {\bibfnamefont {Lukasz}\ \bibnamefont {Cincio}}, \bibinfo
  {author} {\bibfnamefont {Andrew~T.}\ \bibnamefont {Sornborger}}, \ and\
  \bibinfo {author} {\bibfnamefont {Patrick~J.}\ \bibnamefont {Coles}},\
  }\bibfield  {title} {\enquote {\bibinfo {title} {Quantum-assisted quantum
  compiling},}\ }\href {\doibase 10.22331/q-2019-05-13-140} {\bibfield
  {journal} {\bibinfo  {journal} {{Quantum}}\ }\textbf {\bibinfo {volume}
  {3}},\ \bibinfo {pages} {140} (\bibinfo {year} {2019})}\BibitemShut {NoStop}%
\bibitem [{\citenamefont {Schuld}\ \emph {et~al.}(2017)\citenamefont {Schuld},
  \citenamefont {Fingerhuth},\ and\ \citenamefont
  {Petruccione}}]{Schuld2017implementing}%
  \BibitemOpen
  \bibfield  {author} {\bibinfo {author} {\bibfnamefont {Maria}\ \bibnamefont
  {Schuld}}, \bibinfo {author} {\bibfnamefont {M}~\bibnamefont {Fingerhuth}}, \
  and\ \bibinfo {author} {\bibfnamefont {F}~\bibnamefont {Petruccione}},\
  }\bibfield  {title} {\enquote {\bibinfo {title} {Implementing a
  distance-based classifier with a quantum interference circuit},}\ }\href
  {\doibase 10.1209/0295-5075/119/60002} {\bibfield  {journal} {\bibinfo
  {journal} {Europhys. Lett.}\ }\textbf {\bibinfo {volume} {119}} (\bibinfo
  {year} {2017}),\ 10.1209/0295-5075/119/60002}\BibitemShut {NoStop}%
\bibitem [{\citenamefont {Rist{\`e}}\ \emph {et~al.}(2017)\citenamefont
  {Rist{\`e}}, \citenamefont {da~Silva}, \citenamefont {Ryan}, \citenamefont
  {Cross}, \citenamefont {C{\'o}rcoles}, \citenamefont {Smolin}, \citenamefont
  {Gambetta}, \citenamefont {Chow},\ and\ \citenamefont
  {Johnson}}]{riste2017demonstration}%
  \BibitemOpen
  \bibfield  {author} {\bibinfo {author} {\bibfnamefont {Diego}\ \bibnamefont
  {Rist{\`e}}}, \bibinfo {author} {\bibfnamefont {Marcus~P}\ \bibnamefont
  {da~Silva}}, \bibinfo {author} {\bibfnamefont {Colm~A}\ \bibnamefont {Ryan}},
  \bibinfo {author} {\bibfnamefont {Andrew~W}\ \bibnamefont {Cross}}, \bibinfo
  {author} {\bibfnamefont {Antonio~D}\ \bibnamefont {C{\'o}rcoles}}, \bibinfo
  {author} {\bibfnamefont {John~A}\ \bibnamefont {Smolin}}, \bibinfo {author}
  {\bibfnamefont {Jay~M}\ \bibnamefont {Gambetta}}, \bibinfo {author}
  {\bibfnamefont {Jerry~M}\ \bibnamefont {Chow}}, \ and\ \bibinfo {author}
  {\bibfnamefont {Blake~R}\ \bibnamefont {Johnson}},\ }\bibfield  {title}
  {\enquote {\bibinfo {title} {Demonstration of quantum advantage in machine
  learning},}\ }\href {\doibase 10.1038/s41534-017-0017-3} {\bibfield
  {journal} {\bibinfo  {journal} {npj Quantum Information}\ }\textbf {\bibinfo
  {volume} {3}},\ \bibinfo {pages} {16} (\bibinfo {year} {2017})}\BibitemShut
  {NoStop}%
\bibitem [{\citenamefont {Aleksandrowicz}\ \emph {et~al.}(2019)\citenamefont
  {Aleksandrowicz}, \citenamefont {Alexander}, \citenamefont {Barkoutsos},
  \citenamefont {Bello}, \citenamefont {Ben-Haim}, \citenamefont {Bucher},
  \citenamefont {Cabrera-Hern{\'a}dez}, \citenamefont {Carballo-Franquis},
  \citenamefont {Chen}, \citenamefont {Chen}, \citenamefont {Chow},
  \citenamefont {C{\'o}rcoles-Gonzales}, \citenamefont {Cross}, \citenamefont
  {Cross}, \citenamefont {Cruz-Benito}, \citenamefont {Culver}, \citenamefont
  {Gonz{\'a}lez}, \citenamefont {Torre}, \citenamefont {Ding}, \citenamefont
  {Dumitrescu}, \citenamefont {Duran}, \citenamefont {Eendebak}, \citenamefont
  {Everitt}, \citenamefont {Sertage}, \citenamefont {Frisch}, \citenamefont
  {Fuhrer}, \citenamefont {Gambetta}, \citenamefont {Gago}, \citenamefont
  {Gomez-Mosquera}, \citenamefont {Greenberg}, \citenamefont {Hamamura},
  \citenamefont {Havl{\'\i}{\v{c}}ek}, \citenamefont {Hellmers}, \citenamefont
  {Herok}, \citenamefont {Horii}, \citenamefont {Hu}, \citenamefont {Imamichi},
  \citenamefont {Itoko}, \citenamefont {Javadi-Abhari}, \citenamefont
  {Kanazawa}, \citenamefont {Karazeev}, \citenamefont {Krsulich}, \citenamefont
  {Liu}, \citenamefont {Luh}, \citenamefont {Maeng}, \citenamefont {Marques},
  \citenamefont {Mart{\'\i}n-Fern{\'a}ndez}, \citenamefont {McClure},
  \citenamefont {McKay}, \citenamefont {Meesala}, \citenamefont {Mezzacapo},
  \citenamefont {Moll}, \citenamefont {Rodr{\'\i}guez}, \citenamefont
  {Nannicini}, \citenamefont {Nation}, \citenamefont {Ollitrault},
  \citenamefont {O'Riordan}, \citenamefont {Paik}, \citenamefont {P{\'e}rez},
  \citenamefont {Phan}, \citenamefont {Pistoia}, \citenamefont {Prutyanov},
  \citenamefont {Reuter}, \citenamefont {Rice}, \citenamefont {Davila},
  \citenamefont {Rudy}, \citenamefont {Ryu}, \citenamefont {Sathaye},
  \citenamefont {Schnabel}, \citenamefont {Schoute}, \citenamefont {Setia},
  \citenamefont {Shi}, \citenamefont {Silva}, \citenamefont {Siraichi},
  \citenamefont {Sivarajah}, \citenamefont {Smolin}, \citenamefont {Soeken},
  \citenamefont {Takahashi}, \citenamefont {Tavernelli}, \citenamefont
  {Taylor}, \citenamefont {Taylour}, \citenamefont {Trabing}, \citenamefont
  {Treinish}, \citenamefont {Turner}, \citenamefont {Vogt-Lee}, \citenamefont
  {Vuillot}, \citenamefont {Wildstrom}, \citenamefont {Wilson}, \citenamefont
  {Winston}, \citenamefont {Wood}, \citenamefont {Wood}, \citenamefont
  {W{\"o}rner}, \citenamefont {Akhalwaya},\ and\ \citenamefont
  {Zoufal}}]{qiskit}%
  \BibitemOpen
  \bibfield  {author} {\bibinfo {author} {\bibfnamefont {Gadi}\ \bibnamefont
  {Aleksandrowicz}}, \bibinfo {author} {\bibfnamefont {Thomas}\ \bibnamefont
  {Alexander}}, \bibinfo {author} {\bibfnamefont {Panagiotis}\ \bibnamefont
  {Barkoutsos}}, \bibinfo {author} {\bibfnamefont {Luciano}\ \bibnamefont
  {Bello}}, \bibinfo {author} {\bibfnamefont {Yael}\ \bibnamefont {Ben-Haim}},
  \bibinfo {author} {\bibfnamefont {David}\ \bibnamefont {Bucher}}, \bibinfo
  {author} {\bibfnamefont {Francisco~Jose}\ \bibnamefont
  {Cabrera-Hern{\'a}dez}}, \bibinfo {author} {\bibfnamefont {Jorge}\
  \bibnamefont {Carballo-Franquis}}, \bibinfo {author} {\bibfnamefont {Adrian}\
  \bibnamefont {Chen}}, \bibinfo {author} {\bibfnamefont {Chun-Fu}\
  \bibnamefont {Chen}}, \bibinfo {author} {\bibfnamefont {Jerry~M.}\
  \bibnamefont {Chow}}, \bibinfo {author} {\bibfnamefont {Antonio~D.}\
  \bibnamefont {C{\'o}rcoles-Gonzales}}, \bibinfo {author} {\bibfnamefont
  {Abigail~J.}\ \bibnamefont {Cross}}, \bibinfo {author} {\bibfnamefont
  {Andrew}\ \bibnamefont {Cross}}, \bibinfo {author} {\bibfnamefont {Juan}\
  \bibnamefont {Cruz-Benito}}, \bibinfo {author} {\bibfnamefont {Chris}\
  \bibnamefont {Culver}}, \bibinfo {author} {\bibfnamefont {Salvador De
  La~Puente}\ \bibnamefont {Gonz{\'a}lez}}, \bibinfo {author} {\bibfnamefont
  {Enrique De~La}\ \bibnamefont {Torre}}, \bibinfo {author} {\bibfnamefont
  {Delton}\ \bibnamefont {Ding}}, \bibinfo {author} {\bibfnamefont {Eugene}\
  \bibnamefont {Dumitrescu}}, \bibinfo {author} {\bibfnamefont {Ivan}\
  \bibnamefont {Duran}}, \bibinfo {author} {\bibfnamefont {Pieter}\
  \bibnamefont {Eendebak}}, \bibinfo {author} {\bibfnamefont {Mark}\
  \bibnamefont {Everitt}}, \bibinfo {author} {\bibfnamefont {Ismael~Faro}\
  \bibnamefont {Sertage}}, \bibinfo {author} {\bibfnamefont {Albert}\
  \bibnamefont {Frisch}}, \bibinfo {author} {\bibfnamefont {Andreas}\
  \bibnamefont {Fuhrer}}, \bibinfo {author} {\bibfnamefont {Jay}\ \bibnamefont
  {Gambetta}}, \bibinfo {author} {\bibfnamefont {Borja~Godoy}\ \bibnamefont
  {Gago}}, \bibinfo {author} {\bibfnamefont {Juan}\ \bibnamefont
  {Gomez-Mosquera}}, \bibinfo {author} {\bibfnamefont {Donny}\ \bibnamefont
  {Greenberg}}, \bibinfo {author} {\bibfnamefont {Ikko}\ \bibnamefont
  {Hamamura}}, \bibinfo {author} {\bibfnamefont {Vojt{\v{e}}ch}\ \bibnamefont
  {Havl{\'\i}{\v{c}}ek}}, \bibinfo {author} {\bibfnamefont {Joe}\ \bibnamefont
  {Hellmers}}, \bibinfo {author} {\bibfnamefont {{\L}ukasz}\ \bibnamefont
  {Herok}}, \bibinfo {author} {\bibfnamefont {Hiroshi}\ \bibnamefont {Horii}},
  \bibinfo {author} {\bibfnamefont {Shaohan}\ \bibnamefont {Hu}}, \bibinfo
  {author} {\bibfnamefont {Takashi}\ \bibnamefont {Imamichi}}, \bibinfo
  {author} {\bibfnamefont {Toshinari}\ \bibnamefont {Itoko}}, \bibinfo {author}
  {\bibfnamefont {Ali}\ \bibnamefont {Javadi-Abhari}}, \bibinfo {author}
  {\bibfnamefont {Naoki}\ \bibnamefont {Kanazawa}}, \bibinfo {author}
  {\bibfnamefont {Anton}\ \bibnamefont {Karazeev}}, \bibinfo {author}
  {\bibfnamefont {Kevin}\ \bibnamefont {Krsulich}}, \bibinfo {author}
  {\bibfnamefont {Peng}\ \bibnamefont {Liu}}, \bibinfo {author} {\bibfnamefont
  {Yang}\ \bibnamefont {Luh}}, \bibinfo {author} {\bibfnamefont {Yunho}\
  \bibnamefont {Maeng}}, \bibinfo {author} {\bibfnamefont {Manoel}\
  \bibnamefont {Marques}}, \bibinfo {author} {\bibfnamefont {Francisco~Jose}\
  \bibnamefont {Mart{\'\i}n-Fern{\'a}ndez}}, \bibinfo {author} {\bibfnamefont
  {Douglas~T.}\ \bibnamefont {McClure}}, \bibinfo {author} {\bibfnamefont
  {David}\ \bibnamefont {McKay}}, \bibinfo {author} {\bibfnamefont {Srujan}\
  \bibnamefont {Meesala}}, \bibinfo {author} {\bibfnamefont {Antonio}\
  \bibnamefont {Mezzacapo}}, \bibinfo {author} {\bibfnamefont {Nikolaj}\
  \bibnamefont {Moll}}, \bibinfo {author} {\bibfnamefont {Diego~Moreda}\
  \bibnamefont {Rodr{\'\i}guez}}, \bibinfo {author} {\bibfnamefont {Giacomo}\
  \bibnamefont {Nannicini}}, \bibinfo {author} {\bibfnamefont {Paul}\
  \bibnamefont {Nation}}, \bibinfo {author} {\bibfnamefont {Pauline}\
  \bibnamefont {Ollitrault}}, \bibinfo {author} {\bibfnamefont {Lee~James}\
  \bibnamefont {O'Riordan}}, \bibinfo {author} {\bibfnamefont {Hanhee}\
  \bibnamefont {Paik}}, \bibinfo {author} {\bibfnamefont {Jes{\'u}s}\
  \bibnamefont {P{\'e}rez}}, \bibinfo {author} {\bibfnamefont {Anna}\
  \bibnamefont {Phan}}, \bibinfo {author} {\bibfnamefont {Marco}\ \bibnamefont
  {Pistoia}}, \bibinfo {author} {\bibfnamefont {Viktor}\ \bibnamefont
  {Prutyanov}}, \bibinfo {author} {\bibfnamefont {Max}\ \bibnamefont {Reuter}},
  \bibinfo {author} {\bibfnamefont {Julia}\ \bibnamefont {Rice}}, \bibinfo
  {author} {\bibfnamefont {Abd{\'o}n~Rodr{\'\i}guez}\ \bibnamefont {Davila}},
  \bibinfo {author} {\bibfnamefont {Raymond Harry~Putra}\ \bibnamefont {Rudy}},
  \bibinfo {author} {\bibfnamefont {Mingi}\ \bibnamefont {Ryu}}, \bibinfo
  {author} {\bibfnamefont {Ninad}\ \bibnamefont {Sathaye}}, \bibinfo {author}
  {\bibfnamefont {Chris}\ \bibnamefont {Schnabel}}, \bibinfo {author}
  {\bibfnamefont {Eddie}\ \bibnamefont {Schoute}}, \bibinfo {author}
  {\bibfnamefont {Kanav}\ \bibnamefont {Setia}}, \bibinfo {author}
  {\bibfnamefont {Yunong}\ \bibnamefont {Shi}}, \bibinfo {author}
  {\bibfnamefont {Adenilton}\ \bibnamefont {Silva}}, \bibinfo {author}
  {\bibfnamefont {Yukio}\ \bibnamefont {Siraichi}}, \bibinfo {author}
  {\bibfnamefont {Seyon}\ \bibnamefont {Sivarajah}}, \bibinfo {author}
  {\bibfnamefont {John~A.}\ \bibnamefont {Smolin}}, \bibinfo {author}
  {\bibfnamefont {Mathias}\ \bibnamefont {Soeken}}, \bibinfo {author}
  {\bibfnamefont {Hitomi}\ \bibnamefont {Takahashi}}, \bibinfo {author}
  {\bibfnamefont {Ivano}\ \bibnamefont {Tavernelli}}, \bibinfo {author}
  {\bibfnamefont {Charles}\ \bibnamefont {Taylor}}, \bibinfo {author}
  {\bibfnamefont {Pete}\ \bibnamefont {Taylour}}, \bibinfo {author}
  {\bibfnamefont {Kenso}\ \bibnamefont {Trabing}}, \bibinfo {author}
  {\bibfnamefont {Matthew}\ \bibnamefont {Treinish}}, \bibinfo {author}
  {\bibfnamefont {Wes}\ \bibnamefont {Turner}}, \bibinfo {author}
  {\bibfnamefont {Desiree}\ \bibnamefont {Vogt-Lee}}, \bibinfo {author}
  {\bibfnamefont {Christophe}\ \bibnamefont {Vuillot}}, \bibinfo {author}
  {\bibfnamefont {Jonathan~A.}\ \bibnamefont {Wildstrom}}, \bibinfo {author}
  {\bibfnamefont {Jessica}\ \bibnamefont {Wilson}}, \bibinfo {author}
  {\bibfnamefont {Erick}\ \bibnamefont {Winston}}, \bibinfo {author}
  {\bibfnamefont {Christopher}\ \bibnamefont {Wood}}, \bibinfo {author}
  {\bibfnamefont {Stephen}\ \bibnamefont {Wood}}, \bibinfo {author}
  {\bibfnamefont {Stefan}\ \bibnamefont {W{\"o}rner}}, \bibinfo {author}
  {\bibfnamefont {Ismail~Yunus}\ \bibnamefont {Akhalwaya}}, \ and\ \bibinfo
  {author} {\bibfnamefont {Christa}\ \bibnamefont {Zoufal}},\ }\href {\doibase
  10.5281/zenodo.2562110} {\enquote {\bibinfo {title} {Qiskit: An open-source
  framework for quantum computing},}\ } (\bibinfo {year} {2019})\BibitemShut
  {NoStop}%
\bibitem [{\citenamefont {Bergholm}\ \emph {et~al.}(2018)\citenamefont
  {Bergholm}, \citenamefont {Izaac}, \citenamefont {Schuld}, \citenamefont
  {Gogolin},\ and\ \citenamefont {Killoran}}]{bergholm2018pennylane}%
  \BibitemOpen
  \bibfield  {author} {\bibinfo {author} {\bibfnamefont {Ville}\ \bibnamefont
  {Bergholm}}, \bibinfo {author} {\bibfnamefont {Josh}\ \bibnamefont {Izaac}},
  \bibinfo {author} {\bibfnamefont {Maria}\ \bibnamefont {Schuld}}, \bibinfo
  {author} {\bibfnamefont {Christian}\ \bibnamefont {Gogolin}}, \ and\ \bibinfo
  {author} {\bibfnamefont {Nathan}\ \bibnamefont {Killoran}},\ }\bibfield
  {title} {\enquote {\bibinfo {title} {Pennylane: Automatic differentiation of
  hybrid quantum-classical computations},}\ }\href@noop {} {\bibfield
  {journal} {\bibinfo  {journal} {arXiv preprint arXiv:1811.04968}\ } (\bibinfo
  {year} {2018})}\BibitemShut {NoStop}%
\bibitem [{\citenamefont {Luo}\ \emph {et~al.}(2018)\citenamefont {Luo},
  \citenamefont {Liu}, \citenamefont {Zhang},\ and\ \citenamefont
  {Wang}}]{Luo2018yao}%
  \BibitemOpen
  \bibfield  {author} {\bibinfo {author} {\bibfnamefont {Xiuzhe}\ \bibnamefont
  {Luo}}, \bibinfo {author} {\bibfnamefont {Jin-guo}\ \bibnamefont {Liu}},
  \bibinfo {author} {\bibfnamefont {Pan}\ \bibnamefont {Zhang}}, \ and\
  \bibinfo {author} {\bibfnamefont {Lei}\ \bibnamefont {Wang}},\ }\href@noop {}
  {\enquote {\bibinfo {title} {Yao},}\ }\bibinfo {howpublished}
  {\url{https://github.com/QuantumBFS/Yao.jl}} (\bibinfo {year}
  {2018})\BibitemShut {NoStop}%
\bibitem [{\citenamefont {Fingerhuth}\ \emph {et~al.}(2018)\citenamefont
  {Fingerhuth}, \citenamefont {Babej},\ and\ \citenamefont
  {Wittek}}]{fingerhuth2018open}%
  \BibitemOpen
  \bibfield  {author} {\bibinfo {author} {\bibfnamefont {Mark}\ \bibnamefont
  {Fingerhuth}}, \bibinfo {author} {\bibfnamefont {Tomáš}\ \bibnamefont
  {Babej}}, \ and\ \bibinfo {author} {\bibfnamefont {Peter}\ \bibnamefont
  {Wittek}},\ }\bibfield  {title} {\enquote {\bibinfo {title} {Open source
  software in quantum computing},}\ }\href {\doibase
  10.1371/journal.pone.0208561} {\bibfield  {journal} {\bibinfo  {journal}
  {PLOS ONE}\ }\textbf {\bibinfo {volume} {13}},\ \bibinfo {pages} {1--28}
  (\bibinfo {year} {2018})}\BibitemShut {NoStop}%
\bibitem [{\citenamefont {Farhi}\ and\ \citenamefont
  {Harrow}(2016)}]{farhi2016quantum}%
  \BibitemOpen
  \bibfield  {author} {\bibinfo {author} {\bibfnamefont {Edward}\ \bibnamefont
  {Farhi}}\ and\ \bibinfo {author} {\bibfnamefont {Aram~W}\ \bibnamefont
  {Harrow}},\ }\bibfield  {title} {\enquote {\bibinfo {title} {Quantum
  supremacy through the quantum approximate optimization algorithm},}\
  }\href@noop {} {\bibfield  {journal} {\bibinfo  {journal} {arXiv preprint
  arXiv:1602.07674}\ } (\bibinfo {year} {2016})}\BibitemShut {NoStop}%
\bibitem [{\citenamefont {Guerreschi}\ and\ \citenamefont
  {Matsuura}(2019)}]{guerreschi2018qaoa}%
  \BibitemOpen
  \bibfield  {author} {\bibinfo {author} {\bibfnamefont {G.~G.}\ \bibnamefont
  {Guerreschi}}\ and\ \bibinfo {author} {\bibfnamefont {A.~Y.}\ \bibnamefont
  {Matsuura}},\ }\bibfield  {title} {\enquote {\bibinfo {title} {Qaoa for
  max-cut requires hundreds of qubits for quantum speed-up},}\ }\href {\doibase
  10.1038/s41598-019-43176-9} {\bibfield  {journal} {\bibinfo  {journal}
  {Scientific Reports}\ }\textbf {\bibinfo {volume} {9}},\ \bibinfo {pages}
  {6903} (\bibinfo {year} {2019})}\BibitemShut {NoStop}%
\bibitem [{\citenamefont {Zhou}\ \emph {et~al.}(2018)\citenamefont {Zhou},
  \citenamefont {Wang}, \citenamefont {Choi}, \citenamefont {Pichler},\ and\
  \citenamefont {Lukin}}]{zhou2018quantum}%
  \BibitemOpen
  \bibfield  {author} {\bibinfo {author} {\bibfnamefont {Leo}\ \bibnamefont
  {Zhou}}, \bibinfo {author} {\bibfnamefont {Sheng-Tao}\ \bibnamefont {Wang}},
  \bibinfo {author} {\bibfnamefont {Soonwon}\ \bibnamefont {Choi}}, \bibinfo
  {author} {\bibfnamefont {Hannes}\ \bibnamefont {Pichler}}, \ and\ \bibinfo
  {author} {\bibfnamefont {Mikhail~D}\ \bibnamefont {Lukin}},\ }\bibfield
  {title} {\enquote {\bibinfo {title} {Quantum approximate optimization
  algorithm: Performance, mechanism, and implementation on near-term
  devices},}\ }\href@noop {} {\bibfield  {journal} {\bibinfo  {journal} {arXiv
  preprint arXiv:1812.01041}\ } (\bibinfo {year} {2018})}\BibitemShut {NoStop}%
\bibitem [{\citenamefont {Crooks}(2018)}]{crooks2018performance}%
  \BibitemOpen
  \bibfield  {author} {\bibinfo {author} {\bibfnamefont {Gavin~E}\ \bibnamefont
  {Crooks}},\ }\bibfield  {title} {\enquote {\bibinfo {title} {Performance of
  the quantum approximate optimization algorithm on the maximum cut problem},}\
  }\href@noop {} {\bibfield  {journal} {\bibinfo  {journal} {arXiv preprint
  arXiv:1811.08419}\ } (\bibinfo {year} {2018})}\BibitemShut {NoStop}%
\bibitem [{\citenamefont {Tangpanitanon}\ \emph {et~al.}(2019)\citenamefont
  {Tangpanitanon}, \citenamefont {Thanasilp}, \citenamefont {Lemonde},\ and\
  \citenamefont {Angelakis}}]{tangpanitanon2019quantum}%
  \BibitemOpen
  \bibfield  {author} {\bibinfo {author} {\bibfnamefont {Jirawat}\ \bibnamefont
  {Tangpanitanon}}, \bibinfo {author} {\bibfnamefont {Supanut}\ \bibnamefont
  {Thanasilp}}, \bibinfo {author} {\bibfnamefont {Marc-Antoine}\ \bibnamefont
  {Lemonde}}, \ and\ \bibinfo {author} {\bibfnamefont {Dimitris~G}\
  \bibnamefont {Angelakis}},\ }\bibfield  {title} {\enquote {\bibinfo {title}
  {Quantum supremacy with analog quantum processors for material science and
  machine learning},}\ }\href@noop {} {\bibfield  {journal} {\bibinfo
  {journal} {arXiv preprint arXiv:1906.03860}\ } (\bibinfo {year}
  {2019})}\BibitemShut {NoStop}%
\bibitem [{\citenamefont {Jing}\ \emph {et~al.}(2017)\citenamefont {Jing},
  \citenamefont {Shen}, \citenamefont {Dubcek}, \citenamefont {Peurifoy},
  \citenamefont {Skirlo}, \citenamefont {LeCun}, \citenamefont {Tegmark},\ and\
  \citenamefont {Solja{\v{c}}i{\'c}}}]{jing2017tunable}%
  \BibitemOpen
  \bibfield  {author} {\bibinfo {author} {\bibfnamefont {Li}~\bibnamefont
  {Jing}}, \bibinfo {author} {\bibfnamefont {Yichen}\ \bibnamefont {Shen}},
  \bibinfo {author} {\bibfnamefont {Tena}\ \bibnamefont {Dubcek}}, \bibinfo
  {author} {\bibfnamefont {John}\ \bibnamefont {Peurifoy}}, \bibinfo {author}
  {\bibfnamefont {Scott}\ \bibnamefont {Skirlo}}, \bibinfo {author}
  {\bibfnamefont {Yann}\ \bibnamefont {LeCun}}, \bibinfo {author}
  {\bibfnamefont {Max}\ \bibnamefont {Tegmark}}, \ and\ \bibinfo {author}
  {\bibfnamefont {Marin}\ \bibnamefont {Solja{\v{c}}i{\'c}}},\ }\bibfield
  {title} {\enquote {\bibinfo {title} {Tunable efficient unitary neural
  networks (eunn) and their application to rnns},}\ }in\ \href
  {http://dl.acm.org/citation.cfm?id=3305381.3305560} {\emph {\bibinfo
  {booktitle} {Proceedings of the 34th International Conference on Machine
  Learning-Volume 70}}}\ (\bibinfo {organization} {JMLR. org},\ \bibinfo {year}
  {2017})\ pp.\ \bibinfo {pages} {1733--1741}\BibitemShut {NoStop}%
\bibitem [{\citenamefont {Hyland}\ and\ \citenamefont
  {R{\"a}tsch}(2017)}]{hyland2017learning}%
  \BibitemOpen
  \bibfield  {author} {\bibinfo {author} {\bibfnamefont {Stephanie~L}\
  \bibnamefont {Hyland}}\ and\ \bibinfo {author} {\bibfnamefont {Gunnar}\
  \bibnamefont {R{\"a}tsch}},\ }\bibfield  {title} {\enquote {\bibinfo {title}
  {Learning unitary operators with help from u (n)},}\ }in\ \href@noop {}
  {\emph {\bibinfo {booktitle} {Thirty-First AAAI Conference on Artificial
  Intelligence}}}\ (\bibinfo {year} {2017})\BibitemShut {NoStop}%
\bibitem [{\citenamefont {Wang}\ \emph {et~al.}(2019)\citenamefont {Wang},
  \citenamefont {Higgott},\ and\ \citenamefont
  {Brierley}}]{wang2018generalised}%
  \BibitemOpen
  \bibfield  {author} {\bibinfo {author} {\bibfnamefont {Daochen}\ \bibnamefont
  {Wang}}, \bibinfo {author} {\bibfnamefont {Oscar}\ \bibnamefont {Higgott}}, \
  and\ \bibinfo {author} {\bibfnamefont {Stephen}\ \bibnamefont {Brierley}},\
  }\bibfield  {title} {\enquote {\bibinfo {title} {Accelerated variational
  quantum eigensolver},}\ }\href {\doibase 10.1103/PhysRevLett.122.140504}
  {\bibfield  {journal} {\bibinfo  {journal} {Phys. Rev. Lett.}\ }\textbf
  {\bibinfo {volume} {122}},\ \bibinfo {pages} {140504} (\bibinfo {year}
  {2019})}\BibitemShut {NoStop}%
\bibitem [{\citenamefont {Garcia-Escartin}\ and\ \citenamefont
  {Chamorro-Posada}(2013)}]{garcia2013swap}%
  \BibitemOpen
  \bibfield  {author} {\bibinfo {author} {\bibfnamefont {Juan~Carlos}\
  \bibnamefont {Garcia-Escartin}}\ and\ \bibinfo {author} {\bibfnamefont
  {Pedro}\ \bibnamefont {Chamorro-Posada}},\ }\bibfield  {title} {\enquote
  {\bibinfo {title} {Swap test and hong-ou-mandel effect are equivalent},}\
  }\href {\doibase 10.1103/PhysRevA.87.052330} {\bibfield  {journal} {\bibinfo
  {journal} {Physical Review A}\ }\textbf {\bibinfo {volume} {87}},\ \bibinfo
  {pages} {052330} (\bibinfo {year} {2013})}\BibitemShut {NoStop}%
\bibitem [{\citenamefont {Cowtan}\ \emph {et~al.}(2019)\citenamefont {Cowtan},
  \citenamefont {Dilkes}, \citenamefont {Duncan}, \citenamefont {Krajenbrink},
  \citenamefont {Simmons},\ and\ \citenamefont {Sivarajah}}]{cowtan2019qubit}%
  \BibitemOpen
  \bibfield  {author} {\bibinfo {author} {\bibfnamefont {Alexander}\
  \bibnamefont {Cowtan}}, \bibinfo {author} {\bibfnamefont {Silas}\
  \bibnamefont {Dilkes}}, \bibinfo {author} {\bibfnamefont {Ross}\ \bibnamefont
  {Duncan}}, \bibinfo {author} {\bibfnamefont {Alexandre}\ \bibnamefont
  {Krajenbrink}}, \bibinfo {author} {\bibfnamefont {Will}\ \bibnamefont
  {Simmons}}, \ and\ \bibinfo {author} {\bibfnamefont {Seyon}\ \bibnamefont
  {Sivarajah}},\ }\bibfield  {title} {\enquote {\bibinfo {title} {On the qubit
  routing problem},}\ }\href@noop {} {\bibfield  {journal} {\bibinfo  {journal}
  {arXiv preprint arXiv:1902.08091}\ } (\bibinfo {year} {2019})}\BibitemShut
  {NoStop}%
\bibitem [{\citenamefont {Iten}\ \emph {et~al.}(2019)\citenamefont {Iten},
  \citenamefont {Reardon-Smith}, \citenamefont {Mondada}, \citenamefont
  {Redmond}, \citenamefont {Kohli},\ and\ \citenamefont
  {Colbeck}}]{iten2019introduction}%
  \BibitemOpen
  \bibfield  {author} {\bibinfo {author} {\bibfnamefont {Raban}\ \bibnamefont
  {Iten}}, \bibinfo {author} {\bibfnamefont {Oliver}\ \bibnamefont
  {Reardon-Smith}}, \bibinfo {author} {\bibfnamefont {Luca}\ \bibnamefont
  {Mondada}}, \bibinfo {author} {\bibfnamefont {Ethan}\ \bibnamefont
  {Redmond}}, \bibinfo {author} {\bibfnamefont {Ravjot~Singh}\ \bibnamefont
  {Kohli}}, \ and\ \bibinfo {author} {\bibfnamefont {Roger}\ \bibnamefont
  {Colbeck}},\ }\bibfield  {title} {\enquote {\bibinfo {title} {Introduction to
  universalqcompiler},}\ }\href@noop {} {\bibfield  {journal} {\bibinfo
  {journal} {arXiv preprint arXiv:1904.01072}\ } (\bibinfo {year}
  {2019})}\BibitemShut {NoStop}%
\end{thebibliography}
\end{document}